# Effectful Programming in Declarative Languages with an Emphasis on Non–Determinism: Applications and Formal Reasoning

Sandra Dylus

*October, 2019*

# Effectful Programming in Declarative Languages with an Emphasis on Non–Determinism: Applications and Formal Reasoning

Sandra Dylus

# Abstract

This thesis investigates effectful declarative programming with an emphasis on non–determinism as an effect.

On the one hand, we are interested in developing applications using non–determinism as underlying implementation idea. We discuss two applications using the functional logic programming language Curry. The key idea of these implementations is to exploit the interplay of non–determinism and non–strictness that Curry employs.

The first application investigates sorting algorithms parametrised over a comparison function. By applying a non–deterministic predicate to these sorting functions, we gain a permutation enumeration function. We compare the implementation in Curry with an implementation in Haskell that uses a monadic interface to model non–determinism.

The other application that we discuss in this work is a library for probabilistic programming. Instead of modelling distributions as list of event and probability pairs, we model distributions using Curry's built–in non–determinism.

In both cases we observe that the combination of non–determinism and non–strictness has advantages over an implementation using lists to model non–determinism.

On the other hand, we present an idea to apply formal reasoning on effectful declarative programming languages. In order to start with simple effects, we focus on modelling a functional subset first. That is, the effects of interest are totality and partiality. We then observe that the general scheme to model these two effects can be generalised to capture a wide range of effects. Obviously, the next step is to apply the idea to model non–determinism. More precisely, we implement a model for the non–determinism of Curry: non–strict non–determinism with call–time choice. Therefore, we finally discuss why the current representation models call–by–name rather than Curry's call–by–need semantics and give an outlook on ideas to tackle this problem.

# Zusammenfassung (deutsch)

Diese Arbeit beschäftigt sich mit der deklarativen Programmierung mit Effekten und legt dabei besonderen Fokus auf Nichtdeterminismus als Effekt.

Einerseits möchten wir Anwendungen entwickeln, deren zugrundeliegende Implementierungsidee auf Nichtdeterminismus basiert. Wir stellen dazu zwei beispielhafte Anwendungen vor, die in der funktional logischen Programmiersprache Curry implementiert sind. Die Kernidee dieser Implementierungen ist dabei die Kombination von Nichtstriktheit und Nichtdeterminismus, die Curry unterliegen, gewinnbringend auszunutzen.

Für die erste Anwendung untersuchen wir Sortierfunktionen, die über eine Vergleichsfunktion parametrisiert sind, und wenden diese Funktionen auf ein nichtdeterministisches Prädikat an. Dabei entsteht eine Funktion, die Permutationen der Eingabeliste berechnet. Wir vergleichen unsere Implementierung in Curry mit einer Implementierung in Haskell, die den Nichtdeterminismus monadisch modelliert.

Als zweite Anwendung werden wir über eine Bibliothek zur probabilistischen Programmierung diskutieren. Statt der üblichen Modellierung von Wahrscheinlichkeitsverteilungen als Liste von Paaren von Ereignis- und korrespondierenden Wahrscheinlichkeitswerten modellieren wir diese Verteilungen mithilfe von Currys nativem Nichtdeterminismus.

Beide Implementierungen haben durch die Kombination von Nichtdeterminismus und Nichtstriktheit Vorteile gegenüber einer Implementierung, die den Nichtdeterminismus durch Listen repräsentiert.

Andererseits möchten wir eine Möglichkeit schaffen, über die Programme, die wir in effektbehafteten deklarativen Programmiersprachen entwickelt haben, in einem formalen Rahmen zu argumentieren. Dabei fangen wir mit der Teilmenge der rein funktionalen Effekte an, das heißt, wir interessieren uns zunächst für totale und partielle Programme. Die zugrundeliegende Idee zur Modellierung dieser zwei Effekte kann dann auch für weitere Effekte genutzt werden. Als natürlichen nächsten Schritt betrachten wir den Effekt, der bei der Sprache Curry zusätzlich hinzukommt: nicht–strikter Nichtdeterminismus mit *call–time choice* Semantik. Dabei geben wir eine Übersicht darüber, warum die aktuelle Repräsentation *call–by–name* modelliert, sowie erste Ideen, wie die für Curry erforderliche *call–by–need* Semantik modelliert werden könnte.

# Acknowledgement

I am thankful that I could be a member of the group on *Programming Languages and Compiler Construction* led by my advisor Michael Hanus for the last five and a half years.

My most profound thanks deserves Jan for having the greatest impact in my growth as a researcher and as a person. We can look back at 5 years of intensive research corporations including all the highs and lows our profession comes with: paper rejections and acceptances remain stable, some ideas worked out and some did not. In the end, a problem shared is a problem halved, and an achievement shared is a delight doubled. During my first years as PhD student I have a lot of great memories working with Jan and Nikita. The two years I shared my working environment with Nikita, he was my main contact for all occasions at work: lunch, Haskell, teaching, and research in general.

As usually, when great colleagues leave, there are new people coming into your life. In this respect, I thank my colleagues Finn and Niels for keeping me sane, especially in the final year of my PhD. It is with first and foremost joy and appreciation that I look back at our joint work on papers and discussions about research topics in general. I also loved hanging out with you at our lunch and occasional coffee breaks, which we shared with Johannes, Christian and Frank. Moreover, I am grateful for the pleasant working environment on the *7th floor* in general, including the people not mentioned so far: JR, Ina, Renate, Rudolf, and Ulrike as well as the Archaeoinformatics group.

My special thanks to Julia, Jan (*Herr Bracker*), Ole, Marcellus, Kai, Nikita, Niels, and Finn for reading drafts of this thesis and providing me with helpful feedback to improve my writing. And I also like to thank Insa for giving me insights of her work using the proof assistant Coq that led to my own research interest in that area. Additionally, I was glad to have Janis Voigtländer and Thomas Wilke on my examination committee and am, especially, thankful for Janis' expert comments on the topics of this thesis.

I hope it is without any doubt that I enjoyed teaching a lot of students the ideas of functional programming and I am especially proud of all the students, whose thesis I advised: you all did a great job and gave me the perfect opportunity to deepen my understanding of my own research.

Last but not least, the aforementioned coffee breaks would not have been the same without the great team at the *Café Lounge*: it was with great pleasure that I could order my chai latte with oat milk just saying "The usual, please".

# Contents

# Introduction

<div style="text-align: right;">**1**</div>

> *And quickly they will observe that functional programming elegantly admits solutions that are very hard (or impossible) to formulate with the programming vehicle of their high school days.*
>
> — **Edsgar W. Dijkstra**
> (On why to choose Haskell over Java)

Many features that are most commonly associated with functional programming gain presence in mainstream as well as new programming languages in particular. Version 5 of Java (Arnold et al., 2005) brings polymorphism in the form of generics to an object–oriented language that is heavily used in industry; version 8 introduces a more convenient way to write lambda functions. New programming languages like Elm (Czaplicki, 2012) or Reason (*ReasonML* 2019), which are heavily influenced by the functional languages Haskell (Peyton Jones, 2002) and OCaml (Minsky et al., 2013), respectively, fill the gap concerning statically typed, functional languages for web front–end developers. Over the last decades, programmers seek the comfort of a compiler that warns about potentially bad programs before actually running them. Furthermore, they long for abstractions to simplify and generalise their code basis in order to increase the stability and maintainability of their software products. Such abstractions, like higher–order functions, algebraic data types, polymorphism and purity, are highly regarded in the functional programming language community since the introduction of ML in the 70s.

Among functional languages, the superset of declarative languages also includes the paradigm of logic programming. Features associated with logic programming, like non–determinism, free variables and constraint solving, have not caught on mainstream languages such as the functional features we mentioned above — yet. One part of this thesis illustrates the advantages of a functional logic programming language, which combines both of these paradigms, over encoding logic features in a functional language. These advantages we discuss in this thesis are specific to non–strict non–determinism. The first application considers permutation enumeration functions. More precisely, these enumeration functions arise from sorting functions that are parametrised by a comparison function. The trick is to apply such sorting functions to a non–deterministic comparison function that produces all possible rearrangements of the input list, leading to a permutation enumeration function. The advantages of the functional logic programming language become apparent when comparing the implementation against an implementation in a functional language that uses lists to model the non–determinism. The second application is an implementation of a library for probabilistic programming. The key ingredient of probabilistic programs are distributions that are most commonly modeled as pairs of events and their corresponding probabilities. Users of probabilistic programming languages query these distributions in order to get the probability of a subset of the events. An implementation that represents distributions using non–strict non–determinism can have advantages with respect to these queries. These advandates with respect to non-strict non-determinism are already known from property–based testing (Christiansen and Fischer, 2008; Runciman et al., 2008).

Moreover, pure functional programming languages are said to be particularly well–suited for equational reasoning. Most equational reasoning efforts do not bother about disruptive properties like partiality that may occur in programs we want to prove properties about. Partial programs are usually allowed in both of these paradigms and complicate equational reasoning. In case of partiality, we can argue that a total subset is reasonable, because functional programmers are interested in preventing the possibility of run–time errors anyway. In case of a functional logic language, however, partiality comes in a different flavour: non–determinism consists of a non–deterministic choice operator as well as a failure value that behaves as neutral element with respect to the choice operator. More precisely, we would exclude all programs except the ones that would also compile in a functional language. Whereas the restriction concerning partiality in the context of a functional language might be acceptable, it is rather radical to exclude non–deterministic programs when we want to reason about a functional logic language. As the second part of this thesis, we discuss how to model non–strict functional languages with effects like partiality and non–determinism in a proof assistant in order to apply equational reasoning in a computer–assisted approach.

The thesis is structured as follows. We give an introduction to advanced concepts of functional and functional logic programming as well as a beginner–friendly introduction to a dependently typed language we later use as the proof assistance of choice for the formal reasoning (Chapter 2). The advantages of non–strict non–determinism are illustrated using the example of two applications: implementing a permutation enumeration function by means of a sorting function that is applied to a non–determinism predicate (Chapter 3), and developing a library for probabilistic programming that relies on non–determinism instead of lists as internal structure (Chapter 4). We then switch the focus from applications to formal reasoning and present a framework to model non–strict functional programs that come with effects like partiality and non–determinism as well as prove exemplary properties about programs using the framework (Chapter 5). Finally, we give a summary of the presented work (Chapter 6) and collect additional source code in the appendix.

As some contributions presented in this thesis have been already published, the corresponding chapters (Chapter 3–5) end with information about the previously published work. Furthermore, we highlight the changes and additions with respect to the published version, and present related work as well as a summary for each chapter individually.

# Declarative Programming    2

The declarative programming paradigm allows the user to concentrate on finding a solution to a problem without explicitly stating the control–flow of the program but rather specifying which sub–problems to solve first. Representatives of declarative programming are functional and logic programming languages.

The first group of languages consists of functions in the mathematical sense: input arguments uniquely determine the result. This property is especially interesting in the context of purely functional programming languages and called *referential transparency*: every function call with the same arguments always evaluates to the same result regardless of the context and concrete order of evaluation (Horowitz, 1983). A pure programming language does not have any side effects like mutable state, that is, variables are only abbreviations for expressions and not references that are manipulated by updates or other modifications. In this thesis, all declarative programming languages we consider are pure and also statically typed. That is, every expression has a type, which is checked at compile time and does not change within the course of evaluating the overall program.

In this thesis we work with three different languages: Haskell (Peyton Jones, 2002) as representative for a functional language, Curry (Hanus et al., 1995) for functional logic programming, and Coq (Barras et al., 1997) as dependently typed language and representative for interactive proof assistants. We expect the reader to be familiar with the basics of Haskell in order to directly start with more advanced features in the next section. Topics we cover include Haskell's demand–driven evaluation strategy and monadic abstractions. We then move over to the integration of logic features that is Curry; here, the combination of non–determinism and laziness is especially interesting. Lastly, we will take a look at a richer type system for functional programming using the example of the interactive proof assistant Coq. After a quick introduction to its syntax, we will discuss how to use the dependent type system to state and prove properties about functional programs.

## 2.1 Functional Programming

We present all concepts related to pure functional programming in this thesis using the language Haskell. As we assume a basic familiarity of the reader regarding functional programming in Haskell, we will focus on specific and advanced aspects we will make use of. For a more detailed introduction to Haskell, we recommend interested readers to take a look at other sources (Hudak et al., 2007; Hutton, 2016).

First, we illustrate the advantages and subtleties of Haskell's non–strict and especially lazy evaluation strategy using a handful of examples. Next, we show how to work with side effects and other effectful operations that are not allowed otherwise due to Haskell's purity. In this matter, we discuss how to model such effects using monadic abstraction. More precisely, we illustrate how to model partiality and non–determinism using monads. Finally, we generalise the monadic abstraction to use free monads instead, a representation that we will make use of in different parts of this thesis.

If not explicitly stated differently, we use GHC 8.4.3 to compile and run the presented Haskell code. We display the interaction with GHC's REPL using a prompt showing a lambda — λ>— at the start of each command.

### 2.1.1 Non–strictness and Laziness

Haskell's evaluation strategy is call–by–need (Ariola and Felleisen, 1997). The strategy evaluates sub–expressions only when explicitly needed and shared expressions only once. That is, call–by–need combines the advantages of both, call–by–name and call–by–value. Call–by–name semantics behaves non–strict and evaluates expressions only when needed; call–by–value semantics corresponds to a strict evaluation, having the advantage that it evaluates expressions only once. The combination of non–strictness and sharing, which Haskell employs, is often called lazy evaluation.

In order to demonstrate the non–strictness part of Haskell's lazy evaluation, we use the following definition of `head` to project the first element of a list.

```haskell
head :: [a] -> a
head []      = undefined
head (x : _) = x
```

Let us compute the head of a partial list: the head element is defined but the remaining list is not. In Haskell the value `undefined` represents a partial value that produces a run–time error, when evaluation demands the value.

```
λ> head (1 : undefined)
1
```

Non–strictness allows us to work on partial values and, more importantly, and does not evaluate non–demanded sub-expressions. The demand–driven evaluation comes into play not only in case of partial values, but also in case of expensive computations.

The next example uses a function that computes the factorial of a given number as representative of such an expensive computation, and the function `const :: a -> b -> a` that ignores its second and yields its first argument.

```
λ> const 42 (fac 100)
42
```

The evaluation immediately yields 42 as the second argument of `const` is not demanded, thus, not computed.

The second component of lazy evaluation — sharing expressions — is in most cases only observable regarding the performance of programs. We can, however, observe the difference of a shared expression and an expression that needs to be evaluated multiple times by using Haskell's `trace` function. Using `trace :: String -> a -> a` we can print debug messages on the console while evaluating a program. More precisely, the first argument is the message we want to log and the second argument the expression we want to log the message for.

In order to illustrate how `trace` works, consider the following two examples.

```
λ> let log42 = trace "fortytwo" 42 in 103 + log42
fortytwo
145

λ> let log42 = trace "fortytwo" 42 in const 103 log42
103
```

In both cases we want to log the message "fortytwo" when the variable log42 is used, and 42 is the actual value that is used to compute with. The first example logs the message during evaluation and then yields 145 as result. In the second example, we do not observe any logging message, because, again, the second argument of const does not need to be computed.

In order to observe that we shared an expression, we consider the following two expressions that double a value that is traced with a message.

```haskell
test1, test2 :: Int -> Int
test1 n = trace "msg" n + trace "msg" n
test2 n = let x = trace "msg" n in x + x
```

```
λ> test1 42
msg
msg
84

λ> test2 42
msg
84
```

The first example logs the message two times for each call to trace, whereas the second example shares the effectful expression trace "msg" 42 by binding it to a variable x and doubles the pure value 42 only. Although the first example test1 looks like an inlined version of test2, due to Haskell's call–by–need semantics these expressions have different results when used in combination with a side effect like tracing.

### 2.1.2 Monadic Abstractions

In a functional programming language like Haskell, we are used to define functions that map input values to output values. If a program, however, does not only return a value, but additionally has an observable interaction with the outside world (for example through an additional context that needs to be considered), such a program is said to have computational *effects*. As a pure language, Haskell does not allow any side effects conceptually, unless they are explicitly modelled. Such an explicit model becomes visibile at the type–level. For example, Haskell models the interaction with the user through reading input and printing output explicitly with the type IO (Wadler, 1997). Such an explicit model of computational effects capture the necessity to represent the additional context the function interacts with. Haskell's notion of purity allows, however, computational effects like tracing and partiality — using trace and undefined, respectively — that we discussed above. Effects that do not have explicit constructs for propagation and representation are sometimes called ambient or implicit effects (Filinski, 1996); in Haskell these examples, thus, are usually not visibile in the type signature.

In this thesis, we are interested in explicit representations of effects like partiality as well as non–determinism. We can, for example, explicitly model partiality using the following data type.[1]

---
[1] Note that the definition is equivalent to the Maybe type, here, we decide for an aptronym using a custom definition Partial instead.

```
data Partial a = Undefined
               | Defined a
```

These constructors represent undefined and defined values, respectively. As noted above, Haskell also has an implicit model of partiality: the polymorphic value `undefined` can be used without any indication on the type–level. The evaluation of `undefined` yields a run–time error. Consider the following examples that demand the evaluation of `undefined`.

```
λ> head []
*** Exception: Prelude.undefined

λ> head (tail [1])
*** Exception: Prelude.undefined
```

Using the `Partial` type, we can model a function that accesses the head of a list that explicitly yields `Undefined` instead of a run–time error.

```
headPartial :: [a] -> Partial a
headPartial []    = Undefined
headPartial (x:_) = Defined x
```

The representation of the exemplary expressions from above evaluates the undefined value explicitly using the appropriate constructor.

```
λ> headPartial []
Undefined

λ> headPartial (tail [1])
Undefined
```

Note that a corresponding implementation of `tailPartial` would not compose with `headPartial` anymore as in the original example above. Before we talk about this downside of the model, let us take a look at a representation for non–determinism as effect. We model functions that possibly produce several results (non–deterministically) using lists. In order to not confuse the representation of non–determinism using lists with lists that we use as algebraic datatypes in type signatures of functions, we use a type synonym `ND` for the former usage of lists.

```
type ND a = [a]
```

On top of that, we use the following convenience functions to yield a deterministic result and combine two potentially non–deterministic results.

```
det :: a -> ND a
det x = [x]

(?) :: ND a -> ND a -> ND a
(?) = (++)
```

The former function yields a singleton list, whereas the latter corresponds to the concatenation of the two lists. Using this representation of non–determinism and these convenience functions, we define the function `insertND` that non–deterministically inserts a given element at all possible positions of a list.

```
insertND :: a -> [a] -> ND [a]
insertND x []     = det [x]
insertND x (y:ys) = det (x : y : ys) ? map (y:) (insertND x ys)
```

The first rule is deterministic, it yields one list as result that contains just the element `x` we want to insert to the list. The second rule yields at least two results. The first argument of the `(?)`–operator inserts the element in front of the list and yields the deterministic result. For the second argument, we map over all lists for the recursive call `insertND x ys` by inserting the first element `y` to the front of all these resulting lists. Note that the `map` functions transforms elements of type `ND [a]` to `ND [a]`, since `ND` is just a type synonym for lists.

As an example, we non–deterministically insert `1` into the list `[2..5]`. Note that we manipulate the output for the REPL to use set–like parentheses for the lists that correspond to the modelled non–determinism of type `ND`.

```
λ> insertND 1 []
{ [1] }

λ> insertND 1 [2,3,4,5]
{ [1,2,3,4,5] , [2,1,3,4,5] , [2,3,1,4,5] , [2,3,4,1,5] , [2,3,4,5,1] }
```

A commonly used abstraction to model all these computational effects are monads (Moggi, 1989): the most common monadic abstraction is the `IO` type mentioned in the beginning. Using a type constructor class, a monad provides the following two operations.

```
class Monad m where
  return :: a -> m a
  (>>=)  :: m a -> (a -> m b) -> m b
```

A type constructor class allows to define overloaded functions for type constructors like `IO`, `Partial`, and `[]`. Note that we define an instance for `[]` instead of `ND`, because we can define type class instances for data types only, not for type synonyms.[2] That is, we can define type class instances for our modelled effects `Partial` and `[]` as follows.[3]

```
instance Monad Partial where
  return          = Defined
  Defined x >>= f = f x
  Undefined >>= f = Undefined

instance Monad [] where
  return   = det
  xs >>= f = concat (map f xs)
```

We reimplement the definition of `insertND` using `(>>=)` as follows, which leads to a more natural implementation concerning the separation of operation on lists as data structures and lists as model for non–determinism. More precisely, instead of using `map`, the usage of the `(>>=)`–operator gives us access to each list of the non–deterministic result.

---

[2] Note that we can define such instances using `TypeSynonymInstances` if these instances do not overlap with predefined ones.
[3] Strictly speaking, the instance for lists is already predefined.

```haskell
insertND :: a -> [a] -> ND [a]
insertND x []     = return [x]
insertND x (y:ys) = return (x : y : ys)
                  ? (insertND x ys >>= \zs -> return (y:zs))
```

Now recall the example for partiality again and let us define the `tail` function using `Partial`, analogous to `headPartial`.

```haskell
tailPartial :: [a] -> Partial [a]
tailPartial []     = Undefined
tailPartial (_:xs) = Defined xs
```

As already noted above, the composition `headPartial . tailPartial` is not possible although we can compose `head . tail` in the original Haskell code.

```
λ> headPartial (tailPartial [])
   • Couldn't match expected type '[a]'
                 with actual type 'Partial [a0]'
   • In the first argument of 'headPartial', namely '(tailPartial [])'
     In the expression: headPartial (tailPartial [])
     In an equation for 'it': it = headPartial (tailPartial [])
```

The problem of this composition is that the resulting type of `tailPartial`, namely `Partial [a]` is not the type `headPartial` expects as first argument. We can circumvent the typing problem using the operator (>>=) to access the list within the `Partial`–result of `tailPartial`, which yields the expected result `Undefined`, as follows.

```
λ> tailPartial [1] >>= headPartial
Undefined
```

As second example, we use (>>=) to compose a pure and an effectful function. Since the (>>=) operator needs a monadic function as second argument, we use `return` to lift the pure function into the monadic context. Here, we compute the head element of all the lists resulting from the usage of `insertND`.

```
λ> insertND 1 [2,3,4,5] >>= return . head
{ 1 , 2 , 2 , 2 , 2 }
```

Note, however, that the usage of (>>=) to make the composition work can have unintended effects in case the second function does not demand its argument. For example, the expression `const 42 (tail [])` yields 42 and not a run–time error. Hence, we expect the corresponding usage of `tailPartial` to yield `Defined` 42.

```
λ> const 42 (tail [])
42
λ> tailPartial [] >>= return . const 42
Undefined
```

We do not go into more details concerning this unintended behaviour here, but hope that the curious reader awaits the coming chapters eagerly, as we will discuss this model of non–determinism more thoroughly for Haskell in Chapter 3, and again in Chapter 5 when we discuss representations of effects in the proof assistant Coq.

### 2.1.3 Free Monads

Recently, the functional programming community started using a slightly different approach for modelling effects. The overall monadic structure is still the key of the representation. One observation that leads to the other abstraction is that all representations of such effects have two operations in common: one to lift a value into the effect representation (`return`) and one to manipulate the values of an effect (`(>>=)`). This observation finally leads to a monad instance that can interpret all monadic operations in an abstract way: the free monad (Swierstra, 2008). Consider the following data type `Free` that is parametrised by a type constructor `f` and a value type `a`.

```
data Free f a = Pure a
              | Impure (f (Free f a))
```

The general idea behind free monads is the observation that monadic computations are either pure values or impure effects. We represent the impure effect using the type constructor `f` and pure values are of type `a`. The nice property of the `Free` data type is that `Free f` is a monad, if `f` is a functor.

```
instance Functor f => Monad (Free f) where
  return         = Pure
  Pure x    >>= f = f x
  Impure fx >>= f = Impure (fmap (>>= f) fx)
```

We represent impure operations using the functor `f`. More precisely, the functor `f` represents the syntax of the effectful computation, that is, the operations that are added on top of the pure values we usually work with. In case of `Partial`, we have one operation, namely `Undefined` that corresponds to Haskell's `undefined` value associated with partiality. The other constructor `Defined` is already taken care of by `Pure`. Moreover, we observe that `Undefined` does not contain any further values but is a possible value of its own: it is a nullary operation. In contrast, we modelled the binary operation `(?) :: ND a -> ND a -> ND a` for non–determinism that combines two non–deterministic computations. The corresponding functor, thus, needs to make use of the recursive type argument `Free f a`. More concretely, since `Free` already models the constructor for defined and deterministic values using `Pure`, the functor takes care of the values constructed using `Undefined` for `Partial` and `(?)` for `ND`, respectively. The functors corresponding to the nullary operation `Undefined` and to the binary operation `(?)` look as follows.[4]

```
data One a    = One
data Choice a = Choice a a
```

Intuitively, the number of constructors of the functor corresponds to the number of operations the effect introduces and the arguments of each constructor indicate the arity of the corresponding operations. The key idea for `Partial` is that we represent `Undefined` as `Impure One`; together with `Pure` corresponding to `Defined`, we can represent the same programs as before. Note that the functor `Choice` for non–determinism used in combination with `Free` resembles a tree rather than a list. A leaf corresponds to `det` while a branch with two sub–trees t1 and t2 is represented as `Impure (Choice t1' t2')` where t1' and

---
[4]In the former case we follow the same naming conventions as Swierstra (2008).

| Description | Functor | Monadic Values | Free Values |
|---|---|---|---|
| Totality | `Zero` | `Identity x` | `Pure x` |
| Partiality | `One` | `Just x` <br> `Nothing` | `Pure x` <br> `Impure One` |
| Error | `Const e` | `Right x` <br> `Left y` | `Pure x` <br> `Impure (Const y)` |
| Non–determinism | `Choice` | `Leaf x` <br> `Branch t1 t2` | `Pure x` <br> `Impure (Choice t1' t2')` |

**Table 2.1.:** Overview of values represented using the direct interpretation as monad and using `Free` with the corresponding functor

| Description | Monadic Representation | Free Representation |
|---|---|---|
| Totality | `data Identity a = Identity a` | `Free Zero a` |
| Partiality | `data Maybe a = Just a | Nothing` | `Free One a` |
| Error | `data Either b a = Right a | Left b` | `Free (Const b) a` |
| Non–det. | `data Tree a = Lf a | Nd (Tree a) (Tree a)` | `Free Choice a` |

**Table 2.2.:** Overview of monads and the corresponding representation using `Free` and the associated functor

`t2'` are the transformations to `Free Choice` of the initial sub–trees `t1` and `t2`. Table 2.1 gives an overview of the value correspondences between the monadic representation and the representation using `Free` with the associated functor. The monad instance for `Free` demands the type parameter `f` to have a `Functor` constraint. We present the corresponding type class definition in Haskell as well as the definition of the instances for the concrete functors we use in this section in Appendix A.2.

A variety of common monads are isomorphic to a representation using free monads. A counterexample, however, is the list monad; as Swierstra (2008) states, there is no functor `f` such that type `Free f a` is isomorphic to `[a]`.[5] Due to this counterexample, we rather chose a tree encoding to represent non–determinism. In Chapter 5 we restate this isomorphism property and will show that the free monad applied to the functors `One` and `Choice` are isomorphic to `Maybe` and a leaf-labeled binary tree, respectively. Other popular representations are the identity monad and the error monad using the following functors.

```
data Zero a
data Const e a = Const e
```

Using the types as underlying effect, we get the identity monad using `Free Zero` and the error monad can be represented using `Free (Const e)`, where `e` is the type of the error. Table 2.2 gives an overview of different monads and their representation using `Free`.

Our running example from the preceding section for non–deterministically inserting an element at each possibile position in a list looks as follows using a representation based on `Free Choice`.

```
insertFree :: a -> [a] -> Free Choice [a]
insertFree x []     = return [x]
insertFree x (y:ys) = return (x : y : ys)
                   ?? (insertFree x ys >>= \zs -> return (y:zs))
```

---
[5]A proof sketch of this observation can be found in Appendix A.3.

We define the smart constructor for choices (??) as indicated above and can, thus, nearly reuse the implementation from before, because we already rely on the monadic abstraction.

```
(??) :: Free Choice a -> Free Choice a -> Free Choice a
(??) fx fy = Impure (Choice fx fy)
```

Note that the underlying representation of non–determinism changed from lists to trees, but otherwise the functions behave the same. The exemplary call also reveals five resulting lists.

```
λ> insertFree 1 [2..5]
Impure (Choice (Pure [1,2,3,4,5])
       (Impure (Choice (Pure [2,1,3,4,5])
               (Impure (Choice (Pure [2,3,1,4,5])
                       (Impure (Choice (Pure [2,3,4,1,5])
                                       (Pure [2,3,4,5,1])))))))))
```

## 2.2 Functional Logic Programming

The functional logic programming language Curry combines — as the name already suggests — the features of the functional and logic paradigms. A familiarity with Haskell is especially helpful for using Curry, since the syntax is basically the same. Hence, Curry has most of the features users know from Haskell: algebraic data types, higher–order functions, type classes and lazy evaluation. On top of these functional features, Curry adds non–determinism as a built–in effect. There are two maintained implementations of Curry: KiCS2 (Hanus et al., 2014) and PAKCS (Hanus, 2017). In the remainder of this thesis, we use KiCS2 to compile and run all Curry programs presented here if not specifically mentioned otherwise.[6] All interactions with KiCS2's REPL are displayed as verbatim environment with a turnstile — ⊢ — prompt.

### 2.2.1 Non–determinism

While Curry's non–determinism comes in a variety of forms, we are mostly interested in an explicit introduction of non–determinism using a dedicated operator. The binary operator `(?) :: a -> a -> a` non–deterministically yields two computations: its first and second argument. Note that the non–determinism is not visibile in the type signature, it is an implicit effect with respect to Curry's type system. The following expression non–deterministically yields two results.

```
⊢ True ? False
True
False
```

The other function associated with non–determinism is `failed :: a`, which is often used in similar situations as `undefined` in Haskell. The crucial difference to `undefined` in Haskell is that `failed` is rather a silent failure. More specifically, `failed` is a neutral element with respect to `(?)`. The following expressions illustrate this behaviour.

---
[6]More precisely, all source code was tested with version `2.0.0-b14` of KiCS2.

```
⊢ True ? failed
True

⊢ failed ? False
False
```

Due to its polymorphic type, `failed` can also occur nested within a data structure. The same applies for non–determinism in the form of choices constructed by `(?)`.

```
⊢ (1 : failed) ? [1,2]
[1,2]

⊢ (1 ? 2) : ([] ? [3])
[1]
[1,3]
[2]
[2,3]
```

As the REPL evaluates expressions to normal form, the first example just yields one result, because the left argument of the ?–operator evaluates to `failed`. Note that KiCS2 displays a failure within a deterministic and non–deterministic value differently. On the one hand, a failed branch is not printed at all in case of non–determinism like in the first exemplary expression. On the other hand, in case of determinism the REPL displays a head normal form with `!` as representation for `failed` as the following examples show.

```
⊢ 1 : failed
1 : !

⊢ failed : []
[!]
```

The second exemplary expression above yields four values: all non–determinism is pulled to the top–level leading to four different lists. The step–by–step evaluation of occurring non–deterministic computations that are pulled to the top is called *pull–tabbing* (Alqaddoumi et al., 2010). The equation in Figure 2.1 illustrates the pull–tabbing steps from left–to–right; the last line shows the simplified resulting expression.

$$
\begin{aligned}
&(1\ ?\ 2) : ([]\ ?\ [3]) \\
\equiv\ &\{\text{ pull–tabbing } 1\ ?\ 2\ \} \\
&(1 : ([]\ ?\ [3]))\ ?\ (2 : ([]\ ?\ [3])) \\
\equiv\ &\{\text{ pull–tabbing for left–most } []\ ?\ [3]\ \} \\
&((1 : [])\ ?\ (1 : [3]))\ ?\ (2 : ([]\ ?\ [3])) \\
\equiv\ &\{\text{ pull–tabbing for } []\ ?\ [3]\ \} \\
&((1 : [])\ ?\ (1 : [3]))\ ?\ ((2 : [])\ ?\ (2 : [3])) \\
\equiv\ &\{\text{ simplification of list representation }\} \\
&([1]\ ?\ [1,3])\ ?\ ([2]\ ?\ [2,3])
\end{aligned}
$$

**Figure 2.1.:** Step–by–step evaluation of the expression `(1 ? 2) : ([] ? [3])` using pull–tabbing

An intuitive mental model for the non–determinism is the interpretation as tree structure. The REPL allows us to enable `:set choices` to present all values as trees instead of Curry terms. The above expression, for example, yields the following tree.

```
⊢ :set choices
⊢ (1 ? 2) : ([] ? [3])
?
├── L: ?
│       ├── L: [1]
│       └── R: [1,3]
└── R: ?
        ├── L: [2]
        └── R: [2,3]
```

The tree is a one–to–one representation of the simplified expression from above. Overall there are three branches: one top–level branch and two more branches as sub–trees, leading to the four results. The left (L) and right (R) sub–trees of a branch are illustrated by corresponding labels. Given the tree representation of an expression, the REPL computes all possible results using a tree traversal. The default traversal for KiCS2 is a breadth–first search, but depth–first search is also possible. The user can change the behaviour of the tree traversal in the REPL using `:set dfs` and `:set bfs`, respectively. Note that the order of results changes for expressions corresponding to unbalanced trees.

```
⊢ :set dfs                          ⊢ :set bfs
⊢ (1 ? 2) ? 3                       ⊢ (1 ? 2) ? 3
1                                   3
2                                   1
3                                   2
```

As a side note consider the following adaptation of the expression above that changes the parentheses according to associativity. We evaluate the expression using breadth–first search again.

```
⊢ :set bfs
⊢ 1 ? (2 ? 3)
1
2
3
```

That is, the order of results is not stable with respect to associativity in case of breadth–first search. When we consider only the set of results instead of considering the order of the results, then we observe that the ?–operator is associative.

We can now take a look at the implementation of a non–deterministic version of `insert` that we modelled in Haskell using lists in Section 2.1.2.

```haskell
insert :: a -> [a] -> [a]
insert x []     = [x]
insert x (y:ys) = (x : y : ys) ? y : insert x ys
```

As expected, we get 5 results when we run the exemplary expression from the Haskell section.

```
⊢ insert 1 [2,3,4,5]
[1,2,3,4,5]
[2,1,3,4,5]
[2,3,1,4,5]
[2,3,4,1,5]
[2,3,4,5,1]
```

Last but not least, we compute all the head elements of the above expression. Recall that we discussed the missing composability of effectful functions like `insertND` or `headPartial` in the Haskell representation of non–determinism. In case of Curry's built–in non–determinism, the effect is not visible in the type–level, thus, the composition of pure and effectful functions works out of the box.

```
⊢ head (insert 1 [2,3,4,5])
1
2
```

The interesting part about the composition in case of Curry's built–in non–determinism is its non–strict behaviour. Instead of five results — one head element for each produced list like in the Haskell version, the computation yields only two results. Once again, we leave it here to build up the tension for the forthcoming chapters, when we discuss the difference between Curry's built–in non–determinism and the representation using lists in Haskell in more details.

### 2.2.2 Non–strictness

For Haskell we discussed some use–cases for non–strictness. A common programming pattern to exploit non–strictness that we have not introduced so far, is especially interesting for Curry due to its combination with non–determinism. We can use `let`–bindings instead of pattern matching to define a less–strict version of a function. In order to understand this difference, let us consider the following implementation of `fromJustToList` and an alternative implementation `fromJustToListLet`.

```
fromJustToList :: Maybe a -> [a]
fromJustToList (Just x) = x : []

fromJustToListLet :: Maybe a -> [a]
fromJustToListLet mx = let Just x = mx in x : []
```

The second implementation, `fromJustToListLet`, is less strict, because it yields a list constructor, `(:)`, without evaluating its argument first. That is, we can observe the difference when passing `failed` and checking if the resulting list is empty or not.

First, we define `null` to check if a list is empty and observe that the function does not demand the evaluation of the head element or the remaining list, because it only checks the surrounding list constructor.

```
null :: [a] -> Bool
null []      = True
null (_ : _) = False
```

Next, we evaluate the two functions above passing `failed` as argument in the context of `null`.

```
⊢ null (fromJustToList failed)
!

⊢ null (fromJustToListLet failed)
False
```

Due to the pattern matching in the definition of `fromJustToList` the argument `failed` needs to be evaluated and, thus, the function `null` propagates `failed` as return value. The definition of `fromJustToListLet` postpones the evaluation of its argument to the right–hand side: the argument needs to be evaluated only if the computation demands the value `x` explicitly.

The strictness property for `fromJustToList` holds for a definition via explicitly pattern matching using `case ... of` as well. In particular, pattern matching of the left–hand side of a rule desugars to case expressions on the right–hand side.

```
fromJustToListCase :: Maybe a -> [a]
fromJustToListCase mx = case mx of
                            Just x -> [x]

⊢ null (fromJustToListCase failed)
!
```

Note that we can observe the same non–strictness and strictness property, respectively, when using a non–deterministic value instead of using Curry's `failed`–value in all examples.

### 2.2.3 Call–time Choice

As Curry's similarities to Haskell are obvious, we also need to talk about the underlying evaluation strategy. The combination of Haskell's functional features with the logic paradigm in Curry has consequences with respect to lazy evaluation. More precisely, the combination of sharing and non–determinism is particularly interesting. Consider an adaption of the sharing example we used in Haskell that adds two numbers.

```
test1, test2 :: Int -> Int
test1 n = (n ? (n+1)) + (n ? (n+1))
test2 n = let x = n ? (n+1) in x + x
```

Recall that we showed the effect of logging a message whenever an expression is evaluated in Haskell. The key observation was that introducing a let–binding for such an effectful computation, i.e., sharing the computation, behaves differently than an inlined version: in the latter case the evaluation triggers the effect twice but only once when sharing the computation. We observe a similar behaviour for Curry's non–determinism effect. Sharing a non–deterministic computation leads to different results than its inlined version.

```
⊢ test1 42                          ⊢ test2 42
84                                  84
85                                  86
85
86
```

In case of Curry non–determinism emerges as effect: we can observe sharing of computations as it affects the number of results. We take a look at the corresponding tree visualisations for both expressions.

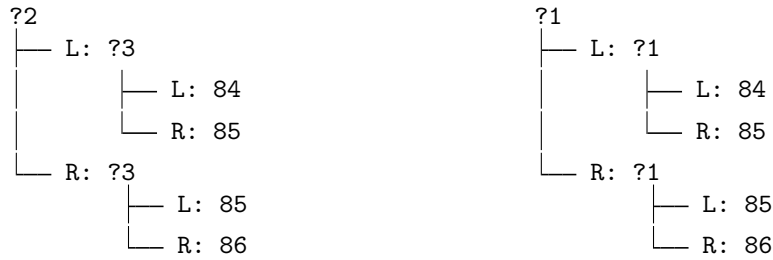

**Figure 2.2.:** Tree representation for `test1 42` (left) and `test2 42` (right)

The tree structure is the same for both expressions. We can, however, observe that the crucial difference lies in the labels for the non–deterministic choices that are displayed as `?1`, `?2` and `?3`, respectively. Note that we simplified the tree visualisation we showed before by removing these labels, because the labels are only of interest when an expression introduces sharing. Let us take a look at a step–by–step evaluation of `test2 42` that introduces sharing in Figure 2.3.

$\quad$ **let** $x = 42\ ?\ 43$ **in** $x + x$
$\equiv\quad$ { inlining of let–binding for $x$: use label $?_1$ }
$\quad (42\ ?_1\ 43) + (42\ ?_1\ 43)$
$\equiv\quad$ { pull–tabbing for left–most $42\ ?_1\ 43$ }
$\quad (42 + (42\ ?_1\ 43))\ ?_1\ (43 + (42\ ?_1\ 43))$
$\equiv\quad$ { pull–tabbing for left–most $42\ ?\_1\ 43$ }
$\quad ((42 + 42)\ ?_1\ (42 + 43))\ ?_1\ (43 + (42\ ?_1\ 43))$
$\equiv\quad$ { pull–tabbing for $42\ ?_1\ 43$ }
$\quad ((42 + 42)\ ?_1\ (42 + 43))\ ?_1\ ((43 + 42)\ ?_1\ (43 + 43))$
$\equiv\quad$ { simplification }
$\quad (84\ ?_1\ 85)\ ?_1\ (85\ ?_1\ 86)$

**Figure 2.3.:** Step–by–step evaluation of the expression `test2 42`

The crucial difference to our step–by–step evaluation before is that we give the `(?)`–operator an explicit label when we inline the let–binding. The resulting expression is once again a one–to–one representation of the tree visualisation above. Now the question arises how we should interpret these labels. The idea is to make consistent choices for each label. That is, when the REPL traverses the tree to compute all results, the algorithm tracks for each label if the traversal continues in the left or right sub–tree. If we reach a label that was already tracked, we take the same direction again: the traversal makes consistent choices for all labels.

A more high–level way to use equational reasoning for expressions with shared computations is to pull the whole let–expression to the top–level. That way we do not need to use labeled choices as the following equation in Figure 2.4 illustrates.

### 2.2.4 Encapsulation

The implementation of `insert` is a nice example for a straightforward realisation of the algorithm's specification due to the usage of non–determinism. When we use such non–

$$\begin{aligned}
&\phantom{\equiv}\ \mathbf{let}\ x = 42\mathbin{?}43\ \mathbf{in}\ x + x \\
&\equiv\quad \{\text{ pull–tabbing } 42\mathbin{?}43 \text{ in let–binding }\} \\
&\phantom{\equiv}\ \mathbf{let}\ x = 42\ \mathbf{in}\ x + x\mathbin{?}\mathbf{let}\ x = 43\ \mathbf{in}\ x + x \\
&\equiv\quad \{\text{ inlining of left–most let–binding for } x\ \} \\
&\phantom{\equiv}\ 42 + 42\mathbin{?}\mathbf{let}\ x = 43\ \mathbf{in}\ x + x \\
&\equiv\quad \{\text{ inlining of let–binding for } x\ \} \\
&\phantom{\equiv}\ 42 + 42\mathbin{?}43 + 43 \\
&\equiv\quad \{\text{ simplification }\} \\
&\phantom{\equiv}\ 84\mathbin{?}86
\end{aligned}$$

**Figure 2.4.:** Alternative representation for step–by–step evaluation of the expression `test2 42`

deterministic functions in our code basis, we cannot regain a pure, deterministic function anymore. Luckily, there exists a mechanism in Curry to encapsulate the non–determinism of computations. Curry provides one implementation of strong encapsulation (Braßel et al., 2004) with the primitive function `allValues :: a -> { a }` that operates on a polymorphic — and potentially non–deterministic — computation and yields a multiset of values. For presentation purposes, we use an abstract view of the result of an encapsulation to emphasise that the order of encapsulated results does not matter. In order to work with encapsulated values, Curry furthermore provides the following two functions to fold and map the resulting multiset.

```
foldValues :: (a -> a -> a) -> a -> { a } -> a
mapValues  :: (a -> b) -> { a } -> { b }
```

We do not discuss the implementation details behind `allValues` here, it it sufficient to assume that an implementation of the above interface uses, for example, lists to represent the encapsulated results. That is, the function `allValues` yields ordinary lists as result. Using lists as underlying representation, let us focus on the usage of the function `allValues` by encapsulating all the results from our previous examples.

```
⊢ allValues (insert 1 [2,3,4,5])
[[1,2,3,4,5], [2,1,3,4,5], [2,3,1,4,5], [2,3,4,1,5], [2,3,4,5,1]]

⊢ allValues (head (insert 1 [2,3,4,5]))
[1, 2]
```

We can also manipulate all the results of a non–deterministic computation in a deterministic way by working directly on the list structure. For example, we can project to the head element of each resulting list.

```
⊢ mapValues head (allValues (insert 1 [2,3,4,5]))
[1, 2, 2, 2, 2]
```

## 2.3 Programming with Dependent Types in Coq

In this section we give an introduction to programming with the interactive theorem prover Coq. In order to write functional programs, Coq provides the specification language *Gallina*. More precisely, Gallina is a dependently typed functional programming language. The theorem prover part of Coq is based on the calculus of inductive construction (Coquand and Paulin, 1988), a derivative of the calculus of construction that was introduced by Coquand and Huet (1986). As Coq is not an automatic theorem prover, it additionally provides a tactic language called *Ltac* that enables the user to interactively construct proofs.

This introduction is structured as follows. We first give an overview on writing functional programs in Coq including common obstacles with regard to Coq's totality restriction as well as how to overcome them. As some of these solutions lead us to dependently typed programming, we take a look at how to formalise properties about programs and how to prove them. In that light, we give a beginner–friendly introduction on how to use Coq's tactic language to write proofs. Note that we do not give a formal definition of the calculus of constructions or other concepts with regard to the implementation of Coq's logic, but suggest the interested reader to study other textbooks for a full–blown introduction (Pierce et al., 2010; Chlipala, 2013). Instead we focus on how to use Coq as a tool to formalise and prove properties about programs.

### 2.3.1 Functional Programming

We can define a lot of common functions we know from functional programming languages like Haskell or Curry one–to–one in Coq. As a code convention, we will start data type and constructor names with lower–case letters as many standard types in Coq follow the same convention. Consider the following data type definition for Peano numbers.

```
Inductive nat : Type :=
| z : nat
| s : nat -> nat.
```

Note that the data type declaration for `nat` is annotated with `Type`. Coq is build upon a sophisticated type hierarchy; for us it is enough to know that we will use `Type` for inductive data types and type synonyms. The syntax to define constructors for data types resembles GADT–style definitions in Haskell, that is, we specify the full type signature for each constructor, as we are used to for function definitions. Furthermore, we list all constructors line–by–line and each line begins with a pipe | as delimiter.

Next, we define a function to check if a natural number `nat` is zero as well as a function for addition on natural numbers using pattern matching. Note that both definitions have explicit type annotations for the argument and return types, although Coq could infer all of them.

```
Definition isZero (p : nat) : bool :=
  match p with
  | z _ => true
  | s _ => false
  end.

Fixpoint add (p1 p2 : nat) : nat :=
  match p1 with
  | z    => p2
  | s p  => s (add p p2)
  end.
```

The pipe symbol | is again used as a delimiter: this time to list all patterns we want to distinguish in a `match ... with`–expression that corresponds to Haskell's `case ... of`–

expressions. Note that the definition of `isZero` uses the predefined `bool` type as result, and that we can group successive arguments of the same type in the type signature. The function definitions above represent two different kinds of functions: `isZero` is a non–recursive function, whereas `add` is a representative for recursive functions. In Coq it is crucial to define recursive functions using the keyword `Fixpoint`. Definitions that are marked as `Fixpoint`s need to pass an additional check: as a total language, all functions in Coq have to terminate. In case of recursive functions, Coq checks if there is one argument that is structurally smaller than the original argument passed to the function for each recursive call. In our example definition `add`, the argument `p` of the recursive call in the second case of the branch is structurally smaller than the original value `p1`, since the pattern match reifies `p1` to `(s p)`.

As an example for a recursive function that Coq's termination checker fails to accept, we try to define the following artificial definition that decreases on one of its arguments at each recursive call, but does not have one designated argument that decreases for each call.

```
Fail Fixpoint test (b : bool) (n m : nat) : bool :=
  match (n,m) with
  | (z,_)     => true
  | (_,z)     => false
  | (s p, s q) => if b then test b p m else test b n q
  end.
```

The keyword `Fail` indicates that the definition was not accepted by Coq without an error. Here, the command has failed with the error message `Cannot guess decreasing argument of fix`. If we want to define such a recursive function that does not follow Coq's restriction, we can use the keyword `Program Fixpoint` and prove the termination of the function afterwards. In this thesis, we will not use capabilities like `Program Fixpoint` to define recursive functions that do not obey Coq's restrictions, so we abstain from going into more details concerning proving termination of recursive functions.

Moving on with our definitions on Peano numbers above, we can introduce a `Notation` that allows us to write more natural expressions like `p1 + p2` for the expression `(add p1 p2)`. We declare the newly introduced operator syntax as left associative a give the syntax a fixity level.

```
Notation "p1 + p2" := (add p1 p2) (left associativity, at level 50).
```

In Coq we can check types, print definitions and evaluate expressions directly in the Coq–file using so–called *Vernacular*–commands like `Check`, `Print` and `Compute` that will print the information on the console. Since we are used to having a REPL from languages like Haskell and Curry, we will write these commands and the printed answer in a REPL–style as follows.

```
Π> Compute (s z + s z).
   = s (s z)
   : nat
```

Let us now consider a polymorphic data type definition as well as polymorphic functions using lists as example. We use the predefined definition for lists defined as follows.

```
Inductive list (A : Type) : Type :=
| nil  : list A
| cons : A -> list A -> list A.
```

As a code convention we use upper–case letters for type variables. In case of polymorphic functions, including constructors of data types, we need to pass the instantiation of the type arguments explicitly. For example, when we try to compute the following expression to normal form, Coq yields an error message.

```
Π> Compute (cons z nil).
Error: The term "z" has type "nat" while it is expected to have type
       "Type".
```

When using Coq, constructors of polymorphic functions need to be applied to more arguments than we are used to from functional languages such as Haskell. More precisely, all constructors of a polymorphic data type — like `list` — have additional type arguments. In our case, `cons` and `nil` have the following types.

```
Π> Check nil.
   nil : ∀ (A : Type), list A
Π> Check cons.
   cons : ∀ (A : Type), A -> list A -> list A
```

The first argument is the type that determines the concrete type instantiation of the constructors `cons` and `nil`. The definition above works when we apply a type like `nat` explicitly or instruct Coq to infer the argument by using an underscore (_).

```
Π> Compute (cons nat z (nil nat)).
    = cons nat z (nil nat)
    : list nat

Π> Compute (cons _ z (nil _)).
    = cons nat z (nil nat)
    : list nat

Π> Compute (cons _ tt (nil _)).
    = cons unit tt (nil unit)
    : list unit
```

Note that the type `unit` we use in the third example has only one constructor `tt`. Instead of applying type arguments explicitly, we can tweak some settings in order to use functions as we are used to in Haskell, such that type arguments are inferred. [7]

Next, we define the recursive functions `length` and `map` as exemplary functions on lists.

```
Fixpoint length (A : Type) (xs : list A) : nat :=
  match xs with
  | nil     => z
  | cons _ ys => s (length ys)
  end.

Fixpoint map (A B : Type) (f : A -> B) (xs : list A) : list B :=
  match xs with
```

---

[7] In particular, we use the option `Set Implicit Arguments` and specific commands like `Arguments nil [_]` to make Coq infer all type arguments if possible.

```
| nil => nil
| cons y ys => cons (f y) (map f ys)
end.
```

A rather obvious property of a combination of these function states that mapping over a list does not change its length. In Coq we can define such a proposition mostly consisting of language features we have used so far.

```
Lemma map_length (A B : Type) (f : A -> B) (xs : list A)
  : length xs = length (map f xs).
```

First, instead of using the `Definition` or `Fixpoint` keyword, we use `Lemma` that has the same purpose as the former. That is, the definition `map_length` is a function. While the arguments of the function `map_length` look as usual, the resulting type involves a dependent type. The symbol = on the right–hand side of the type signature is just a notation for the type `eq` that is defined as follows.

```
Inductive eq (A : Type) (x : A) : A -> Prop :=
| eq_refl : eq x x.
```

The data type `eq` is a propositional type, indicated by the resulting type `Prop`. Intuitively, `Prop` is the type of propositions in Coq. Furthermore, the type of the constructor `eq_refl` that represents the reflexivity property for structural equality is `eq x x`. Here, `eq` is a dependent type: the type of the constructor `eq_refl` uses not only types but the value `x` in its type signature. The type `eq` can only be instantiated with arguments `x` and `y` if they evaluate to the same value. That is, a value of type `eq` is the proof that two expressions are structurally equal.

The type of lemmas like `map_length` corresponds to the proposition we want to prove and the implementation is one concrete proof. That is, if we can implement `map_length` with type `length xs = length (map f xs)`, we have proven the corresponding proposition.

### 2.3.2 Proving in Coq: A Step–by–Step Introduction

For the proof of the proposition `map_length` we need to take a look at the tactic language `Ltac`. We enclose a proof using the commands `Proof` and `Qed` and write all tactics we want to apply between these commands. So, let us start with the proof; we present the tactic or other commands on the left–hand side and illustrate the progress after that command in a verbatim environment on the right–hand side.

```
Proof.                          A : Type
                                B : Type
                                f : A -> B
                                xs : list A
                                ============================
                                length xs = length (map f xs)
```

The current state of the proof shows that we have the types `A` and `B` in scope as well as a function `f : A -> B` and a list `xs : list A`. These variables are *hypotheses* we can use for the proof. Underneath the hypotheses we see the resulting type of `map_length` that represents the current *goal*. We can then use constructs of the tactic language to manipulate hypotheses and the goal in a sensible way until we find the final proof of the goal. As we proceed with the proof, we will use ... to indicate that we do not list all hypotheses, e.g., we will leave out types of identifiers, variables and hypotheses that we not explicitly use in the current goal.

Taking a look at the goal, we realise that we cannot simplify neither the left–hand side of the equation nor the right–hand side, because the definition of `length` distinguishes between the `nil` and `cons` case. In order to argue about these two different cases separately, we start with an induction on `xs`. Using the `induction`–tactic on a list generates two new goals: one for `nil` and one for `cons`.

```
induction xs as [ | y ys H ].        ...
                                     f : A -> B
                                     ============================
                                     length nil = length (map f nil)

                                     ...
                                     H : length ys = length (map f ys)
                                     ============================
                                     length (cons y ys) =
                                     length (map f (cons y ys))
```

We use a more involved version of the `induction`–tactic that additionally supplies the names for the new arguments that need to be introduced. The introduction pattern `[ | y ys H ]` describes the naming conventions for the variables introduced in the two subgoals. We do not supply any variable names for the first goal, because the `nil`–case does not introduce new variables. For the `cons`–subgoal we pass the three names: `y` and `ys` for the two arguments of the `cons`–constructor as well as `H` for the induction hypothesis that Coq generates automatically based on the definition of the `list` data type. More generally, the vertical bar is a separator for the resulting subgoals and we introduce new names depending on the arguments of the corresponding subgoal. Coq generates names for all arguments that are not explicitly introduced by the pattern.

In case of `nil`, we can simplify both sides of the equation using the definition of `length` and `map` by applying the tactic `simpl`.

```
- simpl.                             ...
                                     ============================
                                     z = z
```

We use bullet points like -, +, * to structure the proof and bring the next subgoal in focus. Now we are at the point that we can use the tactic `reflexivity`, which, intuitively, constructs the final expression using the above introduced constructor `eq_refl`. Next, we take a look at the `cons`–case after simplifying both expressions again.

```
- simpl.                             ...
                                     H : length ys = length (map f ys)
                                     ============================
                                     s (length ys) = s (length (map f ys))
```

The induction hypothesis `H` states that the proposition already holds in case of `ys`. That is, we can directly rewrite `length ys` on the left–hand side of the equation in the goal with the right–hand side of the hypothesis.

**22**   **Chapter 2**   Declarative Programming

```
  rewrite -> H.                   ...
                                  ============================
                                  s (length (map f ys)) =
                                  s (length (map f ys))
```

The `rewrite`–tactic gets two arguments: the first argument specifies the direction we want to perform the rewriting in[8] and the second argument indicates which equality hypothesis we want to rewrite in our goal. After rewriting the hypothesis, the final goal can be proven using `reflexivity` again. We then finish the proof by using the keyword `Qed`.

```
  Qed.                            map_length is defined
```

At that point, Coq provides the information that the lemma `map_length` was successfully defined and can be used in future proofs and definitions. In order to give a better overview of the proof in its entirety, we restate the proposition as lemma including the complete proof script.

```
Lemma map_length (A B : Type) (f : A -> B) (xs : list A)
  : length xs = length (map f xs).
Proof.
  induction xs as [ | y ys H ].
  - simpl. reflexivity.
  - simpl. rewrite -> H. reflexivity.
Qed.
```

**Induction Principle**

In the preceding section we proved a lemma about `map` and `length` using induction on the list argument. The `induction`–tactic makes use of the associated induction principle for the corresponding type. An induction principle for a data type is an ordinary function of type `Prop`. Moreover, Coq automatically generates this induction principle for each data type declaration defined using `Inductive`. The induction principle for a inductive type `T` is available as a function named `T_ind`.

Let us take a look at an example: the induction principle for Coq's predefined list datatype that we used in the last section.

```
Π> Check list_ind.
  list_ind : ∀ (A : Type) (P : list A -> Prop),
    P nil ->
    (∀ (e : A) (l : list A), P l -> P (cons e l)) ->
    ∀ (l : list A), P xs
```

Dissecting this type signature gives us the following parts: given a proposition `P` on lists, we need to supply two proofs. A proof of type `P nil` in case of an empty list and a proof of type `P (cons e l)` otherwise, where `e` and `l` are universally quantified. The first proof is the base case of the induction principle for lists, while the second proof is the inductive case. That is, in the second proof we assume that the proposition already holds for an arbitrary but fixed list `l` and can then prove that the proposition also holds for `cons e l`.

---

[8]The direction from left–to–right is the default and does not need to be provided in this case.

As an example, we reprove the lemma above: `map` retains the length of its input list. Instead of using the dedicated `induction`–tactic, we use the induction principle directly. We adapt the above proof as follows.

```
Lemma map_length' (A B : Type) (f : A -> B) (xs : list A)
  : length xs = length (map f xs).
Proof.
  apply list_ind with (l := xs).
  - simpl. reflexivity.
  - intros y ys H. simpl. rewrite -> H; reflexivity.
Qed.
```

When using backward–reasoning, we have to use the function `list_ind` on a function of type `l -> P l`, which is the type of our lemma, when we instantiate the argument list `l` with the concrete list `xs`. Since `list_ind` has three arguments, these arguments are now the new subgoals we need to prove. One of these arguments is the proposition `P : list A -> Prop` of the induction principle, which Coq infers automatically from the goal we apply the function to. In this case, the proposition is the function `fun xs => length xs = length (map f xs)`. That is, we end up with only two subgoals: we need to prove that the proposition holds for an empty and a non–empty list. We can prove the first goal for the `nil`–constructor using the same tactics as before: after simplification of the involved functions, both sides of the equation are already equal. The second goal looks as follows.

```
  ∀ (y : A) (ys : list A),
    length ys = length (map f ys) ->
    length (cons y ys) = length (map f (cons y ys))
```

Given an element `y`, a list `ys`, and the induction hypothesis that the proposition already holds for that `ys`, we need to prove that the proposition holds for `cons y ys` as well. While Coq's `induction` tactic introduces these universally quantified variables for us, the direct variant is not as convenient. That is, we need to introduce these variables ourselves using `intros` to specify names for the quantified variables first, but can then prove the goal using the induction hypothesis like in the first version above.

### 2.3.3 Representing Data Types using Containers

Up to now, the only dependent type we have seen is the equality type we use to state properties for our programs. Another prominent example for the usage of dependent types originates from generic programming (Altenkirch and McBride, 2003; Hinze, 2000): we can encode a variety of polymorphic data types using containers as introduced by Abbott et al. (2003).

A container is described as a product of shapes and a position function. The shape is a type `Shape` and the position type `Pos` is a type function that maps shapes to types. Using these two components, we can define a container extension that gives access to values of type `A`.

```
Inductive Ext (Shape : Type) (Pos : Shape -> Type) (A : Type) : Type :=
| ext : ∀ s, (Pos s -> A) -> Ext Shape Pos A.
```

A container extension `Ext Shape Pos` is then isomorphic to a type constructor `F`, where `F A` represents the polymorphic data type. The position type `Pos` is a dependent type: the type depends on a value of type `Shape`. The general idea is that the position type specialised to a shape `Pos s` describes all the possible positions of a data structure. A container extension of type `Ext Shape Pos A` consists of a function `Pos s -> A` for all shapes `s : Shape` that gives access to the polymorphic components of the data type it describes. That is, given one concrete position of type `Pos s` for a shape `s`, one value of type `A` can be accessed.

As an example, let us consider the polymorphic data type `One` we discussed in Section 2.1.3 in the context of free monads. Recall that the definition was used by Swierstra (2008) and that the name captures the number of representable values quite well.

```
Inductive One (A : Type) : Type :=
| one : One A.
```

Now we want to represent `One` as a container described as a type of shapes and a position type function. The data type `One` has only one constructor, that is, there is only one shape that we need to represent.

```
Definition One_S := unit.
```

Intuitively, the shape type represents the different constructors of a data type. Note that instead of introducing a new type with one value, we reuse Coq's `unit` type. The only value of type `unit` is called `tt`.

The position type function, on the other hand, describes the possible positions of polymorphic arguments for a given constructor. As observed above, `One` has only one constructor. This constructor `one` does not have any polymorphic arguments. More precisely, there are no possible positions for polymorphic arguments for any constructor. In order to represent that there are no possible positions, we use an empty type. An empty type is a data type without any values, that is, we cannot construct values of that type.

```
Inductive Empty : Type := .
Definition One_P (s : One_S) := Empty.
```

Recall that the position type function depends on the corresponding shape. Here, however, the shape does not matter as we do not have any position anyhow.

Using Coq's ability to prove properties about programs, we can show that the container representation is isomorphic to the original data type. More precisely, we define two functions `from_One` and `to_One`, and show that both compositions yield the identity.

```
Definition from_One A (o : One A) : Ext One_S One_P A :=
  ext tt (fun (p : One_P tt) => match p with end).
```

```
Definition to_One A (e : Ext One_S One_P A) : One A :=
  one.
```

The definition of `from_One` uses an empty pattern match, because there are no possible values of type `One_P tt` — recall that the type is just a synonym for `Empty`. Since the argument `p` is of type `Empty` and `Empty` has no constructors to match on, Coq realises that we cannot have a value of type `Empty` in the first place and accepts the definition. That is, we can define a function that yields an arbitrary polymorphic value given an empty type as argument.

```coq
Definition bogus (A : Type) (em : Empty) : A :=
  match em with end.
```

The definition of `to_One`, on the other hand, is straightforward, because there is only one way to construct the value `one`. Note that we do not need to access a polymorphic component of type `A` in order to construct a value of type `One`.

An alternative definition of the above function displays the components of the argument type `Ext One_S One_P A` more explicitly.

```coq
Definition to_One A (e : Ext One_S One_P A) : One A :=
  let '(ext tt pf) := e in one.
```

Since `Ext` has only one constructor, Coq allows a let–binding for such irrefutable patterns with the above syntax using a tick `'`. The function `pf` is of type `One_P tt -> A`, which becomes a function of type `Empty -> A` after inlining the definition of `One_P tt`. As a side note, we could not produce a value of type `A` using the function `pf` if we needed to: we cannot construct the appropriate argument of type `Empty`.

With these definitions at hand, we can prove that `from_One` and `to_One` form an isomorphism. We start with the more involved proof `from_to_One` first.

```coq
Lemma from_to_One : ∀ (A : Type) (e : Ext One_S One_P A),
    from_One (to_One e) = e.
```

```coq
Proof. intros A e; simpl.
```
```
e : Ext One_S One_P A
============================
ext tt (fun p => match p with end) = e
```

First, we introduce all variables on the right–hand side of the type signature. After introducing these quantified variables as hypotheses, we simplify the function applications of `from_One` and `to_One` as they disregard the given argument. Next, we observe that the argument `e` on the right–hand side of the equation is of type `Ext`: it must be constructed using `ext` as well. More precisely, we even know that the first component, the shape, is of type `One_S`, and, hence, the only valid value is `tt`. We destruct `e` directly in its two components: `tt` for the shape and `pf` for the position function using `[]` as a nested destruction for the shape. The pattern `[]` can be used for deconstruction of values with only one constructor and without any arguments.

```coq
destruct e as [ [] pf ].
```
```
s : One_S
pf : One_P s -> A
============================
ext tt (fun p => match p with end) =
ext tt pf
```

Next, we observe that we have the `ext`–constructor on both sides of the equation. By applying the tactic `f_equal`, we use the following predefined lemma about functional equality.

```
Π> Check f_equal.
  f_equal : ∀ (A B : Type) (f : A -> B) (x y : A), x = y -> f x = f y
```

In the case above, we apply the function `ext tt` to (`fun p => match p with end`) on the left–hand side and to `pf` on the right–hand side of the equation.

```
f_equal.                           pf : One_P tt -> A
                                   ============================
                                   (fun p : One_P tt => match p with end) = pf
```

The assumption of the `f_equal`–lemma becomes the new goal: we now need to show that the two position functions are the same. In pure functional programming, we are used to a property called functional extensionality: two functions `f : A -> B` and `g : A -> B` are equal, if they yield the same result for all possible input arguments of appropriate type.

```
Π> Check  functional_extensionality.
  functional_extensionality : ∀ (A B : Type) (f g : A -> B),
    (∀ (x : A), f x = g x) -> f = g
```

The corresponding tactic `extensionality p` introduces a new variable `p` of appropriate type that is used as argument on both sides of the equation.

```
extensionality p.                  pf : One_P tt -> A
                                   p : One_P tt
                                   ============================
                                   match p with end = pf p
```

Finally, we prove the statement by realising that the newly introduced argument `p` of type `One_P tt` is a value of an empty type. Recall that `One_P tt` can be inlined to `Empty`. That is, we finish the proof by trying to destruct the value `p` into its corresponding constructors. Coq realises that there are no values of that type, which proves the last goal.

```
destruct p. Qed.                   from_to_One is defined
```

The second direction of the isomorphism proof is straightforward. After introducing all variables and destructing the argument `o : One A`, we simplify the left–hand side of the equation using the function definitions of `from_One` and `to_One`. The simplification already leads to the value `one : One A` on the left–hand side as well as on the right–hand side.

```
Lemma to_from_One : ∀ (A : Type) (o : One A),
    to_One (from_One o) = o.
Proof.
  intros A o; destruct o; reflexivity.
Qed.
```

# Generating Permutations via Non–deterministic Sorting    3

The most prominent use of non–determinism in combination with sorting is permutation sort: a sorting algorithm that generates all permutations and selects the one that is sorted to yield as result. In this chapter we shed some light on a different combination of sorting functions and non–determinism: we enumerate permutations by applying sorting functions to a non–deterministic comparison function. First, we take a look at implementations of some famous sorting functions in Curry and define a suitable comparison function to enumerate permutations. Thanks to Curry's built–in non–determinism, we can reuse all common sorting functions as they are. We discuss the resulting permutation enumeration functions; the number of results is of special interest here. A selection of questions that we will answer reads as follows.

- Can we enumerate all possible permutations of the input list using any sorting function?
- Is there a sorting function that can enumerate exactly the permutations of the input list?
- Can we visualise how a sorting function enumerates the permutations?

As a quick teaser for these questions, we anticipate that all (correct) sorting functions indeed enumerate every permutation of the input list at least once. However, enumerating every permutation exactly once is a property that not all sorting functions share.

In the second part of this chapter we transfer our implementation in Curry to Haskell. One possible model of non–determinism in a functional language is to use lists to represent all non–deterministic results as we discussed in Section 2.1.2. We go even one step further and generalise all functions to monadic functions and use a more specialised version of monads that have an additional `mplus` function that is especially suitable for non–determinism. Using these monadically lifted functions, we are interested in a multi–set model of non–determinism using lists. We compare the monadic representation of non–determinism with the built–in non–determinism of Curry by means of these sorting functions to enumerate permutations.

The main observation of the comparison is that Curry's built–in non–determinism can be less strict than a naive monadic model in Haskell. This observation is not new, there are other applications that exploit this advantage as well. Using non–determinism to transform sorting functions into permutation functions is an interesting use case of this advantage nonetheless.

In summary, this chapter makes the following contributions.

- We inspect sorting functions implemented in Curry and apply these functions to a non–deterministic comparison function and a list to enumerate permutations of that list.
- When taking a closer look at the resulting lists, we observe that the functions enumerate every possible permutation of the input list at least once.
- Some sorting functions yield duplicate results. We investigate why these duplicates emerge.

- Furthermore, we discover that some sorting functions even enumerate lists that are not a permutation of the input list.

- As a second step, we transfer the sorting functions to Haskell and model non–determinism explicitly using monads.

- We investigate the difference between Curry's non–determinism and the list–based model in Haskell for the resulting permutation enumeration functions.

At the end of the chapter, we discuss a variety of questions that follow from the results and observations presented above.

## 3.1 Non–deterministic Sorting Functions in Curry

We start this chapter with an overview of common sorting functions, which we implement using a version that is parametrised by a comparison function. The main quest of this section is to document the behavior of these sorting functions when applied to a particular comparison function. Hereunto, consider the following non–deterministic comparison function that ignores both its arguments and non–deterministically yields `True` and `False`.

```
coinCmp :: a -> a -> Bool
coinCmp _ _ = True ? False
```

The name `coinCmp` suggests a similarity to the popular `coin` definition as an example for non–determinism; the difference here is that we define a binary function with a non–deterministic result, whereas `coin` is a nullary function.

In the remainder of this section we will apply all sorting functions to the non–deterministic comparison function `coinCmp`. We can already anticipate here that the non–determinism introduced by `coinCmp` will transform the sorting functions with resulting list type `[a]` to functions that produce several lists non–deterministically. That is, the non–determinism of `coinCmp` propagates to the result.

The interesting thing about using these sorting functions with a non–deterministic comparison function like `coinCmp` is that we expect that each permutation of the input list is part of the resulting lists. Why is it reasonable to expect to see all permutations? We expect the original sorting function, say `sort :: (Int -> Int -> Bool) -> [Int] -> [Int]`, to actually sort a list in ascending order when we apply it to the comparison function `(<=)`. If `sort (<=)` is indeed such a sorting function, then every permutation of a sorted list `[1..n]` of length $n$ can be sorted by `sort (<=)`. That is, there are decisions of `(<=)` that result in a sorted list. If we apply `sort coinCmp` to a list, then `coinCmp` tries every possible decision when comparing two elements. Hence, there is a sequence of decisions that leads to a specific permutation of the input list. As we are trying every possible decision, we expect all possible permutations to be part of the result. Otherwise, if a permutation is not part of the result, we can deduce that there is an input list such that `sort (<=)` does not correctly sort this list.

In the following all of our examples use lists of integers, more particular, we use lists of the shape `[1..n]` with $n$ as length of the list. These integers are only meant to be placeholders for concrete values and represent the position of the element within the original list. That is, when we produce permutations of the input list, we can deduce the behaviour for a variety of input lists of other types, because we know where elements end up in the result based on their position in the input list. Moreover, when we describe how a function produces its results, we say that a value is *less than or equal* to another value with respect to the comparison function in question. Considering that `coinCmp` is a comparison function that does not have the usual property of order relations, not every usage of *less than or equal* refers to the natural order on integer values (that is, the relation $\leqslant$). More particular, we will use the same terminology when describing the behaviour of the non–deterministic predicate `coinCmp`. We think, however, these descriptions are easier to read when we use *less than or equal* in this overloaded manner.

### 3.1.1 Insertion Sort

The first sorting function we take a look at is insertion sort. The key idea behind the sorting algorithm is to traverse the input list and insert each element in the right position within the resulting list. That is, the element in focus is inserted in front of the first element of the list that is greater than or equal to the former. In order to implement `insertionSort`, we implement a function `insert` first that does exactly that task: it inserts an element in an already sorted list with respect to the given comparison function.

```
insert :: (a -> a -> Bool) -> a -> [a] -> [a]
insert _ x []                    = [x]
insert p x (y:ys) | p x y        = x : y : ys
                  | otherwise    = y : insert p x ys
```

We then define `insertionSort` using `insert` to sort a list by inserting each element into the already sorted intermediate list.

```
insertionSort :: (a -> a -> Bool) -> [a] -> [a]
insertionSort _ []     = []
insertionSort p (x:xs) = insert p x (insertionSort p xs)
```

Let us test if the function works as expected. The following example sorts a list of integers in ascending order.

  ⊢ `insertionSort (<=) (reverse [1..5])`
  `[1,2,3,4,5]`

As the definition of `insertionSort` is parametric over the comparison function, we can apply different orderings using the same function. Besides a deterministic order specified by `(<=)` or `(>=)`, there is nothing that stops us from using a non–deterministic comparison function instead. Consider the following example, where we apply `insertionSort` to the non–deterministic comparison function `coinCmp`.[1]

  ⊢ `insertionSort coinCmp [1,2,3]`
  `[1,2,3]`  `[1,3,2]`
  `[2,1,3]`  `[3,1,2]`
  `[2,3,1]`  `[3,2,1]`

We see that the expression non–deterministically yields six different results. All of these results are a permutation of the input list `[1,2,3]`. Since a list of length $n$ has $n!$ number of permutations, the example above yields exactly all $3! = 6$ permutations of the list.

---

[1] If the result of an evaluated expression in the REPL yields more than four results, we will display the results side–by–side to ease readability.

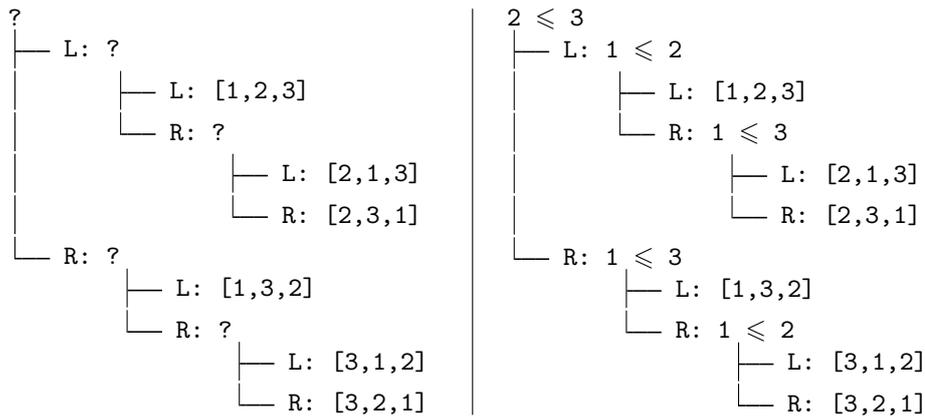

**Figure 3.1.:** Decision tree for `insertionSort coinCmp [1,2,3]` as produced by KiCS2 (left) and in a modified version to highlight compared elements of the list (right)

In order to demystify the generation of permutations using `coinCmp` as predicate, let us revise the implementation of `insert` by inlining the applied predicate `coinCmp`.

```
insertCoin :: a -> [a] -> [a]
insertCoin x []     = [x]
insertCoin x (y:ys) = x:y:ys ? y : insertCoin x ys
```

This variant of `insert` is exactly the definition of a non–deterministic insertion that is used to define a permutation function, for example in the overview of Curry by Hanus (1994). The attentive reader recognises the definition as the definition we discussed in Section 2.2.1 as well as its Haskell equivalent `insertND` we defined using a representation for non–determinism in Section 2.1.2.

As a second step to better understand how `insertSort` computes the permutations of its input list when applied to `coinCmp`, we take a look at the decision tree for the exemplary call above on the left of Figure 3.1. As the comparison function `coinCmp` non–deterministically yields `True` and `False`, the decision tree of our example reflects all possible control flows for `insertionSort` depending on the results of `coinCmp`.

We already introduced the option to draw the decision tree for a non–deterministic expression in Chapter 2. Recall that we do not display labels for the branches if the evaluated expression does not make use of sharing. Moreover, all branches in this decision tree correspond to non–determinism spawned by `coinCmp`.

An interesting insight that we, unfortunately, do not gain from the decision tree is *how* exactly the permutations are computed. In order to make the computation more transparent, we adjust the above decision tree and annotate each branch point with the comparison that takes place. That is, as each branch corresponds to an application of `coinCmp` to arguments, we can annotate the branches with these arguments. The right tree of Figure 3.1 displays the modified decision tree.

Note that the first comparison is between the last two elements of the list: 2 and 3. Thus, `insertionSort` starts by inserting the last element into the empty list, moves one position to the front at each recursive step and inserts the next element to the currently accumulated list. Depending on the decision of `coinCmp` that takes place when evaluating `insert`, the element to be inserted moves one position to the right in the resulting list. For example, the decisions that lead to the permutation [1,2,3] are that 2 and 3 as well as 1 and 2 are already in order with respect to `coinCmp`. The first decision leads to the temporary list `[2,3]`, as 2 and 3 are already in correct order. As next step we want to insert `1`; as `1` and `2` are in the correct order as well, we insert `1` at the front of the list yielding the resulting permutation `[1,2,3]`.

### 3.1.2 Selection Sort

Next, we consider the permutation function derived from selection sort. The key idea of selection sort is to find the minimum of the list and placing it at the front of the resulting list. First, we define a function to find a minimum of a given list parametrised by a comparison function.

```
minList :: (a -> a -> Bool) -> [a] -> a
minList _ [x]      = x
minList p (x:y:ys) = min p x (minList p (y:ys))
```

The function `minList` is partial, because we cannot yield a value in case of the empty list. If the list has one element, we yield that element, otherwise we compare the first element with the minimum of the remaining list. That is, we define a helper function `min`, a parametrised version of a function that takes two arguments and yields the smaller one.

```
min :: (a -> a -> Bool) -> a -> a -> a
min p x y | p x y     = x
          | otherwise = y
```

After picking a minimum, we need to delete that minimum from the list to recursively sort the remaining list. We define `delete` to remove an element from a list.

```
delete :: Eq a => a -> [a] -> [a]
delete _ []     = []
delete x (y:ys) | x == y    = ys
                | otherwise = y : delete x ys
```

Here, it is crucial that we use the comparison function (`==`) to make sure that the element is correctly removed from the list. Now we can define `selectionSort`: we pick the minimum of the input list, place it at the front of the resulting list, and continue with the list without the minimum.

```
selectionSort :: Eq a => (a -> a -> Bool) -> [a] -> [a]
selectionSort _ []         = []
selectionSort p l@(_ : _) = let y = minList p l
                            in y : selectionSort p (delete y l)
```

We are ready to take a look at the resulting permutations.

```
⊢ selectionSort coinCmp [1,2,3]
 [1,2,3]   [2,3,1]
 [1,3,2]   [3,1,2]
 [2,1,3]   [3,2,1]
```

We see that `selectionSort` yields exactly the permutations of the input list like our implementation `insertionSort`. One difference of both implementations becomes apparent when we look at the type: `selectionSort` has an `Eq`–constraint on the elements of the list. As the resulting `Eq`–constraint breaks with the general scheme that we parametrise all functions with a comparison function, we define a function `pickMin` that finds the minimum and yields the remaining list in one traversal of the input list.

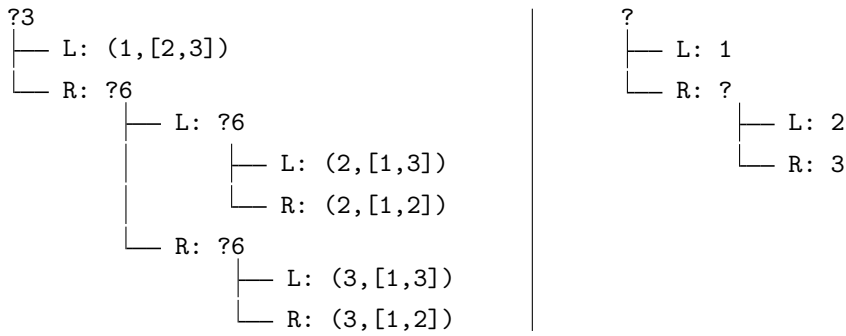

**Figure 3.2.:** Decision trees for the expressions `pickMin coinCmp [1,2,3]` (left) and `minList coinCmp [1,2,3]` (right)

```
pickMin :: (a -> a -> Bool) -> [a] -> (a,[a])
pickMin _ [x]            = (x,[])
pickMin p (x:xs@(_ : _)) = let (m,l) = pickMin p xs
                          in if p x m then (x,xs) else (m,x:l)
```

We adapt the implementation of `selectionSort` to use `pickMin` instead of using a combination of `minList` and `delete` as follows.

```
selectionSortPick :: (a -> a -> Bool) -> [a] -> [a]
selectionSortPick _ []        = []
selectionSortPick p l@(_ : _) = let (m,l') = pickMin p l
                                in m : selectionSortPick p l'
```

Using the adapted implementation `selectionSortPick`, we sort the same example list as before, and get the six expected results.

```
⊢ selectionSortPick coinCmp [1,2,3]
[1,2,3]  [2,3,1]
[1,3,2]  [3,1,2]
[2,1,3]  [3,2,1]
```

Due to the let–bindings used in the implementation of `selectionSortPick` as well as the auxiliary function `pickMin`, the corresponding decision tree is not as easy to look at as for `insertionSort`. Instead, we take a look at the decision tree for `pickMin` and `minList` in Figure 3.2. Note that the decision tree for `pickMin` is an example for the visualisation of sharing: the decision labeled `?6` occurs in each of the two subsequent sub–trees again. This decision originates from the usage of the let–binding for the recursive call in the second rule of `pickMin`. The recursive call `pickMin p xs` binds the minimum, and the list without that minimum to the variables `m` and `l`, respectively. In particular, we cannot use case–expressions without altering the non–strictness property of the original definition using let–bindings. Therefore, in the following, we represent decision trees in a reduced version, where we fuse shared decisions by only displaying the branches that would be considered in a search traversal. Figure 3.3 displays this reduced decision tree — with the additional modification that we label branches with the corresponding comparisons like before. As an interesting aside, note that we cannot give the second argument of the comparison for the first branch — marked with (∗) in Figure 3.3. We do not have this information, because `coinCmp` does not demand its second argument. In the example above the second argument is the expression `pickMin coinCmp [2,3]`; a demand is only necessary if we want to compute the entire list structure of the first solution. We can observe this non–strict behaviour when computing only the head element of the resulting permutations. Consider the following expressions involving a list with `failed` as element.

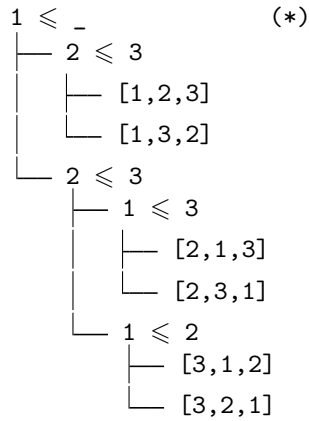

**Figure 3.3.:** Reduced decision tree for `selectionSortPick coinCmp [1,2,3]`

```
⊢ selectionSortPick coinCmp [1,failed]
⊢ head (selectionSort coinCmp [1,failed])
1
```

In theory, the permutations of the list `[1,failed]` are `[1,failed]` and `[failed,1]`; in practice, the expression `failed` is not a value like `42` or `True`, printing `failed` in the REPL propagates the failure to the top–level, that is, there are no results to print. The first expression computes all possible permutations, but the expression does not have any result: `failed` causes the computation to propagate the failure to the top–level. The second expression, however, computes only the head element and yields indeed a solution. That is, in order to yield the first element of the first list, we do not need to compute any further elements of the input list.

### 3.1.3 Bubble Sort

The next sorting function we examine is bubble sort. We define an implementation of the bubble sort algorithm that bubbles the minimum element to the front of a list. Bubbling to the front of the list allows for a more efficient implementation with respect to the selection of the minimum and the remaining list. The following function `bubble` defines the bubbling of the minimum element of the list to the front.

```
bubble :: (a -> a -> Bool) -> [a] -> [a]
bubble _ [x]           = [x]
bubble p (x:xs@(_ : _)) = let (y:ys) = bubble p xs
                          in if p x y then x:y:ys else y:x:ys
```

When used with a deterministic comparison function like (<=), the implementation potentially bubbles more elements than the minimum to the front of the list.

```
⊢ bubble (<=) [2,3,4,1]
[1,2,3,4]
⊢ bubble (<=) [1,3,4,2]
[1,2,3,4]
⊢ bubble (<=) [2,4,3,1]
[1,2,4,3]
⊢ bubble (<=) [1,4,3,2]
[1,2,4,3]
⊢ bubble (<=) [2,1,4,3]
[1,2,3,4]
```

During the traversal of the list, each local minimum of a comparison bubbles towards the front of the list. That is, while the lists in the first two examples have only one element that occurs out of order (1 for the first and 2 for the second example), the next three lists have at least two elements that need to be rearranged. Bubbling the local minimum to the front yields, nevertheless, a sorted list for the last example: the pairs 4 <= 3, 1 <= 4 and 2 <= 1 make sure that 3 and 1 bubble to the right place within the list. This behaviour, however, is rather a lucky coincidence than the normal case for the usage of bubble. The third and fourth lists are counterexamples: when bubbling 1 (in the third example) and 2 (in the fourth example), respectively, to the front of the list, 3 is never picked as local minumum to bubble to the front.

Using a non–deterministic comparison function like coinCmp, bubble enumerates all possible *bubblings*. Consider the following two examples.

```
⊢  bubble coinCmp [1,2,3]
[1,2,3]    [2,1,3]
[1,3,2]    [3,1,2]

⊢ bubble coinCmp [1,2,3,4]
[1,2,3,4]   [2,1,3,4]
[1,2,4,3]   [2,1,4,3]
[1,3,2,4]   [3,1,2,4]
[1,4,2,3]   [4,1,2,3]
```

We can now define `bubbleSort` by means of `bubble`: `bubble` gives us easy access to the minimum of the list that we use as head element and sort the remaining list recursively.

```haskell
bubbleSort :: (a -> a -> Bool) -> [a] -> [a]
bubbleSort _ []         = []
bubbleSort p xs@(_ : _) = let (y:ys) = bubble p xs
                          in y : bubbleSort p ys
```

Finally, the application of a sorting function to the non–deterministic comparison function `coinCmp` yields more results than expected.

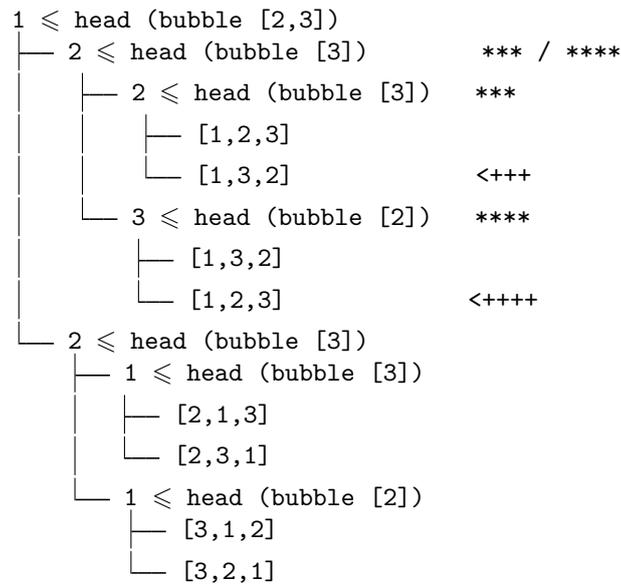

```
1 ⩽ head (bubble [2,3])
├── 2 ⩽ head (bubble [3])         *** / ****
│    ├── 2 ⩽ head (bubble [3])    ***
│    │    ├── [1,2,3]
│    │    └── [1,3,2]             <+++
│    └── 3 ⩽ head (bubble [2])    ****
│         ├── [1,3,2]
│         └── [1,2,3]             <++++
└── 2 ⩽ head (bubble [3])
     ├── 1 ⩽ head (bubble [3])
     │    ├── [2,1,3]
     │    └── [2,3,1]
     └── 1 ⩽ head (bubble [2])
          ├── [3,1,2]
          └── [3,2,1]
```

**Figure 3.4.:** Reduced and modified decision tree for `bubbleSort coinCmp [1,2,3]`

```
⊢ bubbleSort coinCmp [1,2,3]
[1,2,3]    [2,1,3]
[1,3,2]    [2,3,1]
[1,3,2]    [3,1,2]
[1,2,3]    [3,2,1]
```

Instead of six results corresponding to all permutations of the input list, we observe that two permutations occur twice: `[1,2,3]` and `[1,3,2]`. In order to understand why these duplicates occur, we take a look at the decision tree for the example call

   `bubbleSort coinCmp [1,2,3]`

listed in Figure 3.4. Note that the second argument of the comparison function `coinCmp` is, again, an unevaluated expression — we have already seen a similar example when talking about `selectionSort`. The computation of all results in the REPL as well as drawing the decision tree do not demand the evaluation of expressions like `head (bubble [2,3])` until the corresponding list element will be printed.[2]

The new observation about the decision tree for an application of `bubbleSort` is that some results occur twice. The duplicated results are marked with `<+++` and `<++++`, respectively. Note that it is by intention that we marked the first occurence in case of `[1,3,2]` and the second one in case of the result `[1,2,3]`.

Consider the path from the root to the result marked with `<+++`. The crucial parts are the comparisons marked with `***`: we compare 2 and `head (bubble [3])` twice. In a setting where we use a deterministic comparison function, the first path to `[1,3,2]` does not make sense. The decision for the first comparison `2 <= head (bubble [3])` yields `True` — otherwise we would not take the first branch, that is, doing the same comparison a second time should yield `True` again. In the path from the root to the first occurence of `[1,3,2]`, the comparison of 2 and `head (bubble [3])` yields `True` the first and `False` the second time. Note that there is no deterministic comparison function that yields `True` for a comparison first and `False` later. In the following we reference the following property as *behaving consistently*: a predicate behaves consistently, if it yields the same Boolean value for every application to the same values.

---

[2]The actual expression appearing in the evaluation steps is a let–binding that references the head element in the comparison. That is, `let (x:xs) = bubble [2,3] in if 1 <= x then ... else ...` is closer to the expressions occurring during the evaluation.

The other result marked with `<++++` is the second occurence of `[1,2,3]`. Here, the relevant comparisons taken on the path are marked with `****`: `2 <= head (bubble [3])` and `3 <= head (bubble [2])`. Note that the expression `head (bubble [x])` evaluates to `x` for all `x` of appropriate type. In our concrete cases, we have `head (bubble [3]) = 3` and `head (bubble [2]) = 2`. Below we only reference the comparisons with the evaluated function calls to simplify the reasoning. As the second comparison takes place in the second branch of the decision corresponding to `2 <= 3`, we know that the decision was `False`. That is, 2 is not smaller than or equal to 3, 2 is greater than 3. For the result marked with `<++++` we then, however, take the second branch for the decision `3 <= 2`: 2 is greater than 3. Again, this decision does not make sense when using `bubbleSort` with a reasonable comparison function like `(<=)` on integer values. The important property missing here is that the relation described by `<=` needs *connexity*. The connex property describes that any two pairs of elements of appropriate type are comparable with respect to an ordering relation $\leqslant$.

$$\forall xy, x \leqslant y \lor y \leqslant x \qquad \text{(Connexity)}$$

Binary relations that are total orders (like for example the less–than–or–equal–to–relation on integers) fulfil the connex property by definition. This observation implies that the implementation of `bubbleSort` needs to be used with a comparison function corresponding to a binary relation that is a total order, otherwise it cannot sort all lists correctly.

### 3.1.4 Quicksort

In this section we take a look at an implementation of *quicksort* that is known for its particular declarative nature. The general idea behind the quicksort algorithm is to choose a pivot element and split the remaining list into two parts: one with all elements that are smaller than or equal to the pivot element and one with the all elements that do not fulfil this criterion. The algorithm is then recursively applied to these two sublists.

The following function `quickSort` uses the head element of the given list as the pivot element and the function `filter` to split the input list into the two parts.

```haskell
quickSort :: (a -> a -> Bool) -> [a] -> [a]
quickSort _ []     = []
quickSort p (x:xs) = let l1 = filter (\y -> p y x) xs
                         l2 = filter (\y -> not (p y x)) xs
                     in quickSort p l1 ++ x : quickSort p l2
```

Let us take a look at the resulting permutations as well as the number of results for exemplary calls. Note that we use the encapsulation function `allValues` as well as the convenience functions introduced in Section 2.2.4 to count the number of deterministic values in the resulting multiset by means of a helper function `lengthValues`.

```
⊢ quickSort coinCmp [1,2]
[2,1]
[2,1,2]
[1]
[1,2]

⊢ lengthValues (allValues (quickSort coinCmp [1,2,3]))
49

⊢ lengthValues (allValues (quickSort coinCmp [1,2,3,4]))
4225
```

That escalated quickly. The first example call was already odd: for a list with two elements we get four results, of which two are not even permutations of the original list. The second and third example illustrate why we did not print all results for a list with more than two elements: the number of results grows way worse than expected for an algorithm that enumerates permutations. The expected number of results for three elements is $3! = 6$ and $4! = 24$ for four elements, respectively.

A quick look at the two filter calls alone will get us closer to understand what is going on here.

```
⊢ filter (coinCmp 1) [2]
[2]
[]

⊢ filter (not . coinCmp 1) [2]
[]
[2]
```

Both bindings, `l1` and `l2`, evaluate to the same two values, just in reversed order. These two results times two sublists yield four overall results, since all lists for `l1` are combined with all lists of `l2` to yield `l1 ++ x : l2`. In general, the resulting function enumerates the cross product of all subsequences of the input list.

We can again observe that the definition of `quickSort` uses the comparison function `coinCmp` inconsistently, thus, yields the same results for both application of filter. Using a deterministic comparison function `p`, the sublists `l1` and `l2` are disjoint, since we are using `p` for the first and `not . p` for the second list.

We can overcome this drawback of using two `filter` applications by using `split` instead. The function `split :: (a -> Bool) -> [a] -> ([a],[a])` traverses a list only once to split it into two sublists fulfilling exactly the property we use above: the first list contains only elements that fulfil the given predicate and the second list contains the elements that do not. We define `split` as follows; note that it is crucial that split traverses its input list only once.

```
split :: (a -> a -> Bool) -> a -> [a] -> ([a],[a])
split p x l = split' l ([],[])
 where
  split' []     ls                 = ls
  split' (y:ys) (l1,l2) | not (p x y) = split' ys (l1,y:l2)
                       | otherwise   = split' ys (y:l1,l2)
```

Based on `split` we can define a second version of quicksort. We reuse the variable names `l1` and `l2` for the two sublists to indicate that the underlying idea of the algorithm stays exactly the same.

```
quickSortSplit :: (a -> a -> Bool) -> [a] -> [a]
quickSortSplit _ []     = []
quickSortSplit p (x:xs) = let (l1,l2) = split p x xs
                          in quickSortSplit p l1 ++ x : quickSortSplit p l2
```

Let us once again take a look at the resulting permutations.

```
⊢ quickSortSplit coinCmp [1,2,3]
 [2,3,1]   [3,1,2]
 [3,2,1]   [1,2,3]
 [2,1,3]   [1,3,2]
```

As `split` only traverses its input list once, each comparison happens only once. The pivot element is compared to each element of the remaining list, the recursive calls to `quickSortSplit` then pick new pivot elements that are used for comparison. That is, we end up with exactly all permutations of the input list, again.

### 3.1.5 Merge Sort

Last but not least, we take a look at merge sort. The general idea of the sorting algorithm is to divide the input list of length $n$ into $n$ sublists, that is, all sublists are singleton lists. These singleton lists, which are trivially sorted, are merged into new sorted sublists. The algorithm repeats the merging step until only one list remains: the resulting sorted list.

First, we define a function `divideN` to divide the input lists into two lists of approximately the same length. The first list is shortened by one element in case of an odd number of elements.

```
divideN :: [a] -> ([a],[a])
divideN xs = divideN' xs (length xs `div` 2)
 where   divideN' []     _           = ([],[])
         divideN' (y:ys) n | n == 0  = ([],y:ys)
                           | otherwise = let (l1,l2) = divideN' ys (n-1)
                                         in (y:l1,l2)
```

Second, we define a function `merge` that merges two lists based on a comparison function.

```
merge :: (a -> a -> Bool) -> [a] -> [a] -> [a]
merge _ []     l      = l
merge _ (x:xs) []     = x:xs
merge p (x:xs) (y:ys) | p x y     = x : merge p xs (y:ys)
                      | otherwise = y : merge p (x:xs) ys
```

The first two rules cover the case that it is not necessary to merge if one of the lists is empty: we just yield the other one. If both lists are non–empty, we decide based on the comparison function `p` which element to put at the front of the resulting list and merge recursively with the remaining lists. Note that we assume that both list arguments of `merge` are already sorted. Finally, we define the overall sorting function `mergeSort` that uses `divideN` to divide the lists into two sublists that are sorted recursively, and `merge` to merge the sorted sublists.

```
mergeSort :: (a -> a -> Bool) -> [a] -> [a]
mergeSort _ []            = []
mergeSort _ [x]           = [x]
mergeSort p l@(_ : (_ : _)) = let (l1,l2) = divideN l
                              in merge p (mergeSort p l1) (mergeSort p l2)
```

Once again we take a look at an exemplary application of `mergeSort` to the non–deterministic comparison function `coinCmp`.

⊢ mergeSort coinCmp [1,2,3]
[1,2,3]   [1,3,2]
[2,1,3]   [3,1,2]
[2,3,1]   [3,2,1]

We are pleased to observe that the resulting function enumerates exactly all permutations. The procedure followed by `mergeSort` does not make any redundant and, thus, no inconsistent comparisons. The first step of dividing the list into sublists is not based on the predicate but based on the length of the list only. The comparison takes place when using `merge`. As `merge` has the precondition that both argument lists are already sorted, it is sufficient to compare the lists element–by–element. That is, if one head element is smaller than the other head element, we put the former in front of the list and do not need to consider it anymore. The utilisation of the precondition leads to an efficient sorting algorithm: no two elements are compared more than once.

## 3.2 Non–deterministic Sorting Functions in Haskell

After discussing non–deterministic sorting functions in the functional language Curry with built–in non–determinism, we switch to Haskell as an exemplary functional language without non–determinism. We reimplement a selection of the sorting functions introduced in Section 3.1 in Haskell using a naive model of non–determinism based on lists. As we want to test out different models later, we refactor the list–specific implementations to monad–generic implementations for the sorting algorithm.

We will notice a difference between the Curry and the Haskell implementation when testing the sorting functions on concrete lists. This difference is not a new insight, but interesting nonetheless: Curry's non–determinism can exploit non–strictness in a way the common Haskell model of non–determinism using a monadic interface cannot.

### 3.2.1 Modelling Non–determinism

In a pure functional language like Haskell, we can express non–deterministic functions using lists to represent multiple non–deterministic results as we have already introduced and discussed in Section 2.1.2. That is, we reuse the type synonym `ND` in order to distinguish between list values that are used to model non–determinism and list values in the common sense. Recall that we use the monadic operations `return` and `(>>=)` for lists when working with `ND` as well as the convenience operator `(?)` to combine multiple non–deterministic results.

```
instance Monad ND where
  return x = [x]
  xs >>= f = concat (map f xs)
```

```
(?) :: ND a -> ND a -> ND a
(?) = (++)
```

Using the monadic abstraction and the helper function, we can define the non–deterministic comparison function `coinCmpND` — corresponding to the function `coinCmp` that we have used in Curry before, which transfers easily to the list model in Haskell.

```
coinCmpND :: a -> a -> ND Bool
coinCmpND _ _ = [True] ? [False]
```

**Example: Non–deterministic application of filter** Equipped with these auxiliary functions, let us consider the Haskell function `filterND :: (a -> ND Bool) -> [a] -> ND [a]`, which is a non–deterministic extension of the higher–order function `filter`.

```
filterND _ []     = return []
filterND p (x:xs) = p x >>= \b ->
                      if b then filterND p xs >>= \ys -> return (x:ys)
                           else filterND p xs
```

Note that the potentially non–deterministic values occur in the result of the predicate and in the resulting type of the overall function `filterND`; moreover, the input list is a deterministic argument. We need to process the potentially non–deterministic computation resulting from the predicate check `p x` and the recursive call `filterND p xs` using (>>=) to handle each possible value of the computation. The attentive reader notices that the definition of `filterND` is not specific to the specified type `ND`, but works for any monad. That is, since we solely rely on the abstractions provided by monads, we can generalise the type definition. The resulting definition is `filterM :: Monad m => (a -> m Bool) -> [a] -> m [a]`; the implementation stays the same.[3]

When running concrete examples, we then instantiate the monadic contexts with `ND` to illustrate the behaviour of a non–deterministic version.

Since `filter` needs to be applied to a unary predicate, we partially apply `coinCmpND` with 42 in the examples.

```
λ> filterM (coinCmpND 42) [1,2,3]
{ [1,2,3], [1,2], [1,3], [1], [2,3], [2], [3], [] }
```

As a side note, consider the following urge to outsource the duplicate call to `filterM p xs` in both branches of the if–then–else–expression.

```
filterM' :: Monad m => (a -> m Bool) -> [a] -> m [a]
filterM' _ []     = return []
filterM' f (x:xs) = f x >>= \p ->
                      filterM' f xs >>= \ys ->
                      return (if p then x:ys else ys)
```

This transformation, which computes the non–deterministic computation `filterM p xs` only once, is still equivalent to the original implementation of `filterM`.

---

[3]Note that the definition of `filterM` is based on the `Applicative` instead of `Monad` type class now. http://hackage.haskell.org/package/base-4.12.0.0/docs/Control-Monad.html#v:filterM (last accessed: 2019–09–10)

```
λ> filterM' (coinCmpND 42) [1,2,3]
{ [1,2,3], [1,2], [1,3], [1], [2,3], [2], [3], [] }
```

We must be aware, however, that the transformation is only valid because we use the result of `filterM p xs` in both branches of the if–then–else–expression. In the next paragraph we discuss an example that yields different results before and after such a transformation.

**Example: Non–deterministic application of insert**   Consider the following two monadic versions of the function `insert` we defined in Curry.

```haskell
insertM :: Monad m => (a -> a -> m Bool) -> a -> [a] -> m [a]
insertM _ x []     = return [x]
insertM p x (y:ys) = p x y >>= \b ->
                     if b  then return (x:y:ys)
                           else insertM p x ys >>= \zs -> return (y:zs)

insertM' :: Monad m => (a -> a -> m Bool) -> a -> [a] -> m [a]
insertM' _ x []     = return [x]
insertM' p x (y:ys) = p x y >>= \b ->
                      insertM' p x ys >>= \zs ->
                      return (if b then x:y:ys else y:zs)
```

The alternative version `insertM'` computes the potentially non–deterministic computation of the recursive call to `insertM' p x ys` before checking the condition `b` such that it does not behave as the original version of `insertM` anymore.

```
λ> insertM coinCmpND 1 [2,3]
{ [1,2,3], [2,1,3], [2,3,1] }

λ> insertM' coinCmpND 1 [2,3]
{ [1,2,3], [1,2,3], [2,1,3], [2,3,1] }
```

The exemplary calls using the non–deterministic comparison function `coinCmpND` do not yield the same results. When we apply a monadic version of insert to `coinCmpND`, we expect $n + 1$ results for a input list of length $n$ — the same result we observed in Curry. The application `insertM' coinCmpND`, however, yields $2^n$ results.

```
λ> length (insertM' coinCmpND 1 [2,3])
4

λ> length (insertM' coinCmpND 1 [2..4])
8

λ> length (insertM' coinCmpND 1 [2..11])
1024
```

Due to the call to `insertM' p x ys` before checking the Boolean value `b`, we need to evaluate the recursive call, even though we do not need the resulting variable binding `zs` when taking the then–branch. The important insight is that we need to be careful when using the `(>>=)`–operator. In most settings, and the list instance is no exception, `(>>=)` needs to be interpreted as a sequencing operator that is strict in its first argument. That is, if we have an expression `mx >>= f`, we cannot proceed with `f` without evaluating `mx` first.

In order to check the claim about the strictness of (>>=) in case of `ND`, recall that the corresponding `Monad` instance for `ND` is the one for lists based on `concat` and `map`. That is, let us retake a look at the definition of `concat` to see that the resulting function is indeed strict in its argument of type `ND a`.

```
concat :: [[a]] -> [a]
concat []       = []
concat (xs:xss) = xs ++ concat xss
```

The function definition of `concat` makes a case distinction on its first argument. That is, in order to evaluate an expression like

```
insertM' p x ys >>= \zs -> return (if b then x:y:ys else y:zs)
```

we need to evaluate `insertM' p x ys` first. In this example, we trigger the evaluation of the non–deterministic comparison function `coinCmpND` although we do not need the result `zs` if the condition `b` is `True`.

As example, consider the excerpt of a step–wise evaluation of the example from above listed in Figure 3.5. Note that we need to evaluate `filterM' (coinCmpND 42) [1,2,3]` and all recursive calls of `filterM'` that arise during evaluation.

### 3.2.2 Drawing Decision Trees

Thanks to the generic implementation using a monadic interface, we are free to use whatever instance fits our purpose to actually run the sorting functions. For example, we can generate decision trees like in Curry by using a monad that keeps track of all operations and pretty–prints the non–deterministic parts of our computation. As first step to define such a pretty–printing function, we generalise the comparison function `coinCmpND` to `MonadPlus`, which is an extension of the `Monad` type class that introduces an additional function `mplus` to combine monadic computations, and `mzero` as neutral element for the function `mplus`.

```
class Monad m => MonadPlus m where
  mplus :: m a -> m a -> m a
  mzero :: m a
```

The idea of the non–deterministic comparison function `coinCmpND` is to yield two results non–deterministically. In the concrete implementation using lists, we define `coinCmpND` based on singleton lists `[True]` and `[False]` that are combined using the concatenation operator (++). A generalisation using `MonadPlus` replaces the concatenation operator by `mplus`.

```
coinCmp :: MonadPlus m => a -> a -> m Bool
coinCmp _ _ = return True `mplus` return False
```

As second step, we use a monad instance that can interpret all monadic operations in an abstract way: the free monad (Swierstra, 2008) we introduced in Section 2.1.3. As we are interested in pretty–printing the non–deterministic components of our monadic computations, we need a suitable functor to model non–determinism. The important primitive operations of non–determinism are exactly the ones provided by the `MonadPlus` type class: an operator to combine two effectful computations and the failing computation. Note that the simplified version in the introduction (Section 2.1.3) does not have a representation

$$
\begin{aligned}
&\quad\textit{filterM}'\ (\textit{coinCmpND}\ 42)\ [1,2,3] \\
&= \quad\{\ \text{Definition of } \textit{filterM}'\ \} \\
&\quad \textit{coinCmpND}\ 42\ 1 \ggg \lambda p \to \textit{filterM}'\ (\textit{coinCmpND}\ 42)\ [2,3] \\
&\qquad\qquad\qquad \ggg \lambda \textit{ys} \to \textit{return}\ (\textbf{if}\ p\ \textbf{then}\ 1:\textit{ys}\ \textbf{else}\ \textit{ys}) \\
&= \quad\{\ \text{Definition of } \textit{coinCmpND}\ \} \\
&\quad [\textit{True}, \textit{False}] \ggg \lambda p \to \textit{filterM}'\ (\textit{coinCmpND}\ 42)\ [2,3] \ggg \lambda \textit{ys} \to \\
&\qquad\qquad\qquad \textit{return}\ (\textbf{if}\ p\ \textbf{then}\ 1:\textit{ys}\ \textbf{else}\ \textit{ys}) \\
&= \quad\{\ \text{Definition of } (\ggg)\ \} \\
&\quad \textit{filterM}'\ (\textit{coinCmpND}\ 42)\ [2,3] \ggg \lambda \textit{ys} \to \textit{return}\ (\textbf{if}\ \textit{True}\ \textbf{then}\ 1:\textit{ys}\ \textbf{else}\ \textit{ys}) \\
&\quad +\!\!+ \\
&\quad [\textit{False}] \ggg \lambda p \to \textit{filterM}'\ (\textit{coinCmpND}\ 42)\ [2,3] \ggg \lambda \textit{ys} \to \\
&\qquad\qquad\qquad \textit{return}\ (\textbf{if}\ p\ \textbf{then}\ 1:\textit{ys}\ \textbf{else}\ \textit{ys}) \\
&= \quad\{\ \text{Definition of } (\ggg)\ \} \\
&\quad \textit{filterM}'\ (\textit{coinCmpND}\ 42)\ [2,3] \ggg \lambda \textit{ys} \to \textit{return}\ (\textbf{if}\ \textit{True}\ \textbf{then}\ 1:\textit{ys}\ \textbf{else}\ \textit{ys}) \\
&\quad +\!\!+ \\
&\quad \textit{filterM}'\ (\textit{coinCmpND}\ 42)\ [2,3] \ggg \lambda \textit{ys} \to \textit{return}\ (\textbf{if}\ \textit{False}\ \textbf{then}\ 1:\textit{ys}\ \textbf{else}\ \textit{ys}) \\
&\quad +\!\!+ \\
&\quad [\,] \ggg \lambda p \to \textit{filterM}'\ (\textit{coinCmpND}\ 42)\ [2,3] \ggg \lambda \textit{ys} \to \\
&\qquad\qquad\qquad \textit{return}\ (\textbf{if}\ \textit{False}\ \textbf{then}\ 1:\textit{ys}\ \textbf{else}\ \textit{ys}) \\
&= \quad\{\ \text{Definition of } \textit{filterM}'\ \} \\
&\quad (\textit{coinCmpND}\ 42\ 2 \ggg \lambda p \to \textit{filterM}'\ (\textit{coinCmpND}\ 42)\ [3] \ggg \lambda \textit{ys} \to \\
&\qquad\qquad\qquad\qquad \textit{return}\ (\textbf{if}\ p\ \textbf{then}\ 1:\textit{ys}\ \textbf{else}\ \textit{ys})) \ggg \lambda \textit{ys} \to \\
&\quad \textit{return}\ (\textbf{if}\ \textit{True}\ \textbf{then}\ 1:\textit{ys}\ \textbf{else}\ \textit{ys}) \\
&\quad +\!\!+ \ldots +\!\!+ \ldots \\
&= \quad\{\ \text{Definition of } \textit{coinCmpND}\ \} \\
&\quad [\textit{True}, \textit{False}] \ggg \lambda p \to \textit{filterM}'\ (\textit{coinCmpND}\ 42)\ [3] \ggg \lambda \textit{ys} \to \\
&\qquad\qquad\qquad \textit{return}\ (\textbf{if}\ p\ \textbf{then}\ 1:\textit{ys}\ \textbf{else}\ \textit{ys})) \ggg \lambda \textit{ys} \to \\
&\quad \textit{return}\ (\textbf{if}\ \textit{True}\ \textbf{then}\ x:\textit{ys}\ \textbf{else}\ \textit{ys}) \\
&\quad +\!\!+ \ldots +\!\!+ \ldots \\
&= \quad\{\ \text{Definition of } (\ggg)\ \} \\
&\quad (\textit{filterM}'\ (\textit{coinCmpND}\ 42)\ [3] \ggg \lambda \textit{ys} \to \textit{return}\ (\textbf{if}\ \textit{True}\ \textbf{then}\ 1:\textit{ys}\ \textbf{else}\ \textit{ys}) \\
&\qquad +\!\!+ \textit{filterM}'\ (\textit{coinCmpND}\ 42)\ [3] \ggg \lambda \textit{ys} \to \textit{return}\ (\textbf{if}\ \textit{False}\ \textbf{then}\ 1:\textit{ys}\ \textbf{else}\ \textit{ys}) \\
&\qquad +\!\!+ [\,] \ggg \lambda p \to \textit{filterM}'\ (\textit{coinCmpND}\ 42)\ [3] \ggg \lambda \textit{ys} \to \\
&\qquad\qquad\qquad \textit{return}\ (\textbf{if}\ p\ \textbf{then}\ 1:\textit{ys}\ \textbf{else}\ \textit{ys})) \ggg \lambda \textit{ys} \to \\
&\quad \textit{return}\ (\textbf{if}\ \textit{True}\ \textbf{then}\ 1:\textit{ys}\ \textbf{else}\ \textit{ys}) \\
&\quad +\!\!+ \ldots +\!\!+ \ldots \\
&= \quad\{\ \text{Definition of } \textit{filterM}'\ \} \\
&\quad \ldots
\end{aligned}
$$

**Figure 3.5.:** Extract of a step–wise evaluation of `filterM' (coinCmpND 42) [1,2,3]`

for the latter computation. Since we want to print the arguments the non–deterministic comparison function is applied to, we store additional information in the constructor `Choice` as follows.

```haskell
data Sort a = Choice (Maybe (String,String)) a a | Fail deriving Show
```

In order to use `Free Sort` as underlying monad in a non–deterministic application of, for example, `filterM coinCmp`, we need to define a functor instance for `Sort` and a `MonadPlus` instance for `Free Sort`.

```haskell
instance Functor Sort where
 fmap f (Choice id m1 m2 ) = Choice id (f m1) (f m2)
 fmap _ Fail               = Fail

instance MonadPlus (Free Sort) where
 mzero      = Impure Fail
 mplus m1 m2 = Impure (Choice Nothing m1 m2)
```

Note that, initially, we do not have any information about the arguments of the `mplus` operator, so we use `Nothing`. We add information to the structure when we apply the function that introduces non–determinism. For example, we define the non–deterministic function `cmpCoinFree` that stores the string representation of its arguments and non–deterministically yields `True` and `False` as follows.

```haskell
coinCmpFree :: Show a => a -> a -> Free Sort Bool
coinCmpFree x y =
  Impure (Choice (Just (show x,show y)) (return True) (return False))
```

Now we can apply `filterM` to our non–determinism–tracking comparison function `cmpCoinFree` and get a term of type `Free Sort` that contains information about the arguments that need to be compared.

```
λ> filterM (coinCmpFree 42) [1,2]
Impure (Choice (Just ("42","1"))
        (Impure (Choice (Just ("42","2")) (Pure [1,2]) (Pure [1])))
        (Impure (Choice (Just ("42","2")) (Pure [2])   (Pure []))))
```

Since this term representation looks more complicated than helpful, as last step, we define a pretty–printing function for `Free Sort`. The function

```haskell
pretty :: Show a => Free Sort a -> String
```

produces a decision tree similar to the one we got to know from Curry. Now we take a look at the well–arranged decision tree resulting from the above call.

```
λ> putStrLn (pretty (filterM (coinCmpFree 42) [1,2]))
                        +-[1,2]
             +- 42 <= 2  -+
             |          +-[1]
 +- 42 <= 1  -+
             |          +-[2]
             +- 42 <= 2  -+
                        +-[]
```

We will use these drawing capabilities in the next section when we compare our implementation of sorting functions in Haskell with the implementation in Curry.

### 3.2.3 Curry versus Monadic Non–determinism

With this insight about the strictness of (>>=) in mind, we check out the consequences when applying a non–deterministic comparison function to monadic sorting functions. That is, we transform the Curry implementation discussed in Section 3.1 to Haskell.

**Insertion Sort**  As we have just seen the definition of `insertM`, we start with `insertionSort`.

```
insertionSortM :: Monad m => (a -> a -> m Bool) -> [a] -> m [a]
insertionSortM _ []     = return []
insertionSortM p (x:xs) = insertionSortM p xs >>= \ys -> insertM p x ys
```

Note that it is again crucial to introduce potentially non–deterministic values only as result of the comparison function and the result of the function itself. This observation also applies to the definition of `insertM`: the input list `ys` needs to be deterministic. That is, in order to insert the head element `x` into the already sorted tail, we unwrap the monadic context using (>>=) and apply `insertM` to each possible value of the computation `insertionSortM p xs`.

Applying `insertionSortM` to `coinCmpND` and exemplary list values yields the expected permutations, more precisely, exactly the permutations of the input list.

```
λ> insertionSortM coinCmpND [1..3]
{ [1,2,3], [2,1,3], [2,3,1], [1,3,2], [3,1,2], [3,2,1] }

λ> let fac n = if n == 0 then 1 else n * fac (n-1) in
       all (\ n -> length (insertionSortM coinCmpND [1..n]) == fac n)
           [1..10]
True
```

The second example call checks for lists of length 1 to 10, if the number of non–deterministic results is equal to the factorial of the corresponding length, which is indeed the case. Now we have a good feeling that both implementations — the Haskell and the Curry version — compute the same number of results. The interesting question is, however, if they behave the same in all contexts.

Recall that the Curry implementation defines `insertionSortM` using a let–declaration for the recursive call. This recursive call only has to be evaluated if we demand more than one element of the resulting list. In the example below, we call `insertionSortM` on a non–empty list to compute the head element of all non–deterministic results and count the number of non–deterministic results afterwards.

```
λ> map (\ n -> length (insertionSortM coinCmpND [1..n] >>= \ xs ->
                              return (head xs)))
                [5..10]
[120,720,5040,40320,362880,3628800]
```

Again, we have $n!$ non–deterministic results for an input list of length $n$. The result illustrates that all resulting permutations need to be computed to yield the corresponding head element. Next, we compare the behaviour of the Haskell implementation with the Curry implementation `insertionSort`.

```
⊢ map (\ n -> length (allValues (head (insertionSort coinCmp [1..n]))))
               [5..10]
[16,32,64,128,256,512]
```

In Curry we do not need to evaluate all non–deterministic computations to yield the head element. Instead of $n!$ number of non–deterministic results, we only get $2^{(n-1)}$ results for an input list of length $n$. The crucial difference between the Haskell and the Curry implementation with respect to the model of non–determinism is that Haskell's non–determinism is flat, while in Curry non–deterministic computations can occur in arbitrarily deep positions. Here, deep position means that the non–determinism is not visible at the outermost constructor, but hides in the component of a constructor.

Consider the following non–deterministic expression `exp` of type `[Bool]` and its projection to the head element and tail, respectively, in Curry.

```
⊢ let exp = True : ([] ? [False]) in head exp
True
⊢ let exp = True : ([] ? [False]) in tail exp
[]
[False]
```

The list `exp` is non–deterministic in its tail component, the head element is deterministic and the top–level list constructor `(:)` is also deterministic. That is, on the one hand applying `head` to `exp` does not trigger any non–determinism, the evaluation yields a deterministic result, namely `True`. On the other hand the non–determinism appears in the overall result when we project to the tail of the list `exp`. This application yields the two results `[]` and `[False]`. In contrast, we cannot model the same behaviour in Haskell when using a list–based model for non–deterministic computations.

```
λ> let exp = True : (return [] ? return [False]) in head exp
    * Couldn't match expected type '[Bool]'
                with actual type 'ND [Bool]'
    * In the second argument of '(:)', namely
        '(return [] ? return [False])'
      In the expression: True : (return [] ? return [False])
      In an equation for 'exp':
          exp = True : (return [] ? return [False])
```

The error message says that the list constructor `(:)` expects a second argument of type `[Bool]`, but we apply it to an argument of type `ND [Bool]`. Due to the explicit modelling of non–determinism that is visible in the type–level — using `ND` — we cannot construct non–deterministic computations that occur deep in the arguments of constructors like `(:)` out of the box. In contrast, Curry's non–determinism is not visible on the type–level, such that we can use non–determinism expressions in any constructor argument without altering the type of the expression. We can reconcile the computation we want to express with the explicit non–determinism in Haskell by binding the non–deterministic computation first and reuse the list constructor then.

```
λ>  return [] ? return [False] >>= \ nd ->
       let exp = True : nd in return (head exp)
{ True, True }

λ>  return [] ? return [False] >>= \ nd ->
       let exp = True : nd in return (tail exp)
{ [], [False] }
```

In this case, however, the non–determinism is definitely triggered: even though head does not need to evaluate its tail — where the non–determinism occurs, the first argument of (>>=) is evaluated. The overall computation then yields two results. All in all, the main insight here is that the non–determinism in Curry can occur deep within data structure components and gives us the possibility to exploit non–strictness. In contrast, the naive Haskell model using lists can only express flat non–determinism, that is, all possibly deep occurrences of non–determinism are pulled to the top–level.

**Selection Sort** Whereas the application of insertion sort to a non–deterministic comparison function yields the same number of results for the Haskell as well as the Curry implementation, we will now take a look at an example that yields duplicate results: selection sort. We directly define the version of selection sort that uses pickMinM instead of traversing the list twice.

```
pickMinM :: Monad m => (a -> a -> m Bool) -> [a] -> m (a, [a])
pickMinM _ [x]    = return (x,[])
pickMinM p (x:xs) = pickMinM p xs >>= \(m,l) ->
                    p x m >>= \b ->
                    return (if b then (x,xs) else (m, x:l))

selectionSortM :: Monad m => (a -> a -> m Bool) -> [a] -> m [a]
selectionSortM _ [] = return []
selectionSortM p xs = pickMinM p xs >>= \(m,l) ->
                     selectionSortM p l >>= \ys ->
                     return (m:ys)
```

The application of selectionSortM to coinCmpND yields more results than expected, the resulting function enumerates some permutations multiple times.

```
λ> selectionSortM coinCmpND [1,2,3]
{ [1,2,3], [1,3,2], [2,1,3], [2,3,1]
, [1,2,3], [1,3,2], [3,1,2], [3,2,1] }

λ> all (\ n -> length (selectionSortM coinCmpND [1..n]) == 2^($\frac{n(n-1)}{2}$))
                [1..7]
True
```

In fact, we get

$$2^{\frac{n(n-1)}{2}}$$

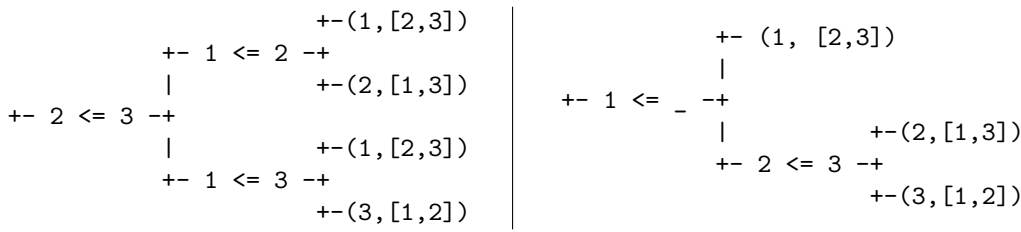

**Figure 3.6.:** Decision trees for the expressions `pickMinM coinCmpND [1,2,3]` in Haskell (left) and `pickMin coinCmp [1,2,3]` in Curry (right)

results for an input list of length $n$. Note that this function grows much faster than the number of permutations $n!$. For example, for $n = 10$ there are $n! = 3628800$ permutations, whereas an application of `selectionSort` to the list `[1..10]` yields

$$2^{\frac{10*9}{2}} = 2^{45} = 35184372088832$$

results.

More generally, for $n \geqslant 7$ we have that $n \leqslant 2^{\frac{n-1}{2}}$ such that we can make the following estimation..[4]

$$n! \leqslant n^n \leqslant 2^{\frac{n-1}{2}n} = 2^{\frac{n*(n-1)}{2}}$$

Since the number of results for `selectionSort` applied to a non–deterministic comparison function differs from the result we got for the Curry implementation, we compare the underlying decision trees. The non–determinism produced by `selectionSort` arises from the usage of `coinCmpND`, which is only evaluated in the auxiliary function `pickMinM`. That is, it is sufficient to take a look at the decision tree for a sub–call of `pickMinM` to detect the different behaviour. We compute the decision tree displayed left in Figure 3.6 by applying a free monad based data type as described in Section 3.2.2. The right side of the figure recapitulates the decision tree when using the Curry implementation.

The monadic version is more strict: the recursive call to `pickMinM` needs to be evaluated in order to apply the predicate `p`. In the Curry version, however, we can already take the `True`–branch for the application of `p` without considering the recursive call first. Thus, the first result `(1, [2,3])` triggers only one non–deterministic decision in Curry. Of course, the number of unnecessarily triggered non–deterministic decisions in the Haskell version increases with each recursive call of `pickMinM`. That is, the number of duplicate results increases with the length of the list.

```
λ> map  (\ n -> length (pickMinM coinCmpND [1..n] >>= return . fst ))
          [1..20]
[1,2,4,8,16,32,64,128,256,512,1024,2048,4096,8192,16384,32768,
65536,131072,262144,524288]
```

More precisely, `pickMinM coinCmpND xs` yields $2^{\text{length xs}}$ results, while the Curry version only yields *length xs* results. Note that the Curry version is what we expect in the first place: picking a minimum with a non–deterministic predicate is basically a function that non–deterministically yields each element of the list.

In the end, `pickMinM` and `pickMin`, respectively, are the functions used to implement the selection sort algorithm and, thus, determine the number of permutations. Whereas

---
[4]We prove the first inequation by induction in Appendix A.1.

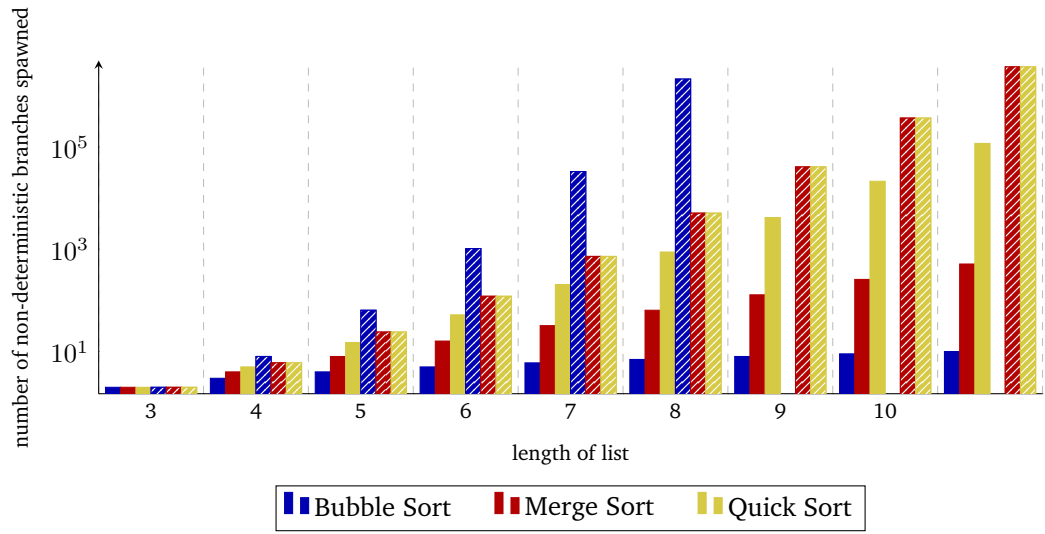

**Figure 3.7.:** Comparison of the number of triggered non–deterministic computations for demanding the head element of all permutations

`selectionSort` yields only the permutations of the input list in Curry, we get duplicate permutations in the Haskell version.

**Other Sorting Algorithms**   The remaining sorting algorithms discussed in Section 3.1 — bubble sort, quick sort and merge sort — yield the same results for the monadic Haskell version as they do in Curry. However, we observe similar effects as with `insertionSortM` in Section 3.2.3 concerning non–strictness. When we demand only the head elements of all permutations, the monadic Haskell versions need to trigger more non–determinism than is necessary in the Curry version. Figure 3.7 visualises the number of triggered non–deterministic computations that are necessary to compute only the head element of all permutations. We observe that all Curry implementations (visualised by the colour-filled bars) compute less non–deterministic computations than all Haskell implementations. One interesting contrast is the behaviour of bubble sort: the Curry version only needs to trigger one non–deterministic computation for each element of the list. That is, the number of non–deterministic computations is linear in the length of the list, whereas the Haskell version triggers $n!$ non–deterministic computations for an input list of length $n$. Note that the evaluation of all permutations for bubble sort needs to trigger $n!$ non–deterministic computations as well, that is, in this case demanding only the head of each permutations is as strict as evaluating all list elements for each permutation.

## 3.3 Future Work

While we started first investigations to compare the strictness behaviour of different sorting functions, we think that a more rigorous investigation might lead to further interesting properties. For example, it would be interesting to analyse the run–time behaviour for the resulting permutation functions. As we only compared the non–deterministic version in Curry and the monadic version in Haskell, it would be interesting to compare the non–deterministic version with deterministic permutation algorithms as well. For example, according to Sedgewick (1977) the classical permutation algorithm based on inserting an element at each possible position in a list was developed independently by Trotter (1962) and

**3.3**   Future Work   |   **51**

Johnson (1963). The implementation of the `permutations :: [a] -> [[a]]` function in Haskell is also based on this approach but has been improved with respect to non–strictness via a mailing list discussion by van Laarhoven (2007). A comparison of the improved permutation algorithm used in Haskell and the best–performing non–deterministic Curry version would be an interesting topic for future work.

A different line of future work could be to focus more on the resulting permutations function. For example, we noticed only by chance that selection sort enumerates all permutation in lexicographic order when evaluated using depth–first search instead of the default behaviour of breadth–first search. It would be interesting to analyse the order of results for all other sorting algorithms as well. This analysis can then be extended to other non–deterministic sorting functions, which are perhaps more strict than `coinCmp`, but yield an interesting property. An example of an alternative non–deterministic comparison function is the following definition of `liftCmp` that lifts a comparison function `cmp` into a monadic context.

```
liftCmp :: (a -> a -> Bool) -> a -> a -> Bool
liftCmp cmp x y = cmp x y ? not (cmp x y)
```

When using the function `liftCmp` with a deterministic predicate like `(<=)` on numbers, the arguments `x` and `y` need to be evaluated, whereas the predicate `coinCmp` yields a non-deterministic result without demanding any of its arguments.

There are lots of other properties related to permutations. For example, enumerating derangements, that is, enumerating all permutations where an element does not appear at its original position, enumerating all permutations of a sublist of a given list, or analysing the resulting order of a permutation with respect to the parity for the transposition from one permutation to the next.

Besides a further investigation of permutation enumeration functions, we are interested if there are other predicate–based functions that are useful when applied to a non–deterministic predicate. The other example next to sorting we have discussed is filtering a list. Namely, we used `filter` to implement quicksort in 3.1.4. We have seen that using `filter` with a non–deterministic predicate yields all sublists of the input list. It might be interesting to search for more predicate–based functions and their non–deterministic counterparts. The first other common functions that come to mind are `takeWhile` and `dropWhile` that drop and take elements from a list as long as the given predicate holds, respectively. We are keen to search for other functions that have interesting and useful non–deterministic counterparts.

## 3.4 Conclusion

In this chapter we implemented a variety of sorting functions parametrised over a comparison function in Curry as well as in Haskell. Instead of ordinary sorting tasks, we applied the sorting functions to a non–deterministic comparison function. The resulting function enumerates permutations of the input list. For the Curry implementation there are several sorting functions that compute only the permutations of the input list without any duplicates: insertion sort, selection sort, merge sort and a variant of quicksort that uses `split` instead of two `filter` calls. We then compared these Curry implementations to their Haskell counterparts. In order to mimic the non–deterministic component that Curry brings along out of the box, we use a monadic lifting to define potentially effectful computations in Haskell. In the end, the Haskell implementations use a monadic lifting of the ordinary, pure sorting function in order to use a non–deterministic comparison function and compute non–deterministic results. One particularly interesting observation was that whereas the Curry version of selection sort computes only the permutations, the Haskell version does not. This observation was the main reason we investigated the difference of both implementations in the first place. The other sorting functions compute the same non–deterministic results: insertion sort, merge sort, and quicksort using `partition` do not compute duplicate results and bubble sort computes duplicates in Curry as well as in Haskell. Although we did not find other differences with respect to the computed permutations, we observed that the Curry version of these implementations can exploit non–strictness better than their Haskell counterparts. As an example, we demanded the head elements of all permutations and counted the number of non–deterministic choices that were triggered to compute the result. The most impressive sorting functions for this example were selection sort and bubble sort implemented in Curry as they only demanded $n$ non–deterministic choices for a list of length $n$. On top of that, none of the Curry implementations need to trigger all $n!$ non–deterministic computations for a list of length $n$, whereas the Haskell implementations trigger at least $n!$ computations. This property for the Curry implementation feels impressive since $n!$ non–deterministic computations corresponds to evaluating all non–deterministic computations that occur for an implementation that yields exactly all permutations. That is, selecting only the head element of the permutations has no effect on the non–determinism that needs to be triggered.

## 3.5 Final Remarks

The basis of the work discussed in this chapter has been published as functional pearl in the Proceedings of the 20th International Conference on Functional Programming (Christiansen et al., 2016). The conference paper motivates the investigation of non–deterministic sorting functions solely based on the monadic Haskell version. The generic monadic interface allows for a more detailed investigation of the behaviour of the resulting permutation functions. We enhance the monadic comparison function `coinCmp` with an additional state in order to mimic a comparison function that meets additional properties like consistency and totality. In case of consistency, for example, the state collects all compared value pairs and the corresponding decision of their comparison in order to repeat that decision when the same values are compared again.

On top of that, the conference paper formulates and proves a theorem stating that no matter which sorting function we use, the corresponding permutation function enumerates all permutations of the input list. In order to prove this statement, we use free theorems as presented by Wadler (1989), which are derived from the type of a function alone.

The novelty of the content presented in this chapter lies in the comparison to a direct implementation in Curry. Instead of modelling non–determinism with monads, we reused Curry's built–in non–determinism. The main insight of this comparison is the advantage of Curry's built–in non–determinism as it can exploit non–strictness that a naive Haskell implementation using a list monad cannot. This monadic lifting in Haskell mimics flat non–determinism, whereas Curry's non–determinism can occur deep in arguments of constructors. Curry's deep non–determinism can have advantages with respect to non–strictness in comparison to this naive monadic lifting in Haskell. A more advanced representation for non–determinism that mimics Curry's behaviour more closely is presented by Fischer et al. (2009). The representation, however, cannot reuse ordinary data structures from Haskell.

While we lift the resulting type of functions to model flat non–determinism, all arguments of constructors as well as functions need to take monadically lifted arguments when we want to model Curry's deep non–determinism.

# Probabilistic Functional Logic Programming

This chapter presents *PFLP*, a library providing a domain specific language for probabilistic programming in the functional logic programming language Curry. Key applications of the probabilistic programming paradigm are probabilistic processes and other applications based on probability distributions. PFLP makes heavy use of functional logic programming concepts and shows that this paradigm is well–suited for implementing a library for probabilistic programming. In fact, there is a close connection between probabilistic programming and functional logic programming. For example, non–deterministic choice and probabilistic choice are similar concepts. Furthermore, the concept of call–time choice as known from functional logic programming coincides with (stochastic) memoisation (De Raedt et al., 2007) in the area of probabilistic programming. We are not the first to observe this close connection between functional logic programming and probabilistic programming. For example, Fischer et al. (2009) present a library for modelling functional logic programs in the functional language Haskell. As they state, by extending their approach to weighted non–determinism we can model a probabilistic programming language.

Besides a lightweight implementation of a library for probabilistic programming in a functional logic programming language, this chapter provides the following contents.

- We investigate the interplay of probabilistic programming with features of a functional logic programming language. For example, we show how call–time choice and non–determinism interact with probabilistic choice.

- We discuss how we utilise functional logic features to improve the implementation of probabilistic combinators.

- We present an implementation of probability distributions using non–determinism in combination with non–strict probabilistic combinators that can be more efficient than an implementation using lists.

- We illustrate that the combination of non–determinism and non–strictness with respect to distributions has to be handled with care. More precisely, it is important to enforce a certain degree of strictness in order to guarantee correct results.

- We reason about laws for two operations of the library that are known as monad laws.

- We present performance comparisons between our library and two probabilistic programming languages.

Note that the current state of the library cannot compete against full–blown probabilistic languages or mature libraries for probabilistic programming as it is missing features like sampling from distributions. Nevertheless, the library is a good showcase for languages with built–in non–determinism, because the functional logic approach can be superior to the functional approach using lists. Furthermore, we want to emphasise that we use non–determinism as an implementation technique to develop a library for probabilistic programming. That is, we are not mainly concerned with the interaction of non–determinism and probabilism as, for example, discussed in the work of Varacca and Winskel (2006) and multiple others. The library we developed does not combine both effects, but provides combinators for probabilistic programming by leveraging Curry's built–in non–strict non–determinism.

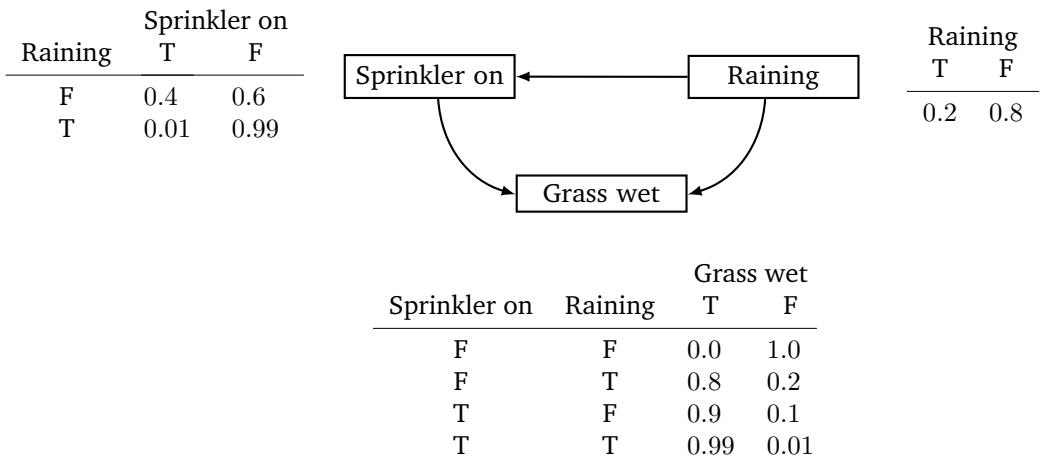

| | Sprinkler on | |
|Raining| T | F |
|---|---|---|
| F | 0.4 | 0.6 |
| T | 0.01 | 0.99 |

| Raining | |
| T | F |
|---|---|
| 0.2 | 0.8 |

| | | Grass wet | |
| Sprinkler on | Raining | T | F |
|---|---|---|---|
| F | F | 0.0 | 1.0 |
| F | T | 0.8 | 0.2 |
| T | F | 0.9 | 0.1 |
| T | T | 0.99 | 0.01 |

**Figure 4.1.:** A simple Bayesian network with associated probability tables

## 4.1 What is Probabilistic Programming

The probabilistic programming paradigm allows the succinct definition of probabilistic processes and other applications based on probability distributions, for example, Bayesian networks as used in machine learning. A Bayesian network (Pearl, 1988) is a visual, graph–based representation for a set of random variables and their dependencies. One of the *hello world* examples of Bayesian networks is the influence of rain and a sprinkler on wet grass. Figure 4.1 shows an instance of this example. A node in the graph represents a random variable, a directed edge between two nodes represents a conditional dependency. Each node is annotated with a probability function represented as a table. The input values are on the left–hand side of the table and the right–hand side of the table describes the possible output and the corresponding probability. The input values of the function correspond to the incoming edges of that node. For example, the node for sprinkler depends on rain, thus, the sprinkler node has an incoming edge that originates from the rain node. The input parameter rain appears directly in the table that describes the probability function for the sprinkler. For the example in Figure 4.1 the interpretation of the graph reads as follows: it rains with a probability of $20\,\%$; depending on the rain, the probability for an activated sprinkler is $40\,\%$ and $1\,\%$, respectively; depending on both of these factors, the grass can be observed as wet with a probability of $0\,\%$, $80\,\%$, $90\,\%$ or $99\,\%$. The network can answer the following exemplary questions.

- What is the probability that it is raining?
- What is the probability that the grass is wet, given that it is raining?
- What is the probability that the sprinkler is on, given that the grass is wet?

The general idea of probabilistic programming has been quite successful. There are a variety of probabilistic programming languages supporting all kinds of programming paradigms. For example, the programming languages Church (Goodman et al., 2008) and Anglican (Wood et al., 2014) are based on the functional programming language Scheme, ProbLog (Kimmig et al., 2011) is an extension of the logic programming language Prolog, Probabilistic C (Paige and Wood, 2014) is based on the imperative language C, and WebPPL (Goodman and Stuhlmüller, 2014), the successor of Church, is embedded in a functional subset of JavaScript. Besides full–blown languages there are also embedded domain specific languages that implement probabilistic programming as a library. For example, FACTORIE (McCallum et al., 2009) is a library for the hybrid programming language Scala, and Erwig and Kollmansberger (2006) present a library for the functional programming language Haskell. We recommend the survey by Gordon et al. (2014) about the current state of probabilistic programming for further information.

## 4.2 An Overview of the Library

In this section we discuss the core of the PFLP library.[1] The implementation is based on a Haskell library for probabilistic programming presented by Erwig and Kollmansberger (2006).

### 4.2.1 Modelling Distributions

One key ingredient of probabilistic programming is the definition of distributions. A distribution consists of pairs of elementary events and their probability. We model probabilities as `Float` and distributions as a combination of an elementary event and the corresponding probability.

```
type Probability = Float
data Dist a = Dist a Probability
```

In a functional language like Haskell, the canonical way to define distributions is to use lists. Here, we use Curry's built–in non–determinism as an alternative to lists to model distributions with more than one event–probability pair. As an example, we define a probabilistic (fair) coin, where `True` represents heads and `False` represents tails, as follows.[2]

```
coin :: Dist Bool
coin = Dist True ½ ? Dist False ½
```

Remember that printing an expression in the REPL evaluates the non–deterministic computations, thus, yields one result for each branch.

```
⊢ 1 ? 2                         ⊢ coin
1                               Dist True ½
2                               Dist False ½
```

It is cumbersome to define distributions explicitly as in the case of `coin`. Hence, we define helper functions for constructing distributions. Given a list of events and probabilities, `enum` creates a distribution by folding these pairs non–deterministically with a helper function `anyOf`.[3]

---

[1] We provide the code for the library at https://www-ps.informatik.uni-kiel.de/~sad/pflp.html.
[2] Here and in the following we write probabilities as fractions for readability.
[3] We shorten the implementation of `enum` for presentation purposes; actually, `enum` only allows valid distributions, e.g., that the given probabilities sum up to `1.0` using an error margin because of the usage of floating point numbers.

**4.2**  An Overview of the Library  |  **57**

```
anyOf :: [a] -> a
anyOf xs = foldr (?) failed xs

enum :: [a] -> [Probability] -> Dist a
enum vs ps = anyOf (zipWith Dist vs ps)
```

The function `anyOf` takes a list and yields a non–deterministic choice of all elements of the list.

As a short–cut, we define a function that yields a `uniform` distribution given a list of events as well as a function `certainly`, which yields a distribution with a single event with a $100\,\%$ probability.

```
uniform :: [a] -> Dist a
uniform xs = let len = length xs in enum xs (repeat 1/len)

certainly :: a -> Dist a
certainly x = Dist x 1.0
```

The function `repeat` yields a list that contains the given value infinitely often. Because of Curry's non–strictness, it is sufficient if one of the arguments of `enum` is a finite list because `zipWith` stops when one of its arguments is empty. We can then refactor the definition of `coin` using `uniform` as follows.

```
coin :: Dist Bool
coin = uniform [True,False]
```

In general, the library hides the constructor `Dist`, that is, the user has to define distributions by using the combinators provided by the library. Hence, the library provides additional functions to combine and manipulate distributions.

In order to work with dependent distributions, the operator (>>>=) applies a function that yields a distribution to each event of a given distribution and multiplies the corresponding probabilities.

```
(>>>=) :: Dist a -> (a -> Dist b) -> Dist b
d >>>= f =  let  Dist x p = d
                 Dist y q = f x
            in Dist y (p * q)
```

Intuitively, we have to apply the function `f` to each event of the distribution `d` and combine the resulting distributions into a single distribution. In a Haskell implementation using lists to model distributions, we would use a list comprehension to define this function. In the Curry implementation, we model distributions as non–deterministic computations, thus, the above rule describes the behavior of the function for an arbitrary pair of the first distribution and an arbitrary pair of the second distribution, that is, the result of `f`.

Using the (>>>=)–operator we can, for example, define a distribution that models flipping two coins. The events of this distribution are pairs whose first component is the result of the first coin flip and whose second component is the result of the second coin flip.

```
independentCoins :: Dist (Bool,Bool)
independentCoins = coin >>>= \c1 -> coin >>>= \c2 -> certainly (c1,c2)
```

In contrast to the example `independentCoins`, we can also use the operator (>>>=) to combine two distributions where we choose the second distribution on basis of the result of the first. For example, we can define a distribution that models flipping two coins, but in this case we only flip a second coin if the first coins yield heads.

```
dependentCoins :: Dist Bool
dependentCoins = coin >>>= \c -> if c then coin else certainly c
```

The implementation of (>>>=) via `let`–bindings seems a bit tedious, however, it is important that we define (>>>=) as introduced above. The canonical implementation performs pattern matching on the first argument but uses a `let`–binding for the result of `f`.

```
(>>>=) :: Dist a -> (a -> Dist b) -> Dist b
Dist x p >>>= f =  let Dist y q = f x
                   in Dist y (p * q)
```

That is, it is strict in the first argument but non–strict in the application of `f`, the second argument. Recall that we discussed the difference between pattern matching and using `let`–bindings in Section 2.2.2. We take a more detailed look at the implementation of (>>>=) later. For now, it is sufficient to keep in mind that (>>>=) yields a `Dist`–constructor without evaluating any of its arguments. In contrast, a definition using pattern matching or a case expression needs to evaluate its argument first, thus, it is more strict.

For independent distributions we provide the function `joinWith` that combines two distributions with respect to a given function. We implement `joinWith` by means of (>>>=).

```
joinWith :: (a -> b -> c) -> Dist a -> Dist b -> Dist c
joinWith f d1 d2 =  d1 >>>= \ x -> d2 >>>= \ y -> certainly (f x y)
```

In a monadic setting, this function is sometimes called `liftM2`. Here, we use the same nomenclature as Erwig and Kollmansberger (2006).

As an example of combining multiple distinct distributions, we define a function that flips a coin $n$ times.

```
flipCoin :: Int -> Dist [Bool]
flipCoin n  | n == 0    = certainly []
            | otherwise = joinWith (:) coin (flipCoin (n-1))
```

When we run the example of flipping two coins, we get four events.

⊢ `flipCoin 2`
Dist [True,True] $\frac{1}{4}$
Dist [True,False] $\frac{1}{4}$
Dist [False,True] $\frac{1}{4}$
Dist [False,False] $\frac{1}{4}$

Recall that `coin` is a non–deterministic choice between `True` and `False` with a uniform probability. That is, applying `joinWith` to `coin` and `coin` combines all possible results of two coin tosses.

### 4.2.2 Querying Distributions

With a handful of building blocks to define distributions available, we now want to query the distribution, that is, calculate the probability of certain events. We provide an operator `(??) :: (a -> Bool) -> Dist a -> Probability` — which we will define shortly — to extract the probability of an event. The event is specified as a predicate passed as first argument. The operator filters events that satisfy the given predicate and computes the sum of the probabilities of the remaining elementary events. It is straightforward to implement the bare filter function on distributions in Curry.

```
filterDist :: (a -> Bool) -> Dist a -> Dist a
filterDist p d = let Dist evnt prb = d
                 in if p evnt then Dist evnt prb else failed
```

The implementation of `filterDist` is a partial identity on the event–probability pairs. Every event that satisfies the predicate is part of the resulting sub-distribution. The function fails for event–probability pairs that do not satisfy the predicate. Similar to the definition of (>>>=) above, we use a let–binding on the right–hand side. In contrast to the definition of (>>>=), we cannot yield a `Dist`–constructor directly. That is, the usage of the `let`–binding does not affect the strictness in comparison to a definition via pattern matching.

```
filterDistP :: (a -> Bool) -> Dist a -> Dist a
filterDistP p (Dist evnt prb) = if p evnt then Dist evnt prb else failed
```

Querying a distribution is a more advanced task in the functional logic approach. In essence, we need to sum up all probabilities that satisfy a given predicate. Remember that we represent a distribution by chaining all event–probability pairs with (?), thus, constructing non–deterministic computations. These non–deterministic computations introduce individual branches of computations that cannot interact with each other. In order to compute the total probability of a distribution, we have to merge these distinct branches. Such a merge is possible by encapsulating the non–deterministic computations. Similar to the *findall* construct of the logic language Prolog, in Curry we encapsulate a non–deterministic computation by using the function `allValues` that we introduced in Section 2.2.4. As a library developer, we can employ this function to encapsulate non–deterministic values and use these values in further computations. However, due to non–transparent behavior in combination with sharing as discussed by Braßel et al. (2004), a user of the library should not use `allValues` at all. The reason for this restriction is, in a nutshell, that innermost and outermost evaluation strategies may cause different results when combining sharing and encapsulation.

With this encapsulation mechanism at hand, we can define the extraction operator (??) as follows.

```
prob :: Dist a -> Probability
prob (Dist _ p) = p

(??) :: (a -> Bool) -> Dist a -> Probability
(??) p dist = foldValues (+.) 0.0 (allValues (prob (filterDist p dist)))
```

First we filter the elementary events by some predicate and project to the probabilities only. Afterwards we encapsulate the remaining probabilities and sum them up. As an example for the use of (??), we may flip four coins and calculate the probability of at least two heads, that is, we check if the list contains at least two `True` values.

```
⊢ (\ coins -> length (filter id coins) >= 2) ?? (flipCoin 4)
0.6875
```

In order to check the result, we calculate the probability by hand. Since there are more events that satisfy the predicate than events that do not, we sum up the probabilities of the events that do not satisfy the predicate and calculate the complementary probability. There is one event where all coins show tails, and four events where one of the coins shows heads and all other show tails.

$$\begin{aligned}
& 1 - (\ P(Tails) \cdot P(Tails) \cdot P(Tails) \cdot P(Tails) \\
& \qquad + 4 \cdot P(Heads) \cdot P(Tails) \cdot P(Tails) \cdot P(Tails)) \\
& = 1 - (0.5 \cdot 0.5 \cdot 0.5 \cdot 0.5 + 4 \cdot 0.5 \cdot 0.5 \cdot 0.5 \cdot 0.5) \\
& = 1 - (0.0625 + 0.25) \\
& = 1 - 0.3125 \\
& = 0.6875
\end{aligned}$$

## 4.3 The Functional Logic Heart of the Library

Up to now, we have discussed a simple library for probabilistic programming that uses non–determinism to represent distributions. In this section we see that we can highly benefit from Curry–like non–determinism with respect to performance when we compare the library's implementation with a list–based implementation. More precisely, when we query a distribution with a predicate that does not evaluate its argument completely, we can possibly prune large parts of the search space. Before we discuss the details of the combination of non–strictness and non–determinism, we discuss aspects of sharing non–deterministic choices. Finally, we discuss details about the implementation of (>>>=)–operator.

### 4.3.1 Call–Time Choice

By default Curry uses call–time choice, that is, variables denote single deterministic choices. When we bind a variable to a non–deterministic computation, one value is chosen and all occurrences of the variable denote the same deterministic choice. Often call–time choice is what you are looking for. For example, the definition of `filterDist` makes use of call–time choice.

```
filterDist :: (a -> Bool) -> Dist a -> Dist a
filterDist p d = let Dist evnt prb = d
                 in if p evnt then Dist evnt prb else failed
```

The variable `d` on the right–hand side corresponds to a single deterministic choice for the input distribution, and not to the non–deterministic computation that was initially passed as second argument to `filterDist`.

Sometimes run–time choice is what you are looking for and call-time choice gets in your way; probabilistic programming is no exception. For example, let us reconsider flipping a

**4.3**  The Functional Logic Heart of the Library  |  **61**

coin $n$ times. We parametrise the function `flipCoin` over the given distribution and define the following generalised function.

```
replicateDist :: Int -> Dist a -> Dist [a]
replicateDist n d | n == 0    = certainly []
                  | otherwise = joinWith (:) d (replicateDist (n-1) d)
```

When we use this function to flip a coin twice, the result is not what we intended.

⊢ replicateDist 2 coin
Dist [True,True] $\frac{1}{4}$
Dist [False,False] $\frac{1}{4}$

Because `replicateDist` shares the variable `d`, we only perform a choice once and replicate deterministic choices. In contrast, top–level nullary functions like `coin` are evaluated every time, thus, exhibit run–time choice, which is the reason why the previously shown `flipCoin` behaves properly.

In order to implement `replicateDist` correctly, we have to enforce run–time choice. We introduce the following type synonym and function to model and work with values with run–time choice behavior.[4]

```
type Plural a = () -> a

plural :: Plural a -> a
plural pl = pl ()
```

We can now use the type `Plural` to hide the non–determinism on the right–hand side of a function arrow. This way, `plural` explicitly triggers the evaluation of `pl`, performing a new choice for every element of the result list. The name `Plural` refers to the notion of *plural semantics* introduced by Clinger (1982) to describe that variables refer to set of values rather than to single values in the context of a non-deterministic call-by-need calculus.

```
replicateDist :: Int -> Plural (Dist a) -> Dist [a]
replicateDist n plDist
  | n == 0    = certainly []
  | otherwise = joinWith (:) (plural plDist) (replicateDist (n-1) plDist)
```

Now, we have to construct a value of type `Plural (Dist Bool)` in order to use the function `replicateDist`. However, we cannot provide a function to construct a value of type `Plural` that behaves as intended. Such a function would share a deterministic choice and non–deterministically yield two functions, instead of one function that yields a non–deterministic computation. The only way to construct a value of type `Plural` is to explicitly use a lambda abstraction as shown in the following example.

⊢ replicateDist 2 (\ () -> coin)
Dist [True,True] $\frac{1}{4}$
Dist [True,False] $\frac{1}{4}$
Dist [False,True] $\frac{1}{4}$
Dist [False,False] $\frac{1}{4}$

---

[4]We adopt the names for the corresponding wrapper type and projection function of the Curry module `Plural` (https://www.informatik.uni-kiel.de/~curry/cpm/DOC/plural-arguments-2.0.0/Plural.html).

Instead of relying on call–time choice as default behavior, we could model `Dist` as a function and make run–time choice the default. In this case, to get call–time choice we would have to use a special construct provided by the library — as it is the case in many probabilistic programming libraries, e.g., *mem* in WebPPL (Goodman and Stuhlmüller, 2014).

On the other hand, ProbLog uses a similar concept to call–time choice, namely, stochastic memoisation, which reuses already computed results. That is, predicates that are associated with probabilities become part of the memoised result. A ProbLog version of `replicateDist` looks as follows.

```
0.5::coin(tt).
0.5::coin(ff).

replicateCoin(0,_,[]).
replicateCoin(N,Dist,[X|Xs]) :-
  N > 0, call(Dist,X), M is N - 1, replicateCoin(M,Dist,Xs).
```

When we query `replicateDist(2,coin,Xs)` we get the following results.

```
replicateCoin(2,coin,[ff, ff]): 0.5
replicateCoin(2,coin,[ff, tt]): 0.25
replicateCoin(2,coin,[tt, ff]): 0.25
replicateCoin(2,coin,[tt, tt]): 0.5
```

We observe that if we flip the same side two times, the resulting probability is not as expected. ProbLog memoises the results of a predicate call — in this case `coin(tt)` and `coin(ff)`, respectively. If a coin was already flipped with `tt` and a probability of $50\%$, then all further coin flips that result in `tt` have probability of $100\%$. Due to stochastic memoisation the coin is not flipped a second time, but is identified as the same coin as before. Thus, stochastic memoisation as used in ProbLog is similar to the extension of tabling in Prolog systems, but adapted to the setting of probabilistic programming that extends predicates with probabilities. Similar to our usage of `Plural` to mimic run–time choice in Curry, we can use a so–called trial identifier, which is basically an additional argument, to circumvent memoisation for a predicate like coin in ProbLog. The difference to `Plural` is that the trial identifier needs to be different for each call to the predicate in order to force re–evaluation.

In the end, we have decided to go with the current modelling based on call–time choice, because the alternative would work against the spirit of the Curry programming language.

There is a long history of discussions about the pros and cons of call–time choice and run–time choice. It is common knowledge in probabilistic programming (De Raedt et al., 2007) that, in order to model stochastic automata or probabilistic grammars, memoisation — that is, call–time choice — has to be avoided. Similarly, Antoy (2005) observes that you need run–time choice to elegantly model regular expressions in the context of functional logic programming languages. Then again, probabilistic languages need a concept like memoisation in order to use a single value drawn from a distribution multiple times.

### 4.3.2 Non–strict Non–determinism

This section illustrates the benefits of the combination of non–strictness and non–determinism with respect to performance. More precisely, in a setting that uses Curry–like non–determinism, non–strictness can prevent non–determinism from being „spawned". Let us consider calculating the probability for throwing only sixes when throwing `n` dice. First we define a uniform distribution for throwing a die as follows.

| # of dice    | 5   | 6   | 7    | 8      | 9    | 10    | 100 | 200 | 300 |
|--------------|-----|-----|------|--------|------|-------|-----|-----|-----|
| Curry ND     | <1  | <1  | <1   | <1     | <1   | <1    | 48  | 231 | 547 |
| Curry List   | 2   | 13  | 72   | 419    | 2554 | 15 394 | –   | –   | –   |
| Curry ND!    | 52  | 409 | 2568 | 16 382 | –    | –     | –   | –   | –   |
| Haskell List | 1   | 5   | 30   | 210    | 1415 | 6538  | –   | –   | –   |

**Table 4.1.:** Overview of running times for the query `allSix n` in ms

```
data Side = One | Two | Three | Four | Five | Six

die :: Dist Side
die = uniform [One,Two,Three,Four,Five,Six]
```

We define the following query by means of the combinators introduced so far. The function `all` simply checks that all elements of a list satisfy a given predicate; it is defined by means of the Boolean conjunction (`&&`).

```
isSix :: Side -> Bool
isSix s = case s of
            Six -> True
            _   -> False

allSix :: Int -> Probability
allSix n = (all isSix) ?? (replicateDist n (\ () -> die))
```

Table 4.1 compares running times[5] of this query for different numbers of dice. The row labeled „Curry ND" lists the running times for an implementation that uses the operator (>>>=). The row „Curry List" shows the numbers for a list–based implementation in Curry, which is a literal translation of the library by Erwig and Kollmansberger. The row labeled „Curry ND!" uses an operator (>>>=!) instead — a strict version of (>>>=), which we will discuss shortly. Finally, we compare our implementation to the original list–based implementation, which the row labeled „Haskell List" refers to. The table states the running times in milliseconds of a compiled executable for each benchmark as a mean of three runs. Cells marked with „–" take more than one minute.

Obviously, the example above is a little contrived. While the query is exponential in both list versions, we can observe a linear running time in the non–deterministic setting.[6] In order to illustrate the behavior of the example above, we consider the following application for an arbitrary distribution `dist` of type `Dist [Side]`.

```
filterDist (all isSix) (joinWith (:) (Dist One 1/6) dist)
```

This application yields an empty distribution without evaluating the distribution `dist`. The trick here is that `joinWith` yields a `Dist`–constructor without inspecting its arguments. When

$$filterDist\ (all\ isSix)\ (joinWith\ (:)\ (Dist\ One\ \tfrac{1}{6})\ dist)$$
$$=\quad \{\ \text{Definition of } joinWith\ \}$$
$$filterDist\ (all\ isSix)$$
$$\qquad (Dist\ One\ \tfrac{1}{6} \ggg (\lambda x \to dist \ggg (\lambda xs \to certainly\ (x:xs))))$$
$$=\quad \{\ \text{Definition of } (\ggg)\ \text{(twice)}\ \}$$
$$filterDist\ (all\ isSix)$$
$$\qquad (\textbf{let}\ Dist\ x\ p = Dist\ One\ \tfrac{1}{6}$$
$$\qquad\qquad Dist\ xs\ q = dist$$
$$\qquad\qquad Dist\ ys\ r = certainly\ (x:xs)$$
$$\qquad \textbf{in}\ Dist\ ys\ (p * (q * r)))$$
$$=\quad \{\ \text{Definition of } filterDist\ \}$$
$$\textbf{let}\ Dist\ x\ p = Dist\ One\ \tfrac{1}{6}$$
$$\quad Dist\ xs\ q = dist$$
$$\quad Dist\ ys\ r = certainly\ (x:xs)$$
$$\textbf{in if}\ all\ isSix\ ys\ \textbf{then}\ Dist\ ys\ (p*(q*r))\ \textbf{else}\ failed$$
$$=\quad \{\ \text{Definition of } certainly\ \}$$
$$\textbf{let}\ Dist\ x\ p = Dist\ One\ \tfrac{1}{6}$$
$$\quad Dist\ xs\ q = dist$$
$$\textbf{in if}\ all\ isSix\ (x:xs)\ \textbf{then}\ Dist\ (x:xs)\ (p*(q*1.0))\ \textbf{else}\ failed$$
$$=\quad \{\ \text{Definition of } all\ \text{and } isSix\ \}$$
$$\textbf{let}\ Dist\ x\ p = Dist\ One\ \tfrac{1}{6}$$
$$\quad Dist\ xs\ q = dist$$
$$\textbf{in if}\ False \wedge all\ isSix\ xs\ \textbf{then}\ Dist\ (x:xs)\ (p*(q*1.0))\ \textbf{else}\ failed$$
$$=\quad \{\ \text{Definition of } (\wedge)\ \}$$
$$\textbf{let}\ Dist\ x\ p = Dist\ One\ \tfrac{1}{6}$$
$$\quad Dist\ xs\ q = d$$
$$\textbf{in if}\ False\ \textbf{then}\ Dist\ (x:xs)\ (p*(q*1.0))\ \textbf{else}\ failed$$
$$=\quad \{\ \text{Definition of } if-then-else\ \}$$
$$failed$$

---
[5]These benchmarks were executed on a Linux machine with an Intel Core i7-6500U (2.50 GHz) and 8 GiB RAM running Fedora 25. We used the Glasgow Haskell Compiler (version `8.0.2`, option `-O2`) and set the search strategy in KiCS2 to depth–first.

[6]Non–determinism causes significant overhead for KiCS2, thus, „Curry ND" does not show linear development, but we measured a linear running time using PAKCS.

**Figure 4.2.:** Simplified evaluation illustrating non–strict non–determinism

we demand the event of the resulting `Dist`, `joinWith` has to evaluate only its first argument to see that the predicate `all isSix` yields `False`. The evaluation of the expression fails without inspecting the second argument of `joinWith`. Figure 4.2 illustrates the evaluation in more detail.

In case of the example `allSix`, all non–deterministic branches that contain a value different from `Six` fail fast — that is, in constant time — due to the non–strictness. Thus, the number of evaluation steps is linear in the number of rolled dice.

We can only benefit from the combination of non–strictness and non–determinism, because we defined (>>>=) with care. Let us take a look at a strict variant of (>>>=) and discuss its consequences.

```
(>>>=!) :: Dist a -> (a -> Dist b) -> Dist b
Dist x p >>>=! f = case  f x of
                    Dist y q -> Dist y (p * q)
```

This implementation is strict in its first argument as well as in the result of the function application. When we use (>>>=!) to implement the `allSix` example, we lose the benefit of Curry–like non–determinism. The row in Figure 4.2 labeled „Curry ND!" shows the running times when using (>>>=!) instead of (>>>=). As (>>>=!) is strict, the function `joinWith`

$$\begin{aligned}
&\mathit{filterDist}\ (\mathit{all\ isSix})\ (\mathit{joinWith}\ (:)\ (\mathit{Dist\ One}\ \tfrac{1}{6})\ \mathit{dist})\\
&=\quad \{\text{ Definition of } \mathit{joinWith}\ \}\\
&\mathit{filterDist}\ (\mathit{all\ isSix})\\
&\qquad (\mathit{Dist\ One}\ \tfrac{1}{6}\ \ggg=!\ (\lambda x \to \mathit{dist}\ \ggg=!\ (\lambda xs \to \mathit{certainly}\ (x:xs))))\\
&=\quad \{\text{ Definition of } (\ggg=!)\ \}\\
&\mathit{filterDist}\ (\mathit{all\ isSix})\\
&\qquad (\mathbf{case}\ (\lambda x \to \mathit{dist}\ \ggg=!\ (\lambda xs \to \mathit{certainly}\ (x:xs)))\ \mathit{One}\ \mathbf{of}\\
&\qquad\qquad \mathit{Dist\ y\ q} \to \mathit{Dist\ y\ }(\tfrac{1}{6}*q))\\
&=\quad \{\text{ Evaluation of the scrutinee }\}\\
&\mathit{filterDist}\ (\mathit{all\ isSix})\\
&\qquad (\mathbf{case}\ \mathit{dist}\ \ggg=!\ (\lambda xs \to \mathit{certainly}\ (\mathit{One}:xs))\ \mathbf{of}\\
&\qquad\qquad \mathit{Dist\ y\ q} \to \mathit{Dist\ y\ }(\tfrac{1}{6}*q))\\
&=\quad \{\text{ Evaluation of } \mathit{dist}\ \text{ as demanded by the definition of } (\ggg=!)\ \}\\
&\ldots
\end{aligned}$$

**Figure 4.3.:** Simplified evaluation illustrating strict non–determinism

has to evaluate both its arguments to yield a result. Figure 4.3 demonstrates how the formerly unneeded distribution `dist` now has to be evaluated in order to yield a value. More precisely, using `(>>>=!)` causes a complete evaluation of `dist`.

Note that an implementation that is similar to `(>>>=)` is *not* possible in a list–based implementation. The bind operator for lists is usually defined by means of a combination of `concat` and `map`, usually named `concatMap`.

```
concatMap :: (a -> [b]) -> [a] -> [b]
concatMap f  []      = []
concatMap f  (x:xs)  = f x ++ concatMap f xs
```

The strict behaviour follows from the definition via pattern matching on the list argument. In contrast to `(>>>=!)` there is, however, no other implementation that is less strict. The pattern matching is inevitable due to the two possible constructors, `[]` and `(:)`, for lists. As a consequence, a list–based implementation has to traverse the entire distribution before we can evaluate the predicate `all isSix`. All in all, that the running times of „Haskell List" in Figure 4.2 cannot compete with „Curry ND" when the number of dice increases.

Intuitively, we expect similar running times for „Curry ND!" and „Curry List" as the bind operator for lists has to evaluate its second argument as well — similar to `(>>>=!)`. However, the observed running times do not have the expected resemblance. „Curry ND!" heavily relies on non–deterministic computations, which causes significant overhead for KiCS2. We do not investigate these differences here but propose it as a direction for future research.

Obviously, turning an exponential problem into a linear one is like getting only sixes when throwing dice. In most cases we are not that lucky. For example, consider the following query for throwing `n` dice that are either five or six.

```
isFiveOrSix :: Side -> Bool
isFiveOrSix s = case s of
                  Five -> True
                  _    -> isSix s

allFiveOrSix :: Int -> Probability
allFiveOrSix n = (all isFiveOrSix) ?? (replicateDist n (\ () -> die))
```

Table 4.2 lists the running times of this query for different numbers of dice with respect to the four different implementations.

| # of dice | 5 | 6 | 7 | 8 | 9 | 10 |
|---|---|---|---|---|---|---|
| Curry ND | 4 | 7 | 15 | 34 | 76 | 163 |
| Curry List | 2 | 13 | 84 | 489 | 2869 | 16 989 |
| Curry ND! | 49 | 382 | 2483 | 15 562 | – | – |
| Haskell List | 2 | 5 | 31 | 219 | 1423 | 6670 |

**Table 4.2.:** Overview of running times of the query `allFiveOrSix n` in ms

As we can see from the running times, this query is exponential in all implementations. Nevertheless, the running time of the non–strict, non–deterministic implementation is much better because we only have to consider two sides — six and five — while we have to consider all sides in the list implementations and the non–deterministic, strict implementation. That is, while the base of the complexity is two in the case of the non–deterministic, non–strict implementation, it is six in all the other cases. As we have observed in the other examples before, we get an overhead in the case of the strict non–determinism compared to the list implementation due to the heavy usage of non–deterministic computations.

### 4.3.3 Leveraging Non–strictness

In this section we discuss our design choices concerning the implementation of the bind operator. We illustrate that we have to be careful about non–strictness, because we do not want to lose non–deterministic results. Most importantly, the final implementation ensures that users cannot misuse the library if they stick to one simple rule.

First, we revisit the definition of (>>>=) introduced in Section 4.2.

```
(>>>=) :: Dist a -> (a -> Dist b) -> Dist b
d >>>= f = let Dist x p = d
               Dist y q = f x
           in Dist y (p * q)
```

We can observe two facts about this definition. First, the definition yields a `Dist`–constructor without matching any argument. Second, if neither the event nor the probability of the final distribution is evaluated, the application of the function `f` is not evaluated either.

We can observe these properties with some exemplary usages of (>>>=). As a reference, we see that pattern matching the `Dist`–constructor of `coin` triggers the non–determinism and yields two results.

```
⊢ (\ (Dist _ _) -> True) coin
True
True
```

In contrast, distributions resulting from an application of (>>>=) behave differently. This time, pattern matching on the `Dist`–constructor does not trigger any non–determinism.

```
⊢ (\ (Dist _ _) -> True) (certainly () >>>= (\ _ -> coin))
True
```

```
⊢ (\ (Dist _ _) -> True) (coin >>>= certainly)
True
```

**4.3** The Functional Logic Heart of the Library | **67**

We observe that the last two examples yield a single result, because the (>>>=)–operator changes the position of the non–determinism. That is, the non–determinism does not reside at the same level as the `Dist`–constructor, but in the arguments of `Dist`. Therefore, we have to be sure to trigger all non–determinism when we compute probabilities. Not evaluating non–determinism might lead to false results when we sum up probabilities. Hence, non–strictness is a crucial property for positive pruning effects, but has to be used carefully.

Consider the following example usage of (>>>=), which is an inlined version of `joinWith` applied to the Boolean conjunction (&&).

```
⊢ (\ (Dist x _) -> x)
    (coin >>>= (\ x -> coin >>>= (\ y -> certainly (x && y))))
False
True
False
```

We lose one expected result from the distribution, because (&&) is non–strict in its second argument in case the first argument is `False`. When the first `coin` evaluates to `False`, (>>>=) ignores the second coin and yields `False` straightaway. In this case, the non–determinism of the second `coin` is not triggered and we get only three instead of four results. The non–strictness of (&&) has no consequences when using (>>>=!), because the operator evaluates both arguments and, thus, triggers the non–determinism.

As we have seen above, when using the non–strict operator (&&), one of the results gets lost. However, when we sum up probabilities, we do not want events to get lost. For example, when we compute the total probability of a distribution, the result should always be `1.0`. The query above, however, has only three results and every event has a probability of `0.25`, resulting in a total probability of `0.75`.

Here is the good news: while events can get lost when passing non–strict functions to (>>>=), probabilities never get lost. For example, consider the following application.

```
⊢ (\ (Dist _ p) -> p)
    (coin >>>= (\ x -> coin >>>= (\ y -> certainly (x && y))))
```
$\frac{1}{4}$
$\frac{1}{4}$
$\frac{1}{4}$
$\frac{1}{4}$

Since multiplication is strict, if we demand the resulting probability, the operator (>>>=) has to evaluate the `Dist`–constructor and its probability. That is, no values get lost if we evaluate the resulting probability. Fortunately, the query operation (??) calculates the total probability of the filtered distributions, thus, evaluates the probability as the following example shows.

```
⊢ not ?? (coin >>>= (\ x -> coin >>>= (\ y -> certainly (x && y))))
```
$\frac{3}{4}$

We calculate the probability of the event `False` and while there are only two `False` events, the total probability is still `0.75`, i.e., three times `0.25`.

All in all, in order to benefit from non–strictness, all operations have to use the right amount of strictness, not too much and not too little. For this reason PFLP does not provide the `Dist`–constructor nor the corresponding projection functions to the user. With this restriction, the library guarantees that no relevant probabilities get lost with respect to non–strictness.

## 4.4 Pitfalls

The preceding section summarised how the library benefits from using functional logic features in its implementation. In the following section, we discuss some pitfalls that may arise due to the decision to implement the library using these features. First, we discuss the usage of non–deterministic computations in events of distributions. As Curry is a functional logic language, it seems quite natural to use non–deterministic functions when working with the probabilistic programming library. Such non–deterministic functions result in non–deterministic computations that we might use as events in modelled distributions. Besides introducing non–deterministic computations, we can also have no value at all due to the usage of `failed`. Hence, the second part of the section considers the usage of partial functions in combination with operators the library provides.

### 4.4.1 Non–deterministic Events

We assume that all events passed to library functions are deterministic, that is, the library does not support non–deterministic events within distributions. In order to illustrate why this restriction is crucial, we consider an example that breaks this rule.

Curry provides free variables using the keyword `free`, that is, computations that non–deterministically evaluate to every possible value of their type. When we revisit the definition of a die, we might be tempted to use a free variable instead of explicitly enumerating all values of type `Side`.

We can define a free variable of type `Side` as follows.

```
side :: Side
side = x where x free
```

This free variable evaluates as follows.

```
⊢ side
One
Two
Three
Four
Five
Six
```

With this information in mind consider the following alternative definition of a die, which is much more concise than explicitly listing all constructors of `Dist`.

```
die2 :: Dist Side
die2 = enum [side] [1/6]
```

We just use a free variable — the constant `side` — and pass the probability of each event as second parameter. Now, let us consider the following query.

⊢ (const True) ?? die2
$\frac{1}{6}$

$(const\ True)\ ??\ die2$
$=$ { definition of $(??)$ }
$foldValues\ (+.)\ 0\ (allValues\ (prob\ (filterDist\ (const\ True)\ die2)))$
$=$ { definition of $die2$ }
$foldValues\ (+.)\ 0\ (allValues\ (prob\ (filterDist\ (const\ True)\ (enum\ [side]\ [\frac{1}{6}]))))$
$=$ { definition of $enum$ }
$foldValues\ (+.)\ 0\ (allValues\ (prob\ (filterDist\ (const\ True)\ (Dist\ side\ \frac{1}{6}))))$
$=$ { definition of $filterDist$ }
$foldValues\ (+.)\ 0\ (allValues\ (prob\ (\textbf{if}\ const\ True\ side\ \textbf{then}\ Dist\ side\ \frac{1}{6}\ \textbf{else}\ failed)))$
$=$ { definition of $const$ }
$foldValues\ (+.)\ 0\ (allValues\ (prob\ (\textbf{if}\ True\ \textbf{then}\ Dist\ side\ \frac{1}{6}\ \textbf{else}\ failed)))$
$=$ { evaluate $\textbf{if}-\textbf{then}-\textbf{else}$ }
$foldValues\ (+.)\ 0\ (allValues\ (prob\ (Dist\ side\ \frac{1}{6})))$
$=$ { definition of $prob$ }
$foldValues\ (+.)\ 0\ (allValues\ \frac{1}{6})$
$=$ { definition of $allValues$ }
$foldValues\ (+.)\ 0\ \{\frac{1}{6}\}$
$=$ { definition of $foldValues$ }
$\frac{1}{6}$

**Figure 4.4.:** Evaluation of a distribution that contains a free variable that is not demanded

The result of this query is $\frac{1}{6}$ and not `1.0` as expected. Consider Figure 4.4 for a step–by–step evaluation of this expression in order to understand better what is going on. This example illustrates that probabilities can get lost if the predicate is not strict enough to pull all non–deterministic values to the outside. Here, the predicate `const True` does not consider the event–component at all, thus, does not trigger `side` to evaluate to all the constructors of `Side`. The definition of `(??)` directly projects to the probability of `die2` instead and throws away all non–determinism left hidden in the event–component of the distribution. Therefore, we lose probabilities we would like to sum up.

As a consequence for PFLP, non–deterministic events within a distribution are not allowed. If users of the library stick to this rule, it is not possible to misuse the operations and lose non–deterministic results due to non–strictness.

One possible approach to overcome this problem is to evaluate the events of a distribution to normal form in order to trigger all the non–determinism that may occur. Changing the library definition accordingly, however, leads to a loss of the advantage with respect to non–strictness. The query `allSix`, for example, heavy relies on the fact that the event is only evaluated as far as needed. A strict evaluation of the event forces the evaluation of the whole list of dice before applying the predicate. In case of our exemplary evaluation with a distribution that only consists of the value `One` in its events, we currently can stop after evaluating the head of the list as the predicate already yields `False`. Using a strict evaluation for the events of a distribution, the whole lists needs to be evaluated and the non–strictness of the predicate does not play a role anymore.

An alternative idea is to only adapt the strictness behaviour of `enum`.

```
enum2 :: [a] -> [Probability] -> Dist a
enum2 xs ps = anyOf (zipWith mkDist xs ps)
 where mkDist x p = (\func y -> Dist y p) $## x
```

The operator (`$##`) evaluates its second argument to normal form, instantiates free variables, and applies its first argument, a function, to the resulting value. If we adapt our example above to use `enum2`, querying with a non–strict predicate does not result in an unexpected probability anymore.

⊢ const True ?? enum2 [side] [$\frac{1}{6}$]
1.0

This approach works out fine, however, we also need to consider that users of the library might define a uniform distribution without using `enum2`. For example, Gibbons and Hinze (2011) define a uniform distribution using the monadic function `return` and a second primitive called `choice`.

```
class Monad m => MonadProb m where
  choice :: Prob -> m a -> m a -> m a

uniformGibbons :: MonadProb m => [a] -> m a
uniformGibbons [x] = return x
uniformGibbons (x : xs) = choice (1 / length (x:xs))
                                (return x)
                                (uniformGibbons xs')
```

In order to translate this definition to our library, we need to define a primitive like `choice`. The basic idea behind `choice` is, given a probability $p$ as first argument, it associates the second argument with probability $p$ and the third argument with probability $1 - p$. In other probabilistic languages this primitive is usually called `flip` or `bernoulli`. We can define this function using `enum2`.

```
flip :: Probability -> a -> a -> Dist a
flip p x y = enum2 [x,y] [p, 1 - p]
```

We can then translate the above definition of **uniformGibbons** into our library using `flip` as follows.

```
uniform2 :: [a] -> Dist a
uniform2 [x] = certainly x
uniform2 (x : xs) = uniform2 xs >>>= \ xs' ->
                    flip (frac 1 (length (x:xs))) x xs'
```

Since the only way to define `flip` with our library is to use the predefined function `enum2`, the strictness adaption in the definition of `enum2` is enough to trigger the evaluation of free variables used as argument of `uniform2`.

⊢ const True ?? uniform2 side
1.0

Alternatively, instead of providing the function `uniform :: [a] -> Dist a` as combinator in our library, we could instead provide a variant `uniformND :: a -> Dist a` that actually expects to be called with a non–deterministic argument.

```
uniformND :: a -> Dist a
uniformND evnts = uniform xs
 where xs = allValues evnts
```

4.4 Pitfalls | 71

The function encapsulates the non–deterministic events and uses this list of events to build a uniform distribution using the function `uniform`, which is then only used internally.

### 4.4.2 Partial Functions

Besides not using non–deterministic values constructed by `(?)` for events, users have to be cautious about using `failed` as well. When using the bind operator `(>>>=)`, the second argument is a function of type `a -> Dist b`, that is, constructs a new distribution. As we have discussed before distributions need to sum up to a probability of $1.0$, and the distributions we create via `(>>>=)` are no exception. This restriction is violated if we use partial functions, which implicitly yield `failed`, as second argument of `(>>>=)`. Recall the definition `coin` that describes a uniform distribution of type `Bool`, and consider the function `partialPattern` that depends on `coin`, but maps `False` to `failed`.

```
partialPattern :: Dist Bool
partialPattern = coin >>>= (\b -> case b of
                                    True  -> certainly True
                                    False -> failed)
```

Due to the partial pattern matching in `partialPattern`, the resulting distribution does not sum up to $1.0$ anymore, thus, violates the rule for a valid distribution. By performing a query with the predicate `const True` we can observe this property.

```
repl > (const True) ?? partialPattern
```
$\frac{1}{2}$

We only allow to filter distributions when a probability is computed using `(??)`, but not in any other situation. In the current implementation this restriction on functions when using `(>>>=)` is neither statically nor dynamically enforced, but a coding convention that users should keep in mind and follow when working with the library. When this restriction is strictly followed, the user has the guarantee that the library works as expected.

## 4.5 Monad Laws

When we comply with the restrictions we have discussed above, the operators `(>>>=)` and `certainly` allow us to formulate probabilistic programs as one would expect. However, there is one obvious question that we did not answer yet. We did not check whether the operator `(>>>=)` together with `certainly` actually forms a monad as the name of the operator suggests. The monad typeclass and its operations usually obey three laws. That is, we have to check whether the following three laws hold for all distributions $d$ and all values $x$, $f$, and $g$ of appropriate types.

(1) $d \ggg certainly \equiv d$

(2) $certainly\ x \ggg f \equiv f\ x$

(3) $(d \ggg f) \ggg g \equiv d \ggg (\lambda y \to f\ y \ggg g)$

In the previous section we have already observed that the equality stated in (1) does not hold in general. For example, we have seen that there is a context that is able to distinguish the left–hand from the right–hand side. For instance, while the expression

$$(\lambda(Dist\ x\ p) \to True)\ coin$$

72   Chapter 4   Probabilistic Functional Logic Programming

yields `True` twice, the expression

$$(\lambda(\mathit{Dist}\ x\ p) \to \mathit{True})\ (\mathit{coin} \ggg \mathit{certainly})$$

yields `True` only once. In Curry semantics based on sets (see for example work by Mehner et al. (2014)), and not on multisets, the two sides of the equality would be the same. Notwithstanding, in a hypothetical multiset-based semantics the user could still not observe the difference between the two expressions because she does not have access to the `Dist`–constructor. The user cannot pattern match on a `Dist`–constructor, but only use the combinator (??) to inspect a distribution.

In order to discuss the validity of the monad laws more rigorously, we apply equational reasoning to check whether the monad laws might fail. That is, we use rewriting rules with respect to function definitions as well as primitives like if-then-else and let to check the resulting terms for equality.

**The first monad law**   Let $d :: \mathit{Dist}\ \tau$ then we reason as follows about the first monad law (1).

$d \ggg \mathit{certainly}$
$=$   { Definition of $(\ggg)$ }
**let** $\mathit{Dist}\ x\ p = d$
    $\mathit{Dist}\ y\ q = \mathit{certainly}\ x$
**in** $\mathit{Dist}\ y\ (p * q)$
$=$   { Definition of $\mathit{certainly}$ }
**let** $\mathit{Dist}\ x\ p = d$
    $\mathit{Dist}\ y\ q = \mathit{Dist}\ x\ 1.0$
**in** $\mathit{Dist}\ y\ (p * q)$
$=$   { Inlining of $\mathit{Dist}\ y\ q = \mathit{Dist}\ x\ 1.0$ }
**let** $\mathit{Dist}\ x\ p = d$ **in** $\mathit{Dist}\ x\ (p * 1.0)$
$=$   { Definition of $(*)$ }
**let** $\mathit{Dist}\ x\ p = d$ **in** $\mathit{Dist}\ x\ p$
$\stackrel{?}{=}$
$d$

Does the last step hold in general? It looks good for the deterministic case with $d = \mathit{Dist}\ \mathit{evnt}\ \mathit{prb}$.

**let** $\mathit{Dist}\ x\ p = \mathit{Dist}\ \mathit{evnt}\ \mathit{prb}$ **in** $\mathit{Dist}\ x\ p$
$=$
$\mathit{Dist}\ \mathit{evnt}\ \mathit{prb}$

However, the equality **let** $\mathit{Dist}\ x\ p = d$ **in** $\mathit{Dist}\ x\ p \equiv d$ does not hold in general. For instance, let us consider the case $d = \mathit{failed}$.

**let** $\mathit{Dist}\ x\ p = \mathit{failed}$ **in** $\mathit{Dist}\ x\ p$
$=$
$\mathit{Dist}\ \mathit{failed}\ \mathit{failed}$
$\neq$
$\mathit{failed}$

That is, using an equality based on the resulting terms, the left–hand side is more defined then the right–hand side if $d = \mathit{failed}$.

Because the user cannot access the `Dist` constructor, they cannot observe this difference. The user can only compare two distributions by using the querying operator (`??`). Therefore, in the following we show that the monad laws hold if we consider a context of the form `pred ?? d` where `pred` is an arbitrary predicate. Recall that we defined the operator (`??`) as follows.

```
(??) :: (a -> Bool) -> Dist a -> Probability
(??) pred d = foldValues (+.) 0.0 (allValues (prob (filterDist pred d)))
```

Fortunately, the monad laws already hold if we consider the context *filterDist pred* for an arbitrary predicate $pred :: a \to Bool$. Therefore, we show that the following equalities hold for all distributions $d$, and all values $x$, $pred$, $f$, and $g$ of appropriate types.

(1) *filterDist pred* $(d \ggg certainly) \equiv$ *filterDist pred* $d$

(2) *filterDist pred* $(certainly\ x \ggg f) \equiv$ *filterDist pred* $(f\ x)$

(3) *filterDist pred* $((d \ggg f) \ggg g)) \equiv$ *filterDist pred* $(d \ggg (\lambda y \to f\ y \ggg g))$

First we show that equation (1) holds. We reason as follows for all distributions $d :: Dist\ \tau$ and predicates $pred :: \tau \to Bool$.

$\quad$ *filterDist pred* $(d \ggg certainly)$
$=\quad$ { Reasoning above }
$\quad$ *filterDist pred* (**let** *Dist x p* $= d$ **in** *Dist x p*)
$=\quad$ { Definition of *filterDist* }
$\quad$ **let** *Dist y q* $=$ (**let** *Dist x p* $= d$ **in** *Dist x p*)
$\quad$ **in if** $(pred\ y)$ **then** $(Dist\ y\ q)$ **else** *failed*
$=\quad$ { Inline **let**–declaration }
$\quad$ **let** *Dist y q* $= d$
$\quad$ **in if** $(pred\ y)$ **then** $(Dist\ y\ q)$ **else** *failed*
$=\quad$ { Definition of *filterDist* }
$\quad$ *filterDist pred d*

The (`>>>=`)–operator defers the pattern matching to the right–hand side via a `let`–expression. This so–called lazy pattern matching causes the monad laws to not hold without any context. However, because `filterDist` introduces a lazy pattern matching via a `let`–expression as well, observing two distributions via `filterDist` hides the difference between the two sides of the equation.

**The second monad law** For the second monad law (2), we reason as follows for all $x :: \tau_1$, and all $f :: \tau_1 \to Dist\ \tau_2$.

$\quad$ *certainly x* $\ggg f$
$=\quad$ { Definition of ($\ggg$) }
$\quad$ **let** *Dist y p* $=$ *certainly x*
$\qquad$ *Dist z q* $= f\ y$
$\quad$ **in** *Dist z* $(p * q))$
$=\quad$ { Definition of *certainly* }
$\quad$ **let** *Dist y p* $=$ *Dist x* $1.0$
$\qquad$ *Dist z q* $= f\ y$

in $Dist\ z\ (p * q))$
$=$ { Inlining of $Dist\ y\ p = Dist\ x\ 1.0$ }
let $Dist\ z\ q = f\ x$ in $Dist\ z\ (1.0 * q)$
$=$ { Definition of $(*)$ }
let $Dist\ z\ q = f\ x$ in $Dist\ z\ q$
$\stackrel{?}{=}$
$f\ x$

Here we observe the same restrictions as before, for example, if $f$ yields *failed* for any argument $x$ the equality does not hold. Once again, we consider the context *filterDist pred* for all $pred :: \tau_2 \to Bool$ to reason that the user cannot observe the difference.

*filterDist pred* (let $Dist\ z\ q = f\ x$ in $Dist\ z\ q$)
$=$ { Definition of *filterDist* }
let $Dist\ x\ p =$ (let $Dist\ z\ q = f\ x$ in $Dist\ z\ q$)
in if $(pred\ x)$ then $(Dist\ x\ p)$ else *failed*
$=$ { Inline let–declaration }
let $Dist\ x\ p = f\ x$
in if $(pred\ x)$ then $(Dist\ x\ p)$ else *failed*
$=$ { Definition of *filterDist* }
*filterDist pred* $(f\ x)$

Fortunately, the second monad law holds as well in the context of *filterDist*.

**The third monad law** In order to complete the discussion of the monad laws, we finally consider the associativity of $(\ggg)$ (3). For all $d :: Dist\ \tau_1$, $f :: \tau_1 \to Dist\ \tau_2$, and $g :: \tau_2 \to Dist\ \tau_3$ we reason as follows.

$(d \ggg f) \ggg g$
$=$ { Definition of $(\ggg)$ }
let $Dist\ x1\ p1 = d \ggg f$
$\quad Dist\ y1\ q1 = g\ x1$
in $Dist\ y1\ (p1 * q1)$
$=$ { Definition of $(\ggg)$ }
let $Dist\ x1\ p1 =$ let $Dist\ x2\ p2 = d$
$\quad\quad\quad\quad\quad\quad\quad\quad Dist\ y2\ q2 = f\ x2$
$\quad\quad\quad\quad\quad\quad$ in $Dist\ y2\ (p2 * q2)$
$\quad Dist\ y1\ q1 = g\ x1$
in $Dist\ y1\ (p1 * q1)$
$=$ { Simplifying nested let–expressions }
let $Dist\ x2\ p2 = d$
$\quad Dist\ y2\ q2 = f\ x2$
$\quad Dist\ x1\ p1 = Dist\ y2\ (p2 * q2)$
$\quad Dist\ y1\ q1 = g\ x1$
in $Dist\ y1\ (p1 * q1)$
$=$ { Inlining $Dist\ x1\ p1 = Dist\ y2\ (p2 * q2)$ }
let $Dist\ x2\ p2 = d$
$\quad Dist\ y2\ q2 = f\ x2$
$\quad Dist\ y1\ q1 = g\ y2$
in $Dist\ y1\ ((p2 * q2) * q1)$

$\quad=\quad$ { Renaming of $y2$ $q2$ }
**let** $Dist\ x2\ p2 = d$
$\quad\quad Dist\ x1\ p1 = f\ x2$
$\quad\quad Dist\ y1\ q1 = g\ x1$
**in** $Dist\ y1\ ((p2 * p1) * q1)$
$\quad=\quad$ { Associativity of $(*)$ }
**let** $Dist\ x2\ p2 = d$
$\quad\quad Dist\ x1\ p1 = f\ x2$
$\quad\quad Dist\ y1\ q1 = g\ x1$
**in** $Dist\ y1\ (p2 * (p1 * q1))$
$\quad=\quad$ { Adding local definition for $Dist\ y1\ (p1 * q1)$ }
**let** $Dist\ x2\ p2 = d$
$\quad\quad Dist\ x1\ p1 = f\ x2$
$\quad\quad Dist\ y1\ q1 = g\ x1$
$\quad\quad Dist\ y2\ q2 = Dist\ y1\ (p1 * q1)$
**in** $Dist\ y2\ (p2 * q2)$
$\quad=\quad$ { Using nested **let**–expressions }
**let** $Dist\ x2\ p2 = d$
$\quad\quad Dist\ y2\ q2 =$ **let** $Dist\ x1\ p1 = f\ x2$
$\quad\quad\quad\quad\quad\quad\quad\quad\quad Dist\ y1\ q1 = g\ x1$
$\quad\quad\quad\quad\quad\quad\quad$ **in** $Dist\ y1\ (p1 * q1)$
**in** $Dist\ y2\ (p2 * q2)$
$\quad=\quad$ { Definition of $(\ggg\!\!=)$ }
**let** $Dist\ x2\ p2 = d$
$\quad\quad Dist\ y2\ q2 = f\ x2 \ggg\!\!= g$
**in** $Dist\ y2\ (p2 * q2)$
$\quad=\quad$ { Definition of $(\ggg\!\!=)$ }
$d \ggg\!\!= (\lambda x \to f\ x \ggg\!\!= g)$

This reasoning shows that the associativity law actually holds without any additional context. All in all, *certainly* and $(\ggg\!\!=)$ form a valid monad from the user's point of view.

## 4.6 Case Studies

After presenting the basic combinators of the library and motivating the advantages of modelling distributions using non–determinism, we implement some exemplary applications. First, we start with an example already motivated in the introduction of this section: Bayesian networks. We define a simple Bayesian network and corresponding queries using our library. The second case study concerns examples that have been characterised as challenging for probabilistic logic programming by Nampally et al. (2018), who use the example to discuss the expressiveness of probabilistic logic programming and its cost with respect to performance. These examples focus on properties of random strings and their probabilities. In the third case study we model the famous secret santa problem in three different versions using our library. Furthermore, we show benchmarks of these examples and compare them with the probabilistic languages ProbLog and WebPPL. These comparisons confirm the advantages of non–strict non–determinism with respect to performance. In Appendix A.4 we provide the code written in ProbLog and WebPPL that corresponds to the Curry code we discuss in the following sections.

## 4.6.1 Bayesian Network

As mentioned in the beginning, Bayesian networks are a popular example for probabilistic programming. Since each node in a Bayesian network corresponds to a function that yields a probability, the implementation of a Bayesian network fits perfectly in the setting of a functional language. In the following we implement the example shown in Figure 4.1. For each node of the acyclic graph we define a function that yields a distribution. In the example, each node performs a binary decision: it either rains or not, the sprinkler is activated or deactivated and the grass is either wet or not. Thus, the return type of all three functions is `Dist Bool`. Since all functions yield Boolean distributions, we use the combinator `flip :: Probability -> Dist Bool`, again. Recall that the function `flip` yields `True` with the probability given as the first argument and `False` with the complementary probability.

```
flip :: Probability -> Dist Bool
flip p = enum [True,False] [p, 1 - p]
```

Now we can start with the simplest function: the node representing rain depends on no other variable in the graph, that is, has no input arguments. Furthermore, the function yields `True` with a probability of $20\,\%$ and `False` with a probability of $80\,\%$, respectively, according to the graph.

```
raining :: Dist Bool
raining = flip 0.2
```

The nodes representing the sprinkler and the grass depend on other variables; the sprinkler depends on the rain, and the grass on the sprinkler as well as the rain. That is, the function `sprinklerOn` takes one, and the function `grassWet` two arguments according to the graph representation.

```
sprinklerOn :: Bool -> Dist Bool
sprinklerOn False = flip 0.4
sprinklerOn True  = flip 0.01

grassWet :: Bool -> Bool -> Dist Bool
grassWet False False = flip 0.0
grassWet False True  = flip 0.8
grassWet True  False = flip 0.9
grassWet True  True  = flip 0.99
```

Notice that the implementation is merely a copy of the table presented in Figure 4.1. Next we want to define some queries on top of our network. The model needs to be used in a certain way: the output of the function `raining` is a parameter for `sprinklerOn` and both are arguments of `grassWet`. In order to query on this model more easily, we define a record data type where each field represents a variable of the model.

```
data GrassModel = Model { isRaining :: Bool
                        , isSprinklerOn :: Bool
                        , isGrassWet :: Bool }
```

Now we define a distribution that yields a `GrassModel` using the functions `raining`, `sprinklerOn`, and `grassWet`, accordingly.

```
grassModel :: Dist GrassModel
grassModel = raining         >>>= \r ->
             sprinklerOn r   >>>= \s ->
             grassWet s r    >>>= \g ->
             certainly (Model r s g)
```

Let us now perform the first query to check if our model is correct. We can ask the model for the probability of wet grass given that it is raining. Remember that we perform queries on the model by using the operator (??), which expects a predicate as first argument. We use a conjunction to check that it is raining and that the grass is wet.

```
grassWetAndRain :: Probability
grassWetAndRain = (\m -> isRaining m && isGrassWet m) ?? grassModel
```

⊢ grassWetAndRain
0.16038

The probability for wet grass and raining is $16.04\,\%$. If we reexamine the graph once again, we can see that this probability cannot be easily read off the graph.

The query above answers a question that follows the dependency flow: we want to know something about `grassWet` depending on one of its arguments, `raining`. However, we can also ask questions about `raining` depending on `grassWet`. For example, what is the probability that it is raining given that the grass is wet? In order to answer this question, it is not enough to query our model as above. The question corresponds to a conditional probability, that is, we need to compute the probability of the conjunction of the events and divide it by the probability of the given condition. For the query in question, the conjunction of events corresponds to the probability that it is raining and that the grass is wet — that is, the query we performed above. The divisor is the probability that the grass is wet, without considering any other side–conditions.

⊢ grassWetWhenRain / (isGrassWet ?? grassModel)
0.35768768

Since it is quite common to calculate conditional probabilities — and we will need it again later — we define some combinators for calculating conditional probabilities. The first convenience function checks for a list of predicates, if all predicates hold for the given distribution.

```
allProb :: [a -> Bool] -> Dist a -> Probability
allProb ps dx = (\ x -> all (\p -> p x) ps) ?? dx
```

Based on `allProb` we then define a function `condProb` that implements a conditional probability based on two lists of predicates: the first list of predicates describes the probability we are actually interested in, the second list gives the side–conditions that should apply to the query.

```
condProb :: [a -> Bool] -> [a -> Bool] -> Dist a -> Probability
condProb ps1 ps2 dx = allProb (ps1 ++ ps2) dx / allProb ps2 dx
```

Note that we pass the concatenation of the predicates to `allProb`, which resembles the conjunction of these predicates. As an example, in the case of the query above we calculate the probability that it is raining with the side–condition that we already know that the grass is wet as follows.

```
⊢ condProb [isRaining] [isGrassWet] grassModel
0.35768768
```

The combinator `condProb` and its usage resembles the formula that is used in probability theory; the conditional probability would be expressed as $P\left(Raining = True \mid GrassWet = True\right)$. The lists of predicates correspond to the listing of variables and their bindings. For example, the predicate `isRaining` corresponds to $Raining = True$.

### 4.6.2 Random Strings

In order to compare our library with other approaches for probabilistic programming, we reimplement two examples about random strings that have also been implemented in ProbLog. The ProbLog implementation can be found online.[7] We generate random strings of a fixed length over the alphabet $\{a, b\}$ and calculate the probability that this string is a palindrome and contains the subsequence $bb$, respectively.

First we define a distribution that picks a character uniformly from the alphabet $\{a, b\}$.

```
pickChar :: Dist Char
pickChar = uniform ['a','b']
```

Based on `pickChar` we define a distribution that generates a random string of length n, that is, picks a random char n times. We reuse `replicateDist` to define this distribution.

```
randomString :: Int -> Dist String
randomString n = replicateDist n (\ () -> pickChar)
```

In order to compute the probability that a random string is a palindrome and contains a subsequence $bb$, respectively, we define predicates that test these properties for a given string. A string is a palindrome, if it reads the same forwards and backwards. The following predicate, thus, checks if the reverse of a given string is equal to the original string. We use a tail–recursive implementation of `reverse` to benefit from a run time complexity that is linear in the length of the list instead of quadratic.

```
reverse :: [a] -> [a]
reverse xs = rev xs []
 where rev ys acc = case ys of
                    [] -> acc
                    (z:zs) -> rev zs (z:acc)

palindrome :: String -> Bool
palindrome str = str == reverse str
```

The predicate that checks if a string contains two consecutive $b$s can be easily defined via pattern matching.

---
[7]https://dtai.cs.kuleuven.be/problog/tutorial/various/04_nampally.html

```
consecutiveBs :: String -> Bool
consecutiveBs bs = case bs of
                     []          -> False
                     ('b':'b':_) -> True
                     (_  : bs  ) -> consecutiveBs bs
```

Now we are ready to perform some queries. What is the probability that a random string of length 5 is a palindrome?

⊢ `palindrome ?? (randomString 5)`
$\frac{1}{4}$

What is the probability that a random string of length 10 contains two consecutive *b*s?

⊢ `consecutiveBs ?? (randomString 10)`
`0.859375`

In general the approach to query using `palindrome` and `consecutiveBs` is quite naive and, thus, inefficient because all strings of the given length have to be enumerated explicitly. Due to this inefficiency, the ProbLog homepage introduces a more efficient version for both problems. In the following, we will discuss the alternative implementation to compute the probability for a palindrome only. This more efficient version has arguments for the index of the front and back position, picks characters for both ends and then moves the position towards the middle. That is, instead of naively generating the whole string of length $n$, this version checks each pair of front and back position first and fails straightaway, if they do not match. If the characters do match, the approach continues by moving both indices towards each other. In Curry an implementation of this idea looks as follows.

```
palindromeEfficient :: Int -> Dist (Bool, String)
palindromeEfficient n = palindrome' 1 n

palindrome' :: Int -> Int -> Dist (Bool,String)
palindrome' n1 n2 | n1 == n2  = pickChar >>>= (\ c -> certainly (True,[c]))
                  | n1 > n2   = certainly (True, [])
                  | otherwise = pickChar >>>= \c1 ->
                                pickChar >>>= \c2 ->
                                palindrome' (n1+1) (n2-1) >>>= \ (b,cs) ->
                                certainly (c1 == c2 && b, c1 : cs ++ [c2])
```

The interesting insight here is that, thanks to the combination of non–determinism and non–strictness, the evaluation of the first query based on `palindrome` behaves similar to the efficient variant in ProbLog. At first, it seems that the query performs poorly, because the predicate `palindrome` needs to evaluate the whole list due to the usage of `reverse`. The good news is, however, that the non–determinism is only spawned if we evaluate the elements of that list, and the elements still evaluate non–strictly, when explicitly triggered by (`==`). More precisely, because of the combination of `reverse` and (`==`), the evaluation starts by checking the first and last characters of a string and only continues to check more characters — and spawn more non-determinism — if they match. If these characters do not match, the evaluation fails directly and does not need to check any more characters. In a nutshell, we get a version competitive with the efficient implementation although we used a naive generate and test approach.

Last but not least, we want to emphasise that the restrictions with respect to using non–determinism (including usage of `failed` due to partial functions) discussed in Section 4.3 did not affect the reimplementation of these examples. As the examples shown here are taken from the probabilistic programming literature, we are confident that the restrictions do not have consequences for the programmability regarding common applications of probabilistic programming.

### 4.6.3 Secret Santa

Most tutorials on probabilistic programming include an example that models the classical Monty Hall game to show the probabilities for several game scenarios. In this section we will take a look at another problem that probably most people have already heard of. We model the preparation for a game of secret santa.[8] Secret santa is a famous western christmas tradition in which a group of people organise to exchange gifts. The main idea is that each person is randomly assigned to another person and that these assignments declare who has to give a gift to whom. More precisely, these assignments are not symmetric by default; they can, however, be symmetric by chance. Furthermore, a person cannot pick themself as gift receiver. As the assignments are usually drawn from a hat with names of all the participants on a piece of paper inside, these draws do not all end up in a valid game constellation. In this section we will take a look at the probability that the name picking phase yields an invalid game constellation.

We start by defining each person as an `Int` value and a `Hat` as merely a list of such `Person`s.

```
type Person = Int
type Hat = [Person]
```

As already noted, an assignment results in an invalid game constellation, if one person draws themself. We incorporate these possible outcomes in a data type `SecretSanta` that represents an invalid game using `Failed` and a valid constellation using `Success`. Since we are interested in the assignments of each person, the `Success`–constructor contains the list of `SantaAssignment`s as argument.

```
data SecretSanta = FailedGame | Success [SantaAssignment]
data SantaAssignment = Assignment { santa :: Person, person :: Person }
```

A `SantaAssignment` always consists of a secret santa and a person receiving the gift.

We define a game of secret santa as function that takes the number of participants as argument and yields a `Hat` with numbers 1 to the number of participants, if there are more than one participant.

```
santaGame :: Int -> Hat
santaGame n | n > 1 = [1..n]
            | otherwise = error "invalid game"
```

It will hopefully come with no surprise that each person in the hat can be drawn with the same probability. Thus, we define `pickFromHat` to yield a uniform distribution for a given hat.

---

[8]This example was motivated by an episode of *Numberphile* hosted by Dr. Hannah Fry. See online via https://youtu.be/5kC5k5QBqcc

```haskell
pickFromHat :: Hat -> Dist Person
pickFromHat = uniform
```

In order to make reasonable use of `pickFromHat`, we need a function that actually keeps track which persons are still left in the hat after a pick. That is, we define `pPicks` that yields a distribution of a potential `SantaAssignment` and the remaining hat with the list of persons. We pass a hat and the person that draws from the hat as arguments. As the hat might be empty, we wrap the result as an optional value, yielding `Nothing` for an empty hat.

```haskell
pPicks :: Person -> Hat -> Dist (Maybe (SantaAssignment, [Person]))
pPicks _ []       = certainly Nothing
pPicks p ps@(_:_) =
  pickFromHat ps >>>= \p' ->
  certainly (Just (Assignment p p', delete p' ps))
```

Now, for a naive round of draws from the hat, we let each person draw from the hat without interference and all draws are only visible at the end when everybody already picked another person.

```haskell
pickRound :: Hat -> Dist SecretSanta
pickRound []       = certainly FailedGame
pickRound xs@(_:_) = pickRound' xs xs [] >>>= \ arrs ->
                     certainly (Success arrs)
 where
  pickRound' []     _   arrs = certainly arrs
  pickRound' (p:ps) hat arrs = pPicks p hat >>>= \ (Just (arr,hat')) ->
                                pickRound' ps hat' (arr:arrs)
```

We can try out our implementation for a small game with three people.

```haskell
santa1 :: Dist SecretSanta
santa1 = pickRound (santaGame 3)
```

```
⊢ santa1
(Dist (Success [(Assignment 3 3),(Assignment 2 2),(Assignment 1 1)]) 1/6)
(Dist (Success [(Assignment 3 2),(Assignment 2 3),(Assignment 1 1)]) 1/6)
(Dist (Success [(Assignment 3 3),(Assignment 2 1),(Assignment 1 2)]) 1/6)
(Dist (Success [(Assignment 3 1),(Assignment 2 3),(Assignment 1 2)]) 1/6)
(Dist (Success [(Assignment 3 2),(Assignment 2 1),(Assignment 1 3)]) 1/6)
(Dist (Success [(Assignment 3 1),(Assignment 2 2),(Assignment 1 3)]) 1/6)
```

We can see that all games are marked as `Success`. If we take, however, a closer look at the assignments for each of the games, we see that there are multiple invalid assignments. This problem is not that surprising, because `pickRound` lets each person pick without interference, that is, we do not check if the pick is valid or not. As mentioned in the beginning, a pick is invalid if a person draws themselves from the hat.

```haskell
isFailedAssign :: SantaAssignment -> Bool
isFailedAssign secret = santa secret == person secret
```

Using the predicate `isFailedAssign`, we can normalise the result from `pickRound` in order to find and mark invalid games. That is, if any assignment of a `SecretSanta` game is invalid, the whole game fails.

```
normFailedGame :: SecretSanta -> SecretSanta
normFailedGame FailedGame        = FailedGame
normFailedGame game@(Success arrs)
  | any isFailedAssign arrs = FailedGame
  | otherwise               = game
```

We check our implementation once again using `normFailedGame` to mark invalid games.

```
santa1' :: Dist SecretSanta
santa1' = pickRound (santaGame 3) >>>= \game -> certainly (normFailedGame game)
```

⊢ santa1'
(Dist FailedGame $\frac{1}{6}$)
(Dist FailedGame $\frac{1}{6}$)
(Dist FailedGame $\frac{1}{6}$)
(Dist (Success [(Assignment 3 1),(Assignment 2 3),(Assignment 1 2)]) $\frac{1}{6}$)
(Dist (Success [(Assignment 3 2),(Assignment 2 1),(Assignment 1 3)]) $\frac{1}{6}$)
(Dist FailedGame $\frac{1}{6}$)

This result looks better. Next, we query for the probability that a game is invalid.

```
isFailedGame :: SecretSanta -> Bool
isFailedGame FailedGame  = True
isFailedGame (Success _) = False
```

⊢ isFailedGame ?? santa1'
$\frac{2}{3}$

Only a third of all game constellations are valid. That is, if every person picks from the hat before we check for invalid game constellations, every three games we need to do it all again from the beginning.

There are usually two possible alternative procedures to avoid invalid game constellations. For the first alternative we check after each pick if this pick invalidates the game, i.e., if a person picked themselves. There is, however, a problem with this alternative: there is no definite ending, a person could end up picking themselves every time they try again. A modified version of this idea is the second alternative. Instead of picking a new name after an invalid pick, we modify the hat before each pick such that invalid picks cannot happen in the first place. That is, if a person p picks from the hat, we (temporarily) delete this person from the hat and add it again before the next person picks. We do this for every person that picks from the hat.

```
pickRoundWOFailed :: Hat -> Dist SecretSanta
pickRoundWOFailed []       = certainly FailedGame
pickRoundWOFailed xs@(_:_) = pickRound' xs xs []
 where
  pickRound' []     _   as = certainly (Success as)
  pickRound' (p:ps) hat as =
```

```
    pPicks p (delete p hat) >>>= \ mAssign ->
    maybe (certainly FailedGame)
          (\(a,_) -> pickRound' ps (delete (person a) hat) (a:as))
          mAssign
```

Note that for this solution to work, it is crucial that `pPicks` can handle an empty hat as well. The only possible invalid game that could happen with this setup is that the last person might pick themself. In this case the hat only contains the name of this person, which will be deleted before the pick. Thus, the person tries to pick from an empty hat which leads to an invalid game.

With this alternative picking procedure, we can reduce the amount of replays as only every fourth game will end in an invalid constellation.

⊢ isFailedGame ?? (pickRoundWOFailed (santaGame 3))

$\frac{1}{4}$

Instead of manipulating the hat before each pick, we can make an additional check and pick a second time as necessary. After a pick, we check if the person picked themself and if so, they make a second pick without putting themself back into the hat. Note that similar to the alternative `pickRoundWOFailed` the hat can be empty for the second try, if the last person to pick can only pick themself.

```
pickAndCheckRound :: Hat -> Dist SecretSanta
pickAndCheckRound [] = certainly FailedGame
pickAndCheckRound xs@(_:_) = pickRound' xs xs []
 where
  pickRound' []      _   assigns = certainly (Success assigns)
  pickRound' (p:ps) hat assigns =
    pPicks p hat >>>= \ mAssign ->
    maybe (certainly FailedGame)
          (\ (assign,newHat) ->
             if person assign == p
                then pPicks p newHat >>>= \ mAssign2 ->
                     maybe (certainly FailedGame)
                           (\ (assign2,newHat2) ->
                              pickRound' ps (p:newHat2) (assign2:assigns))
                           mAssign2
                else pickRound' ps newHat (assign:assigns))
          mAssign
```

⊢ isFailedGame ?? (pickAndCheckRound (santaGame 3))

$\frac{1}{4}$

We can take this idea one step further and repeat the process until we end up with a valid pick. However, as we can only model discrete and not continuous distributions, we need to set a limit for the number of retries we do.

```
pickAndRepeatRound :: Int -> Hat -> Dist SecretSanta
pickAndRepeatRound _ [] = certainly FailedGame
pickAndRepeatRound limit xs@(_:_) = pickRound' limit xs xs []
```

```
  where
   pickRound' _     []     _   assigns = certainly (Success assigns)
   pickRound' limit (p:ps) hat assigns
    | limit == 0 = certainly FailedGame
    | limit >  0 =
      pPicks p hat >>>= \ mAssign ->
      maybe (certainly FailedGame)
            (\ (assign,newHat) ->
              if person assign == p
                then pickRound' (limit - 1) (p:ps) hat assigns
                else pickRound' limit ps newHat (assign:assigns))
            mAssign
```

Running this implementation with an increasing value for the number of retries reveals that the overall probability for an invalid game converges to $25\%$.

⊢ isFailedGame ?? (pickAndRepeatRound 1 (santaGame 3))
$\frac{2}{3}$

⊢ isFailedGame ?? (pickAndRepeatRound 5 (santaGame 3))
0.2802211934156378

⊢ isFailedGame ?? (pickAndRepeatRound 10 (santaGame 3))
0.2509723287280479

⊢ isFailedGame ?? (pickAndRepeatRound 20 (santaGame 3))
0.2500009536026171

When passing 1 as limit to `pickAndRepeatRound`, the implementation is equivalent to using the function `pickRound` and normalising the results like we did in the beginning. Note that both approaches are equivalent because no picker is allowed to retry when picking themself.

### 4.6.4 Performance Comparisons

Up to now, the only performance comparisons we discussed were for different implementations of our library in Curry and Haskell. These comparisons showed the advantage of using non–strict non–determinism concepts in combination with the right amount of laziness for the implementation of the library. Next we want to take a look at the comparison with the full–blown probabilistic programming languages ProbLog and WebPPL. ProbLog is a probabilistic extension of Prolog that is implemented in Python. WebPPL is the successor of Church; in contrast to Church it is not implemented in Scheme but in JavaScript.

In order to try to measure the execution of the programs only, we precompiled the executable for the Curry programs. As Python is an interpreted language, a similar preparation was not available for ProbLog. However, we used ProbLog as library in order to call the Python interpreter directly. ProbLog is mainly implemented in Python, which allows users to import ProbLog as a Python package.[9] For WebPPL, we used *node.js* to run the JavaScript program as a terminal application. All of the following running times are the mean of $1000$ runs as calculated by the Haskell tool bench[10] that we use to run the benchmarks.

---
[9]https://dtai.cs.kuleuven.be/problog/tutorial/advanced/01_python_interface.html

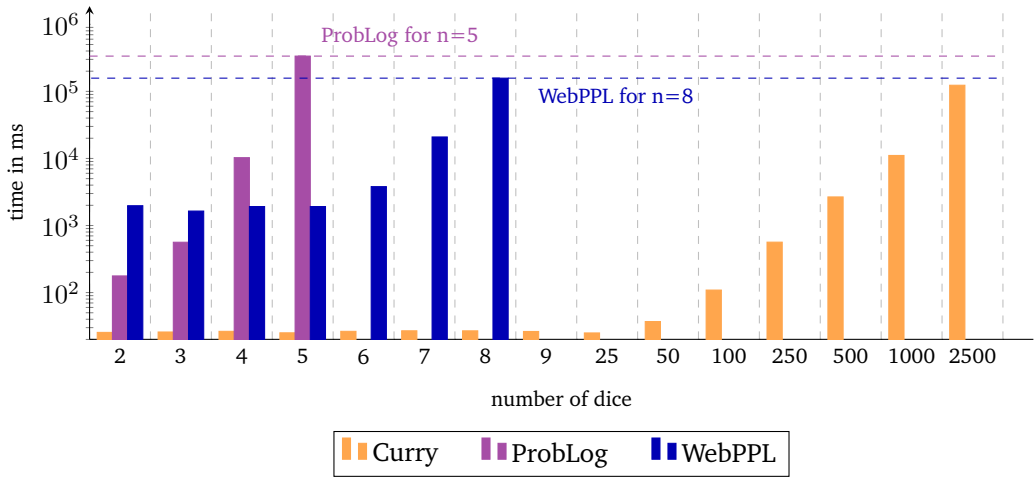

**Figure 4.5.:** Getting only sixes when rolling $n$ dice

We compare the running times based on the dice rolling example presented in Section 4.3.2 and all examples from the previous section.

**Dice Rolling** As discussed before, non–strict non–determinism performs pretty well for the dice rolling example, as a great deal of the search space is pruned early. That is, Figure 4.5 shows an impressive advantage of our Curry library in comparison with ProbLog and WebPPL. The x–axis represents the number of rolled dice and we present the time in milliseconds in logarithmic scale on the y–axis.

In order to demonstrate that our library outperforms ProbLog and WebPPL by several orders of magnitude for this example, we also run the Curry implementation for bigger values of $n$ that eventually had the same running time as the last tested value for the other languages. The right part of Figure 4.5 shows the running times for $25$ to $5000$ dice. We can see that our library can compute the probability for getting only sixes for $2500$ dice in roughly the same time as ProbLog for $5$ dice. The running times for WebPPL seem very bad in the beginning, but after a few throws it becomes obvious that there is a constant overhead. Nevertheless, whereas WebPPL computes the probability for $8$ dice, our library can compute the probability for $2500$ dice in roughly the same time.

**Palindrome** In order to back up the results of the previous example, Figure 4.6 shows benchmarks for implementations of the naive and the efficient versions of the palindrome generation in Curry, ProbLog and WebPPL. The x–axis represents the length of the generated palindrome and, once again, we present the time in milliseconds in logarithmic scale on the y–axis.

The figure uses dashed bars for the efficient version of the algorithm and a solid filling for the naive algorithm. The naive algorithm scales pretty bad in ProbLog and WebPPL. The Curry version is still applicable up to a string length of $30$ as its running time is similar to all three efficient versions. Overall, the efficient versions all perform in a similar time range, but WebPPL shows a slight performance advantage for an increasing length of the string.

**Secret Santa** The modelling of the secret santa problem is the first example, where it becomes apparent that our library in Curry has no chance to compete against other probabilistic languages, if the problem to solve cannot benefit from early pruning. In case of the secret santa problem, most of the computed assignments become invalid in the last pick. That is, all versions of our implementation need to traverse nearly the whole search space in order to compute the probability for an invalid game. More precisely, the predicate `isFailedGame` that we use in the query to compute the probability behaves similar in all implementations; in all implementations the whole list of assignments needs to be traversed in most of the cases. Figure 4.7 shows a comparison of the running times for the simplest secret santa model that does not check the picks early, but only checks the validity at the end. The tests are parametrised over the number of players that participate in the hat picking process, which is the value labeled by the x–axis.

---

[10] https://hackage.haskell.org/package/bench

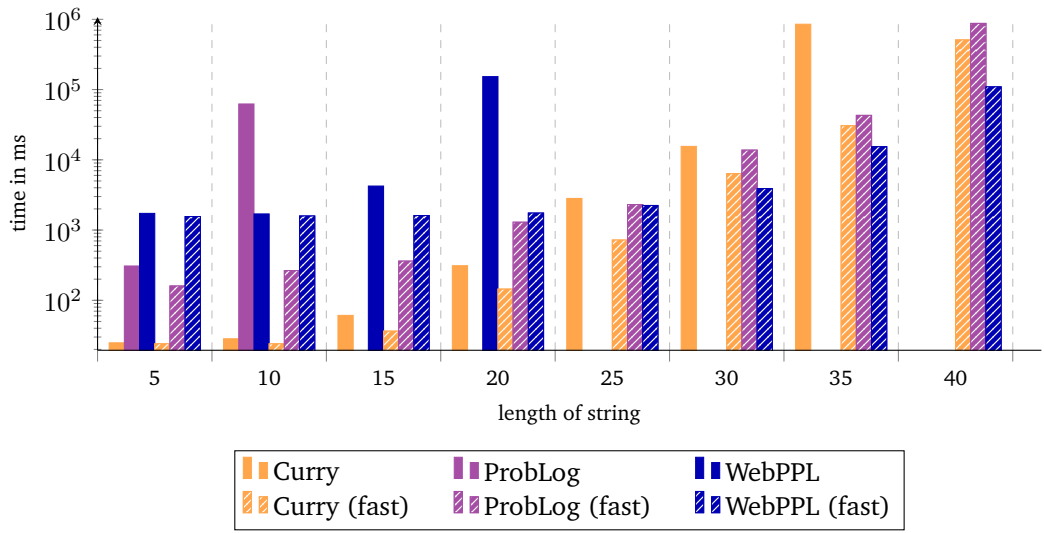

**Figure 4.6.:** Palindrome computation for a string of length $n$

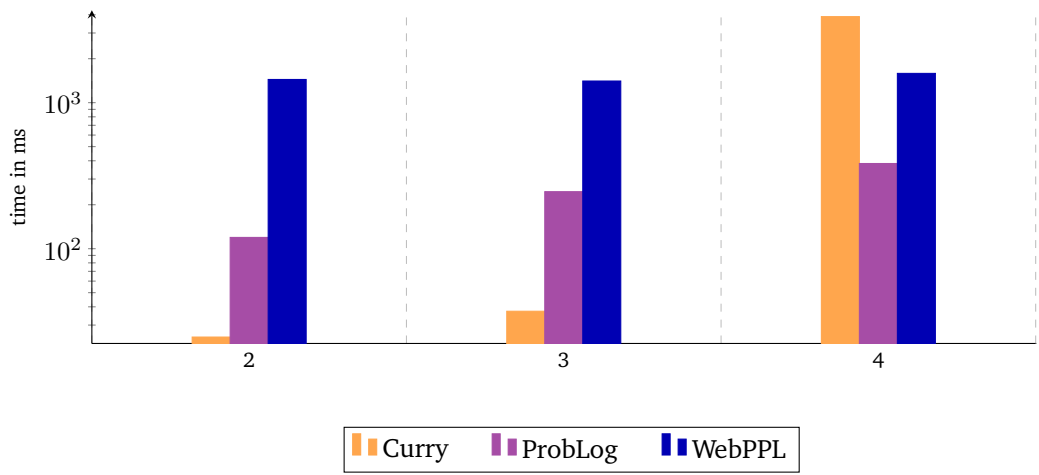

**Figure 4.7.:** Comparisons of running times for the secret santa model for an increasing number of players

At first it seems as our library can outperform the other languages again, but the exponential growth of the search space becomes already visible for a small number of players. The second visual comparison, Figure 4.8, shows the running times for the optimised implementation that does not allow a player to pick themself by construction.

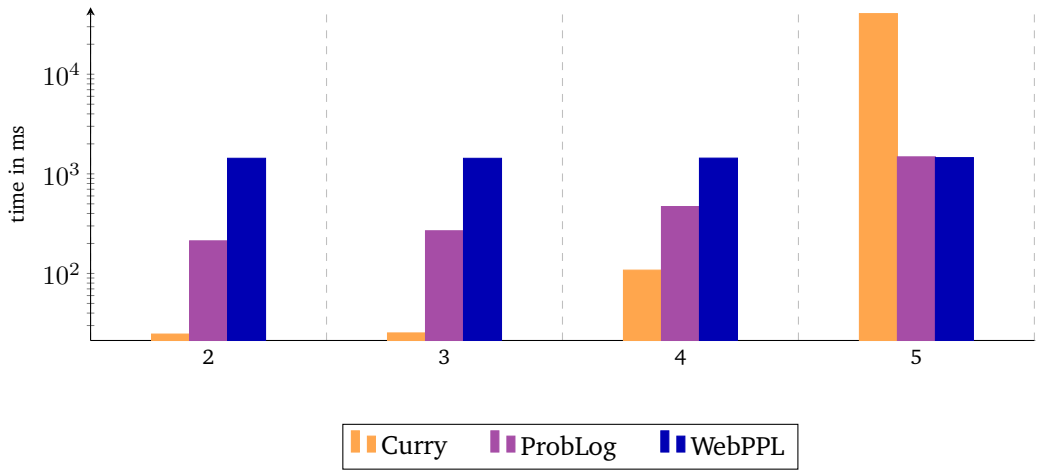

**Figure 4.8.:** Comparisons of running times for the secret santa model for an increasing number of players using an optimised strategy

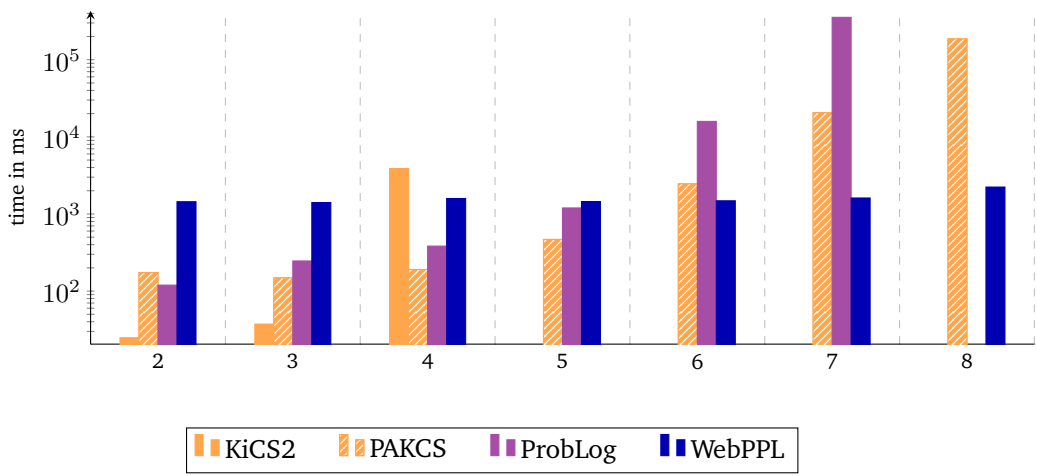

**Figure 4.9.:** Comparisons of running times for the secret santa model for an increasing number of players using PAKCS instead of KiCS2

The search space grows a bit slower, such that we can perform the query for a greater number of players as before. However, our library still performs bad in comparison to ProbLog and WebPPL. In order to show that the bad performance is not specific to our library, but to KiCS2's performance when a non–trivial amount of non–determinism is involved, we present the performance of PAKCS for the santa problems in comparison to ProbLog and WebPPL in Figure 4.9. We observe that PAKCS (orange with patterns) has a performance overhead in comparison to KiCS2 when we run our examples for a small number of players. That is, the less non–determinism is used within the program, the better the performance of KiCS2 in comparison to PAKCS and vice versa. Nevertheless, PAKCS shows a better performance for a greater number of players.

This case study shows that optimising KiCS2's performance with respect to non–determinism is an important topic for future research. However, we emphasise the suitability of a functional logic language like Curry for well–chosen examples and queries that explicitly use non–determinism in such a way that we can prune a great deal of the search space early.

**Bayesian Network** In order to complete the performance comparisons, we include the running times for three queries of the Bayesian network example in Figure 4.10. However, as the model is quite simple, the computational complexity to compute the queried probabilities is negligible. All we can deduce from these performance comparisons is the overhead of the individual language. In this case, Curry and our library performs best followed by ProbLog, WebPPL has a more noticeable overhead that we already discussed before.

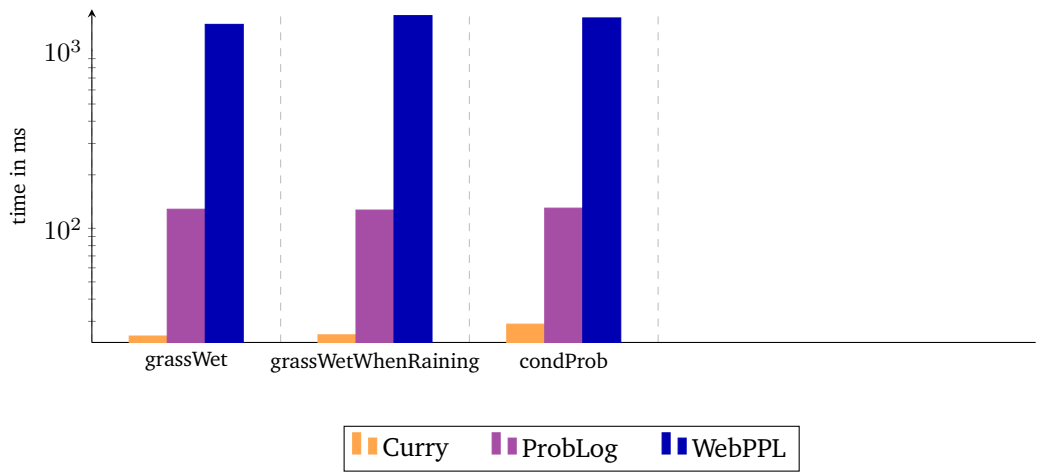

**Figure 4.10.:** Comparisons of running times for Bayesian reasoning examples

## 4.7 Related Work

The approach of the presented library for probabilistic programming in Curry is based on the work by Erwig and Kollmansberger (2006), who introduce a Haskell library that represents distributions as lists of event–probability pairs. Their library also provides a simple sampling mechanism to perform inference on distributions. Inference algorithms come into play because common examples in probabilistic programming have an exponential growth and it is not feasible to compute the whole distribution. Similarly, Ścibior et al. (2015) present a more efficient implementation using a DSL in Haskell. They represent distributions as a free monad and inference algorithms as an interpretation of the monadic structure. Thanks to this interpretation, the approach is competitive to full–blown probabilistic programming languages with respect to performance. In this work we focus on modelling distributions and have not implemented any sampling mechanism.

The benefit with respect to the combination of non–strictness and non–determinism is similar to the benefit of property–based testing in Curry (Christiansen and Fischer, 2008) and in Haskell using Curry–like non–determinism (Runciman et al., 2008). In property–based testing, sometimes we want to generate only test cases that satisfy a precondition. With Curry–like non–determinism the precondition can prune the search space early, while a list–based implementation has to generate all test cases and filter them afterwards. Both applications, probabilistic programming and property–based testing, are examples, where built–in non–determinism outperforms list–based approaches as introduced by Wadler (1985). In comparison to property–based testing, here, we observe that we can even add a kind of monadic layer on top of the non–determinism that computes additional information and still preserve the demand–driven behavior. However, the additional information has to be evaluated strictly — as it is the case for probabilities — otherwise we might lose non–deterministic results.

There are other more elaborated approaches to implement a library for probabilistic programming. For example, Kiselyov and Shan (2009) extend their library for probabilistic programming in OCaml with a construct for lazy evaluation to achieve similar positive effects. They use lazy evaluation for a concrete application based on importance sampling as well as for exact inference (calculating the total probability of a given distribution). Their implementation is a shallow embedding in OCaml that reimplements some features that our library already exploits from its functional logic host language. That is, due to the combination of non–strictness and non–determinism, we can efficiently calculate the total probability of the resulting distribution without utilising sampling or explicitly implementing non–strict probabilism for the examples shown here.

## 4.8 Future Work

As future work, we see a high potential for performance improvements for the Curry compiler KiCS2. The presented library for probabilistic programming serves as a starting point for further studies of functional logic features in practical applications. For example, we would expect the running times of the strict implementation based on non–determinism to be approximately as efficient as a list–based implementation. As the numbers in Section 4.3 show, the list approach is, however, considerably faster. Furthermore, a more detailed investigation of the performance of non–determinism in comparison to a list model is an interesting topic for itself.

The library's design does not support the use of non–determinism in events of a distribution. In case of deep non–determinism, we have to be careful to trigger all non–determinism when querying a distribution as shown in Section 4.4.1. Hence, the extension of the library with an interface using non–determinism on the user's side is an idea worth studying.

Last but not least, we see an opportunity to apply ideas and solutions of the functional logic paradigm in probabilistic programming. For instance, Christiansen et al. (2010) investigate free theorems for functional logic programs. As their work considers non–determinism and sharing, adapting it to probabilistic programming should be easy. As another example, Braßel (2009) presents a debugger for Curry that works well with non–determinism. Hence, it should be possible to reuse these ideas in the setting of probabilistic programming as well.

## 4.9 Conclusion

We have implemented a lightweight library for probabilistic programming for the functional logic programming language Curry. Such a library proves to be a good fit for a functional logic language, because both paradigms share similar features. While other libraries need to reimplement features specific to probabilistic programming, we solely rely on core features of functional logic languages.

The key idea of the library is to use non–determinism to model distributions. We discussed design choices as well as the disadvantages and advantages that result from this approach. In the end, the library provides non–strict probabilistic combinators in order to avoid spawning unnecessary non–deterministic computations. These non–strict combinators have benefits in terms of performance due to early pruning. Using combinators that are too strict leads to a loss of these performance benefits. Fortunately, the user does not have to worry about using the right amount of strictness as long as they only use the provided combinators. There are, however, two restrictions, the user has to follow when using the library. If the user does not follow these restrictions, a program may behave unexpectedly, in particular, the usual monad laws do not hold. Events may not be non–deterministic and the function passed as second argument of (>>>=)–operator may not be partial. Considering these restrictions, we showed that the library obeys the expected monad laws with respect to querying a distribution.

## 4.10 Final Remarks

The basis of this chapter has been published previously. The general introduction of the library and its implementation ideas have been published in the Proceedings of the 20th International Symposium on Practical Aspects of Declarative Languages (Dylus et al., 2018). An extended version includes reasoning about the monad laws, a small case study containing the random string examples as well as performance results for these examples with comparisons to WebPPL and ProbLog. This extended version was published in Theory and Practice of Logic Programming (Dylus et al., 2020). The chapter at hand contains additional material. First, the case study contains three parts: the aforementioned random string examples, a Bayesian network example, and the development of several ways to model the secret santa problem. Second, with the new examples come also new performance comparisons: we reimplemented all case studies in ProbLog as well as WebPPL and compared their performance. Last but not least, we also show the proof for the third monad law that was not included in either version before.

# Formal Reasoning About Effectful Non–Strict Programs

<div style="text-align:right">5</div>

> *There is no deep theoretical reason why this program should be rejected […].*
>
> — **Adam Chlipala**
> (Associate Professor of Computer Science)

This chapter moves the focus from applications using effects like non–determinism to an approach to model effectful programs in order to apply equational reasoning using a proof assistant. Equational reasoning is a popular and common tool to prove properties about effectful functional programs and, thus, prominently used by Haskell enthusiasts (Jeuring et al., 2012; Gibbons and Hinze, 2011; Hutton and Fulger, 2008). In particular, Danielsson et al. (2006) argue that reasoning about total programs only and expecting the result to carry over to partial programs is *morally correct*. Nevertheless, since we are interested in proving properties about partial or other effectful programs more explicitly, we cannot apply equational reasoning as it is and need a model to represent such effects.

In this chapter we present an approach to model effectful non–strict functional programs in the proof assistant Coq and prove exemplary properties about functions used in functional and functional logic languages. Our running example is often used as a finger exercise for proofs in a total setting. The main insight of our approach is that a lot of these properties carry over to partial languages: these properties still hold if partial values or non–deterministic computations are at play. Our approach allows to reason about a whole class of effects, and not only about a concrete effect like partiality or non–determinism. The interesting outcome of this approach is that one proof is enough to prove a proposition for a whole class of effects. In the following we refer to this insight as proving effect–generic properties.

After discussing the general idea for a framework to model effectful non–strict functional programs, we emphasise one effect occurring in Haskell: partiality. We present our approach using the example of the associativity of Haskell's function `append` to concatenate two lists.

Last but not least we give an outlook on first ideas to model Curry programs. The underlying effect of Curry is non–determinism in combination with laziness, that is, non–strictness and sharing. As we have seen in previous chapters, the combination of non–determinism and non–strictness behaves with respect to Curry's call–time choice semantics, that is, variables denote values and not computations. We show how to model non–determinism and discuss the problem of modelling a language with sharing. That is, we make more clear that our framework currently models non–strict program semantics, but does not handle lazy semantics. More precisely, we can model call–by–name but not call–by–need.

In summary, this chapter makes the following contributions.

- We present a generic model of effectful non–strict functional programs in the proof assistant Coq.

- In order to define a valid representation of effectful programs, we take the reader on an introductory tour of Coq's intricacies.

- We implement a model of effectful data structures that pleases Coq's termination checker.

- We argue that our approach is especially beneficial when it comes to effect–generic properties: instead of individually proving a property for each new effect of interest, we can prove such a property for a variety of effects once and for all.

- We present first ideas on modelling Curry's non–determinism and discuss arising obstacles when trying to model call–time choice.

- In the light of these obstacles, we discuss that our current framework works well for call–by–name semantics, but does not work for call–by–need semantics out–of–the box.

- Moreover, we show some examples that illustrate why this distinction comes up when modelling Curry's non–determinism but not for partiality.

## 5.1 A Generic Model For Effectful Non–Strict Programs

In this section we will present our general approach to model effectful non–strict programs in a dependently–typed language like Coq. We illustrate why previous work is insufficient for our goals and does not tackle the same problem as we do, respectively. Since the reasons for the insufficiency of the previous work is due to rather technical details of dependently–typed languages and the general idea of propositions as types (Wadler, 2015), we give a detailed explanation to illustrate which steps are necessary to model our approach in Coq. That is, this section tries to explain the obstacles and their solutions in Coq in a beginner–friendly way.

### 5.1.1 Representation of Partial Programs

The first obstacle that arises when we want to prove properties about Haskell programs in Coq is the fact that Coq is a total language while often Haskell programs are partially defined. That is, we cannot translate partial Haskell programs into Coq as they are, we need to represent the partial parts of the programs more explicitly. In the context of this thesis, partial programs or function definitions are terminating but do not yield a defined output for all inputs.

Let us take a look at an example. One of the first partial functions that comes to mind is the `head` function on lists that we defined in Section 2.1.1. In case of the empty list the function is not defined as there is no element to return.

```
head :: [a] -> a
head []    = undefined
head (x:_) = x
```

In a total language like Coq, the corresponding head function usually takes an additional argument to take care of the undefined behaviour in case of the empty list. The following implementation is part of Coq's predefined library for lists.

```
Definition hd (A : Type) (default : A) (l : list A) : A =
  match l with
  | nil      => default
  | cons x _ => x
  end.
```

Alternatively, we can indicate the potentially undefined behaviour by changing the result type to `option A`. In case of a non–empty list we wrap the head element into the `Some`–constructor, and use `None` as resulting value for an empty list. This alternative implementation is predefined under the name `head`.

```
Inductive option (A : Type) :=
| None : option A
| Some : A -> option A.

Definition head (A : Type) (l : list A) : option A =
  match l with
  | nil      => None
  | cons x _ => Some x
  end.
```

The latter version suites our problem a bit better than the first: there is not always a good candidate in scope that can be used as default value, the latter version fits a general setting better. However, the latter version is not compositional. Consider the following example that uses the result of `head` as an argument.

```
Fail Definition exampleList (A : Type) : list A :=
  cons (head (cons 1 nil)) nil.
```

We cannot use the result as an element of the newly constructed list, we get a type error.

> *The term "cons (head nil) nil" has type "list (option ?A)" while it is expected to have type "list A".*

The list constructor `cons` expects the head element of the list to match its type with all the remaining elements of the list. Since we want to construct a list of type `list A`, the head element needs to be of type `A`; the function `head`, however, yields a value of type `option A`.

As Haskell is a non–strict language, it is crucial that we allow partial values at top–level as well as within the components of data structures, that is, within arguments of constructors, in order to model Haskell's partiality correctly. Due to non–strictness, it is possible that an undefined value is never demanded during evaluation. The Haskell expression `head (1 : undefined)`, for example, does not trigger the `undefined` value in the tail of the list, because `head` only demands the first list constructor and ignores the second argument. In order to have an aptronym for the representation of partial values, we define a new data type `partial` that captures the notion that a value can be undefined or defined more clearly than the predefined type `option` like we have already seen for Haskell in Section 2.1.2.

```
Inductive partial (A : Type) :=
| undefined : partial A
| defined   : A -> partial A.
```

When using a data type like lists we need to adapt the definition of the data type itself by lifting all arguments to allow partial values a well.

```
Inductive List (A : Type) :=
| nil : List A
| cons : partial A -> partial (List A) -> List A.
```

The data type `List` represents a Haskell list in Coq as all constructor arguments may be partial. When defining functions on lists, we need to consider that the top–level list constructor may already be undefined. The definition of functions like `head` adapts to the settings as follows: all arguments as well as result types need to be lifted and the implementation needs to handle the partial values accordingly.

```
Definition head (A : Type) (pxs : partial (List A)) : partial A :=
  match pxs with
  | undefined  => undefined
  | defined xs => match xs with
                  | nil       => undefined
                  | cons px _ => px
                  end
  end.
```

We cannot pattern match on the list directly as in the original definition. In order to access the list, we need to take a look at the `partial`–layer first. In case of an `undefined` input list, the overall result is `undefined` as well. If the input list is `defined`, we can reimplement the original behaviour of the Haskell function, that is, yielding the first element in case of a non–empty list and yielding `undefined` in case of an empty list.

Let us take a look at another list function in Haskell and its counterpart in our model for Coq: concatenation of two lists.

```
(++) :: [a] -> [a] -> [a]
(++) []       ys = ys
(++) (x : xs) ys = x : (xs ++ ys)
```

The concatenation function on lists is a recursive and total function and can, thus, be translated to Coq as it is. However, since we want to mix partial and total functions and, thus, increase the compositionality of both possible effects, we translate the append function using the same scheme as above.

```
Fail Fixpoint append (A : Type) (pxs pys : partial (List A))
  : partial (List A) :=
  match pxs with
  | undefined  => undefined
  | defined xs => match xs with
                  | nil         => pys
                  | cons pz pzs => defined (cons pz (append pzs pys))
                  end
  end.
```

The function `append` is recursive and, as far as we are concerned, the first argument is structurally decreasing in each recursive call. However, Coq does not accept the above definition and rejects it with the following message.

*The command has indeed failed with message: Cannot guess decreasing argument of fix. Recursive call to append has principal argument equal to "pzs" instead of a subterm of "pxs".*

What is going on here? Coq cannot retrace that the value pzs is actually a subterm of pxs; without that connection the recursive call on pzs is not allowed. The problem is our lifted data type definition of List. Due to the lifting, the List becomes a so–called *nested inductive* type, because the type we want to define appears nested in another inductive type. In this case the other inductive type is partial.

We fix the problem by splitting the function into a non–recursive and a recursive part. The recursive part is working only on List A and produces a partial (List A), while the non–recursive function expects and yields a value of type partial (List A), and mainly calls the recursive function after unwrapping the partial–layer.

```
Definition append (A : Type) (pxs pys : partial (List A))
  : partial (List A) :=
  match pxs with
  | undefined => undefined
  | defined xs =>
    let fix append' xs pys :=
        match xs with
        | nil         => pys
        | cons pz pzs => defined (cons pz (match pzs with
                                           | undefined  => undefined
                                           | defined zs => append' zs pys
                                           end))
        end
    in append' xs pys
  end.
```

The recursive function is the local function append' that is introduced using `let fix`. In order to make the recursive call we need to unpack an additional partial–layer of the variable pzs. Since we are now reconstructing the nested inductive structure using a nested recursion, Coq accepts the definition.

In the remainder of this chapter we will define the two functions necessary to translate a recursive Haskell function on top–level to reuse the definitions in proofs. The recursive function that works directly on the nested inductive type will be suffixed like the local function definition append'. That is, the following code shows how the above code can be split into two top–level functions.

```
Fixpoint append' (A : Type) (xs : List A) (pys : partial (List A))
  : partial (List A) :=
  match xs with
  | nil         => pys
  | cons pz pzs => defined (cons pz (match pzs with
                                     | undefined  => undefined
                                     | defined zs => append' zs pys
                                     end))
  end.
```

```
Definition append (A : Type) (pxs pys : partial (List A))
  : partial (List A) :=
  match pxs with
  | undefined  => undefined
  | defined xs => append' xs pys
  end.
```

### 5.1.2 Generalisation Attempt

Up to now, we have seen an encoding of Haskell programs in Coq that lifts all Haskell values to `partial` values. The overall goal is to not only reason about Haskell's partiality effect but a whole class of effects. For example the effects we enumerated and discussed in Section 2.1.2: I/O, tracing, error, or non–determinism. That is, we do not want to restrict our model in Coq to one effect only, but generalise it to arbitrary effects.

The first idea that comes to mind to generalise the above encoding is to change the concrete type `partial` by adding a type parameter. Hence, we can regain the above encoding by instantiating the type parameter with `partial`. Consider the following parametrised definition of `List`.

```
Fail Inductive List (M : Type -> Type) (A : Type) :=
| nil : List M A
| cons : M A -> M (List M A) -> List M A.
```

Similar to `partial` that corresponds to the Maybe monad that we know from Haskell, we will be mostly interested in other monadic instantiations and follow the idea of monadic abstractions we discussed in Section 2.1.2. Thus, we name the type parameter `M` to indicate that it is a placeholder for a monadic type. This representation was already suggested by Abel et al. (2005), who translated Haskell code into Agda (Norell, 2009) code in order to prove propositions about Haskell functions. Their approach, however, is not applicable in recent versions of Agda nor Coq anymore. When we try to compile the above definition, we get the following error message.

> *The command has indeed failed with message: Non strictly positive occurrence of "List" in "M A → M (List M A) → List M A".*

Coq does not allow to define a type definition like `List` because of its *strict positivity requirement*. What is going on this time?

As a first step to understand what is going on here, we will reduce the above definition to the simplest definition that still triggers the same error message.

```
Fail Inductive NonStrictlyPos :=
| con : (NonStrictlyPos -> nat) -> NonStrictlyPos.
```

The data type `NonStrictlyPos` has no type parameters and only one unary constructor `con`. The argument of the constructor `con` is a function from `NonStrictlyPos` to `nat`. The domain of this function is the crucial part of the definition and the origin of the error message. As so often when using Coq, Chlipala (2013) is a great resource introducing typical obstacles beginners may face when using Coq for the first time. Chlipala makes the following comment concerning Coq's strict positivity requirement.

> *We have run afoul of the strict positivity requirement for inductive definitions, which says that the type being defined may not occur to the left of an arrow in the type of a constructor argument.*

Inductive type definitions obey the strict positivity requirement, if the recursive occurrences of the type only occur strictly positive in all argument types for all constructors. That is, in case of `NonStrictlyPos` we need to inspect the types of the arguments of its only constructor `con`. The only argument of `con` is the function type `NonStrictlyPos -> nat`.

---

**Excursus: Positive and Negative Occurrences**  According to the error message Coq produces when we try to compile the data type `List`, the type definition violates a strict positivity restriction; but what is a positive occurence? Taking the quote of Chlipala to a more formal ground, the strictly positive occurrence of a type is described as follows: a type $\tau$ occurs strictly positively in a type $\tau_1 \to \cdots \to \tau_n \to \tau$, if and only if $\tau$ does not occur in any of the types $\tau_i$ with $i$ ranging from $1$ to $n$. Although this description might answer the question we raised to some extent, it raises new ones as well: how do we distinguish between *positive* and *strictly positive* occurrences, and can an occurrence be *negative* as well?

For that matter, let us consider an arbitrary type $\tau = t_1 \to \cdots \to t_n$, where all $t_i$ are types as well. The type arguments $t_1$ to $t_n$ appear in positive or negative positions. A type argument $t_i$ has a negative position if it appears to the left of an odd number of type arrows. That is, starting from an inner (nested) position, each type arrow flips a negative position to a positive and vice versa. In particular, a type argument $t_i$ has a positive position if it appears to the left of an even number of type arrows or no type arrows at all. Moreover, a type $t$ occurs strictly positively in $t_1 \to \cdots \to t_n \to t$ if $t$ occurs in no $t_i$. For more information, Blanqui et al. (2002) give an inductive definition of the set of negative and positive positions.

Here, we consider some examples to clarify the definition of negative, positive and strictly positive positions. Consider the following examples of possible types in Coq, where `A` is a type variable.

(1) `A -> nat`

(2) `nat -> A`

(3) `(A -> nat) -> nat`

(4) `((A -> nat) -> nat) -> nat`

The type `A` occurs positively in (2) and (3), and negatively in (1) and (4). The occurrence in (2) is also strictly positive.

In dependently typed languages, the strict positivity requirement plays an important role for the definition of inductive data types. An inductive type `t` obeys the strict positivity requirement if the recursive occurrences of `t` are strictly positive in the types of all the arguments of its constructors.[1]

Consider once again the following exemplary type definitions in Coq.

(1) `Inductive T := con : T -> T -> T`

(2) `Inductive T := con : (T -> T) -> T`

---

[1] The definition by Blanqui et al. also takes care of mutual inductive types where the recursive occurrence is only implicit.

(3) `Inductive T := con : (nat -> T) -> T`

(4) `Inductive T := con : (nat -> T) -> (T -> T) -> T`

The inductive type `T` obeys the strict positivity requirement in (1) and (3), and violates it in (2) and (4). In (1) we need to check the following arguments of the constructor `con`: both arguments are of type `T` and both types are strictly positive. The argument in question for (3) is `nat -> T`, where `T` does not occur to the left of an arrow, thus, the type is strictly positive as well. In (2) we need to inspect the type `T -> T`; here, `T` occurs to the left of an arrow, thus, in a negative position. The overall definition, then, does not fulfil the requirement. The inductive type in (4) is a combination of (2) and (3), thus, the type of the first argument of `con` fulfils the requirement, whereas the second argument violates it again.

---

Let us think again about the definition of `NonStrictlyPos`. With the excursus above, we now know that `NonStrictlyPos` occurs "left of an arrow" in the first argument of the `con`–constructor, namely in the function type `NonStrictlyPos -> nat`. Due to this non strictly positive occurence, the overall data type definition of `NonStrictlyPos` is non strictly positive as well, that is, fails the strict positivity requirement for data type definitions. In contrast, the following definition is not problematic.

```
Inductive StrictlyPos :=
| con : StrictlyPos -> (nat -> StrictlyPos) -> StrictlyPos.
```

Here, both arguments of the constructor `con` are strictly positive: the first argument is of type `StrictlyPos`, thus, trivially strictly positive; and in the second argument's type `StrictlyPos` does not occur to the left, but to the right of an arrow, fulfilling the positivity requirement as well.

The above excursus explains the origin of the error message Coq presents us when we try to define data types like `NonStrictlyPos` and `List`, but two question still remain open.

1. Why is the strict positivity of data type definitions required in Coq?
2. Why is our definition `List` non strictly positive as well, although there is no function arrow in the argument types of `cons`?

**Why does Coq require the strict positivity of data type definitions?**

Let us for a moment assume that the definition of `NonStrictlyPos` was allowed. We then define the following function.[2]

```
Definition applyFun (t : NonStrictlyPos) : nat :=
  match t with
  | con f => f t
  end.
```

The function `applyFun` takes a value `t` of type `NonStrictlyPos` and applies the function `f` inside the argument of the `con`–constructor to the value `t` itself. A problematic example usage of this function is the expression `applyFun (con applyFun)`. Reducing the expression by using the definition of `applyFun` yields `applyFun (con applyFun)` again. We can apply this reduction infinitely often, thus, we have constructed an expression that does not terminate. Taking a step back, we observe that the data type `NonStrictlyPos` has some resemblance with a specialised instance of the fixpoint combinator `Mu`.

---

[2]We do not use code highlighting to distinguish the following hypothetical code from valid Coq code.

```
Fail Inductive Mu A :=
| mu : (Mu A -> A) -> Mu A.
```

Note that we cannot define the type `Mu` in Coq. The connection to the fix–point combinator explains why Coq needs to restrict such a definition: we know that all Coq programs need to terminate and we need to use a special keyword `Fixpoint` or `fix` to indicate recursive functions in order to apply termination checks with respect to recursion. If we could define a data type like `Mu`, our definition of `applyFun` is an example that shows how we can define non–terminating programs without using the explicit fix–point combinator that Coq already provides for recursive constructions. That is, data types like `Mu` introduce the capability to express general recursion and Coq's logic becomes inconsistent. In order to ensure the consistency of Coq's logic, the strict positivity requirement needs to be enforced for data type definitions.

**Why does `List` not fulfil the strict positivity requirement?**

Now we know what Coq is nagging about when the error message about a non strictly positive occurrence for data type definitions appears. However, it is still not obvious why the definition of `List` falls under the same category. Recall our attempt to define the data type `List`.

```
Fail Inductive List (M : Type -> Type) (A : Type) :=
| nil : List M A
| cons : M A -> M (List M A) -> List M A.
```

The type `List` does not appear to the left of an arrow in the argument types of its constructor `cons`, but Coq still rejects the definition because of the usage of the type `M (List M A)`. However, Coq is of course on the right track rejecting such a definition. While a concrete instantiation using `partial` instead of the type parameter `M` is accepted, since `List` does not appear to the left of an arrow, we cannot guarantee that all usages of `List` obey this requirement. For example, consider the following definition that we might use to instantiate the type parameter with.

```
Definition Cont R A := (A -> R) -> R.
```

The type `Cont` represents the continuation monad. Now let us instantiate the type parameter `M` in the definition of `List` with `Cont R` for an arbitrary `R`, a concrete monad. The following type definition `ListCont` inlines the definition of `Cont` in the constructor corresponding to `cons`.

```
Fail Inductive ListCont R A :=
| nilC  : ListCont R A
| consC : ((A -> R) -> R) -> ((ListCont R A -> R) -> R) -> ListCont R A.
```

Due to the inlining, it now becomes apparent that the type `ListCont` appears indeed to the left of an arrow in one of the type arguments of its constructor. More precisely, `ListCont` appears to the left of an arrow in the second argument of the `consC`–constructor.

Summarising our findings, the type definition `List` defined above allows arbitrary type constructors as instances for its type parameter `M`. Since the type parameter is arbitrary, it is not safe to use this definition for all potential instances of `M`. The strict positivity restriction might be violated for a concrete instation of `M`, for instance, for the concrete instantiation of `Cont` as we demonstrated in the definition of `ListCont`. Since we cannot guarantee by definition that the data type definition `List` is only used with instantiations of `M` that obey the strict positivity requirement, Coq rejects the definition.

### 5.1.3 Free Monad and Containers

Let us summarise the situation so far. We want to model effectful data types and functions, for example representing non–strict partial programs, in a generic way, but cannot use a type constructor variable to represent this generic effect in Coq due to the strict positivity requirement. If we use, however, a concrete effect, like `partial`, the strict positivity is guaranteed again. That is, we want to represent effects using a concrete data type representation in order to satisfy Coq's requirement. Recall that the effects we want to represent are all a superset of pure values. Partiality, for example, adds `undefined` to the set of `defined` values. The other mechanism the representation of effects needs to offer is a way to apply functions to pure values; in case of `partial` we have used pattern matching to unwrap the pure values in order the define `head` and `append`. Fortunately, monadic abstractions as discussed in Section 2.1.2 give us exactly these capabilities. As in Haskell, we define a monad as type constructor class parametrised over `M` that allows to define functions `return : A -> M A` and `bind : M A -> (A -> M B) -> M B`.

The main goal is to retain the generality to model a wide range of arbitrary monads, but to use a concrete data type in order to fulfil the strict positivity requirement. We can achieve this goal by using a data type that represents all strictly positive types in a constructive manner. Fortunately, Abbott et al. (2003) introduced the notion of containers to represent strictly positive types; their main insight is that all strictly positive types can be expressed using containers. A container is described as a product of shapes and a position function. The shape is a type `Sh` and the position type `Pos` is a type function that maps shapes to types. Recall that we defined a container extension that gives rise to a functor in Section 2.3.3 as follows.

```
Inductive Ext (Sh : Type) (Pos : Sh -> Type) A :=
| ext : ∀ (s : Sh), (Pos s -> A) -> Ext Pos A.
```

A container extension `Ext S P` is then isomorphic to a functor `F`. Using the following definition of `fmap` for `Ext`, we show in Appendix A.5 that `Ext` fulfils the functor laws.

```
Definition fmap (Shape : Type) (Pos : Shape -> Type) (A B : Type)
   (f : A -> B) (e: Ext Shape Pos A) : Ext Shape Pos B :=
   match e with
   | ext s pf => ext s (fun p => f (pf p))
   end.
```

In Section 2.3.3 we defined an exemplary concrete instance of a container for the data type `One` and proved both structures are isomorphic, which we will recap shortly. Since we are mainly interested in monads, we can go even further and use monadic containers as proposed by Altenkirch and Pinyo (2017) and Uustalu and Veltri (2017). However, we decided against using monadic containers directly, as they, as far as we know, cannot be implemented in a constructive way, but are an extension of Ext that are modelled using a type class.

Since we know that all monads have a constructor that represents *pure* values, the effect we want to model is everything that we gain on top of these defined values. We define this property more explicitly using a free monad (Swierstra, 2008) as introduced in Section 2.1.3: we either have a pure or an effectful value.

```
Fail Inductive Free F A :=
| pure   : A -> Free F A
| impure : F (Free F A) -> Free F A.
```

As the effect is once again represented by a type constructor variable F that has the data type to be defined as argument, Coq does not accept the definition. Fortunately, the type parameter F needs to be a functor in order to make a Free F a monad. Or to it put differently, given a functor F the free monad construction lifts F to a monad. Now we can apply the insight about representing strictly positive types as containers again: since container extensions are isomorphic to functors, there are a variety of functors that we can represent using containers. We can change the definition of Free as follows to fulfil Coq's strict positivity requirement.

```
Inductive Free (Sh : Type) (Pos : Sh -> Type) A :=
| pure   : A -> Free Sh Pos A
| impure : Ext Sh Pos (Free Sh Pos A) -> Free Sh Pos A.
```

The definition uses the container extension instead of a generic functor F, that is, we are still generic: not over all possible type constructors but over all strictly positive types. Note, however, that we do not need the indirection using Ext explicitly, we instead define Free by inlining the definition of Ext in the constructor impure to make its usage more convenient.

```
Inductive Free (Sh : Type) (Pos : Sh -> Type) A :=
| pure   : A -> Free Sh Pos A
| impure : ∀ s, (Pos s -> Free Sh Pos A) -> Free Sh Pos A.
```

With this definition of Free at hand, we are well–suited to define a lifted version of List as we wished to do at the beginning of this section.

```
Inductive List (Sh : Type) (Pos : Sh -> Type) A :=
| nil : List Sh Pos A
| cons : Free Sh Pos A -> Free Sh Pos (List Sh Pos A) -> List Sh Pos A.
```

We also define smart constructors Nil and Cons as notations that wrap the list in the additional pure–constructor to indicate that the top–level expression is defined.

```
Notation Nil := (pure nil).
Notation Cons fx fxs := (pure (cons fx fxs)).
```

**Representing Partiality using Free Monads and Containers**

In Section 2.3.3 we already defined the container extension for One as Ext $One_S$ $One_P$ and proved that both these structures form an isomorphism using the functions from_One and to_One. The definitions looked as follows.

```
Definition One_S := unit.
Definition One_P (s : One_S) := Empty.

Definition from_One A (o : One A) : Ext One_S One_P A :=
  ext tt (fun (p : One_P tt) => match p with end).

Definition to_One A (e : Ext One_S One_P A) : One A :=
  one.
```

Next up, we can also show that the resulting construction `Free One_S One_P` is isomorphic to the monadic structure `partial`. First, we define the conversion functions `to_partial` and `from_partial` analogous to the setup for `One` above.

```
Definition to_partial A (fx : Free One_S One_P A) : partial A :=
 match fx with
 | pure x     => defined x
 | impure _ _ => undefined
 end.

Definition from_partial A (p : partial A) : Free One_S One_P A :=
  match p with
  | undefined => let '(ext s pf) := from_One one in impure s pf
  | defined x => pure x
  end.
```

Based on these conversion functions, we prove two lemmas that state that both possible compositions of these functions are the identity.

```
Lemma from_to_partial : ∀ (A : Type) (fx : Free One_S One_P A),
    from_partial (to_partial fx) = fx.
Proof.
  intros A fx. destruct fx as [x | [] pf]; simpl.
  - reflexivity.
  - do 2 f_equal. extensionality p. destruct p.
Qed.

Lemma to_from_partial : ∀ (A : Type) (p : partial   A),
    to_partial (from_partial p) = p.
Proof.
  intros A p. destruct p; reflexivity.
Qed.
```

**Representing Totality using Free Monads and Containers**

In order to see a second effect, we find some inspiration by Abel et al. (2005) who are interested in two effects: partiality and totality. Since we have already discussed partiality, we will now take a look at how to model totality. Of course, totality is, strictly speaking, not an effect, but describes the absence of any additional effects, we have pure values only. A suitable monad to represent totality is the identity monad; in order to match our naming scheme used for partiality, we name the data type `total`.

```
Inductive total (A : Type) :=
| totality : A -> total A.
```

When using a free monad to represent `total`, we only need to give a functor for the additional effect on top of pure values. Since totality does not add any effects and describes pure values only, we need to model this absence of effects. In terms of defining values using `Free F` that represent total values, we only want to be able to use the `pure`–constructor. The `impure`–constructor, on the other hand, should not be available, it should be impossible to construct a value using `impure`. That is, we need a functor without any constructors. When an inductive type has no constructors, there is no way to construct a value of that type. Once again, we use the same naming scheme as Swierstra (2008) and define the type constructor `Zero` that has no values.

```
Inductive Zero (A : Type) := .
```

Note that, in contrast to the `Empty` type, `Zero` is a type constructor as it has an additional type parameter `A`. Similar to the definition of `One`, we do not use this type parameter.

In order to represent `Zero` as container, we once again need to define the corresponding shape and position type function. Since `Zero` has no constructors, it has no shapes and no polymorphic values that we might want to access. For both types we can reuse `Empty`.

```
Definition Zero_S := Empty.
Definition Zero_P (s : Zero_S) := Empty.
```

For both conversion functions, `from_Zero` and `to_Zero`, the definition simply matches on the non–existent value to define a function with the wanted type.

```
Definition from_Zero A (u : Zero A) : Ext Zero_S Zero_P A :=
  match u with end.

Definition to_Zero A (e : Ext Zero_S Zero_P A) : Zero A :=
  let '(ext s pf) := e in match s with end.
```

It is quite trivial to prove that the conversion functions form an isomorphism as the following reasoning shows.

```
Lemma from_to_Zero : ∀ (A : Type) (e : Ext Zero_S Zero_P A),
    from_Zero (to_Zero e) = e.
Proof.
  intros A e. destruct e as [[] pf].
Qed.

Lemma to_from_Zero : ∀ (A : Type) (z : Zero A),
    to_Zero (from_Zero z) = z.
Proof.
  intros A z. destruct z.
Qed.
```

The next step is to define the conversion functions `to_total` and `from_total` to show that using the container representation for `Zero` in combination with `Free` yields the original monad `total` again.

```
Definition to_total A (fx : Free ZeroS ZeroP A) : total A :=
  match fx with
  | pure x     => totality x
  | impure s _ => match s with end
  end.

Definition from_total A (t : total A) : Free ZeroS ZeroP A :=
  match t with
  | totality x => pure x
  end.
```

In fact, similar as for the case of the partiality monad, we can even show that the conversion functions form an isomorphism as well.

```
Lemma from_to_total : ∀ (A : Type) (fx : Free ZeroS ZeroP A),
    from_total (to_total fx) = fx.
Proof.
  intros A fx. destruct fx as [x | [] pf]; reflexivity.
Qed.

Lemma to_from_total : ∀ (A : Type) (t : total A),
    to_total (from_total t) = t.
Proof.
  intros A t. destruct t; reflexivity.
Qed.
```

**Other Containers and Limitations**

We will later see that monads like tree and list, which are commonly used when modelling non–determinism, can be represented using our approach as well. There are also common monads from Haskell that have corresponding effects that are interesting to reason about. For example, the writer and state monad that are used to model tracing and I/O–interactions with the user, respectively. We can model these effects using `Free` and a suitable container representation of the underlying effect.

Since we use containers to ensure Coq that we do not define a potentially non–strictly positive type, the class of strictly positive types is the natural limitation when using our approach. An example for a monad that we cannot represent with our approach is the continuation monad. The continuation monad `Cont` was an exemplary instantiation of the type constructor variable `M`, such that our initial try to define `List` became non–strictly positive. The goal was to convince Coq that we won't use an instantiation like `Cont`, so it makes sense that we indeed cannot define a data type that is isomorphic to `Cont`.

In the end, we have to keep in mind that our approach using `Free` as representative for a generic monadic parameterisation can only represent monads that correspond to strictly positive types.

### 5.1.4 Working with Free to Define Lifted Functions

With the definition of `List` using `Free` to model generic effectful lists at hand, we can take a look at the definition of the function `append` again. Analogous to the definition of append at the end of Section 5.1.1, we need to split the function into two parts: one part works on pure `List`s and takes care of pattern matching while the second part handles `Free`–lifted arguments only.

For the remainder of this chapter, we will use `Free` and data types like `List` as type constructors of kind `Type -> Type` when used with a generic container, that is, we do not pass the corresponding shape and position function as argument. Note, however, that this convention is a simplification for better readability only and the resulting definition would not compile in Coq.

In order to compare the differences of using `Free` to our concrete representation using `partial`, recall the definition of `append` and `append'`, respectively.

```
Fixpoint appendP' (A : Type) (xs : List A) (pys : partial (List A))
  : partial (List A) :=
  match xs with
  | nil        => pys
  | cons pz pzs => Cons pz (match pzs with
                            | undefined  => undefined
                            | defined zs => appendP' zs pys
                            end)
  end.

Definition appendP (A : Type) (pxs pys : partial (List A))
  : partial (List A) :=
  match pxs with
  | undefined  => undefined
  | defined xs => appendP' xs pys
  end.
```

For better readability, we added the suffix `P` for this version of `append` that uses `partial`. Let us now try to transfer the general idea of `appendP'` to define a version using `Free`.

```
Fail Fixpoint append' A (xs : List A) (fys : Free (List A)) :=
 match xs with
 | nil         => fys
 | cons fz fzs => pure (cons fz (match fzs with
                                 | pure zs    => append' zs fys
                                 | impure s pf => _ (* what to do here? *)
                                 end))
 end.
```

A case distinction on the concrete constructors of `partial` was quite simple, in the case of `Free` and its representation using `Ext`, however, the second constructor `impure` has recursive occurrences of values of type `Free`. That is, a case distinction like in the definition of `appendP'` needs to be defined as a recursive function when using `Free`.

Recall that the basic idea is to model monadically lifted data types, that is, `Free` is a representative for a monad. The case distinction we are using in the definition of `appendP'` is the monadic *bind* function (`(>>=)`) in disguise. Hence, we need to define bind for `Free` in order to transfer the above definition successfully.

```
Fixpoint free_bind A B (fx : Free A) (f : A -> Free B) : Free B :=
 match fx with
 | pure x      => f x
 | impure s pf => impure s (fun p => free_bind (pf p) f)
 end.
```

The definition of `free_bind` distinguishes between the two possible constructors `pure` and `impure`; in the former case we have a value `x` in place that `f` can be applied to. In the latter case we have variable bindings `s` and `pf` for the shape and position function to work with. Note that the position function applied to a position yields a `Free`–value again, that is, `pf` has type `Pos s -> Free A`. The trick is that we just need to reconstruct the effect that we have seen, characterised by its shape `s`, and apply the function `f` recursively for all recursive occurrences of the effect that we can access using the position function `pf`.

With the definition of `free_bind` at hand, we can go ahead and start the second try to define `append'`: now we know that we need to use `free_bind` instead of an explicit case distinction. More precisely, the usage of `free_bind` mimics the evaluation to head normal form, which is exactly how pattern matching in Haskell behaves when using a case distinction.

```
Fail Fixpoint append' A (xs : List A) (fys : Free (List A)) :=
  match xs with
  | nil         => fys
  | cons fz fzs => Cons fz (free_bind fzs (fun zs => append' zs fys))
  end.
```

Unfortunately, the definition of `append'` is not accepted by Coq, because it cannot guess the decreasing argument of the provided fixpoint function. More precisely, Coq gives the following error message when we provide the information that the list `xs` is supposed to be the decreasing argument.

> *Recursive call to append' has principal argument equal to "zs" instead of "fzs".*

What is the problem now?

---

**Excursus: Definition of recursive, higher–order functions**   Once again Coq is not content with the way we define our programs. In this case, the fault is not with `append`, but with the definition of `free_bind`. As a recursive, higher–order function that uses the same function as its functional argument for each recursive call, we need to convince Coq that the function indeed stays the same. That way, Coq comprehends that the overall function terminates, if it does not depend on the function we pass as higher–order argument. That is, the problem is not specific to `free_bind` and our usage of `Free`, but a general problem for recursive, higher–order functions. Consider, for example, the definition of `map` for ordinary lists. Note that the constructors `nil` and `cons` used in this excursus are exclusively used as names for the constructors of the predefined `list` type.

```
Fixpoint map A B (f : A -> B) (xs : list A) : list B :=
 match xs with
 | nil       => nil
 | cons x xs => cons (f x) (map f xs)
 end.
```

Next, we give a definition for rose trees, consisting of a `leaf`–and `branches`–constructor. The `branches`–constructor uses a list to allow for arbitrary branching of the tree.

```
Inductive rose A :=
| leaf     : A -> rose A
| branches : list (rose A) -> rose A.
```

When we then try to define a map–like function for `rose` as well, we might be tempted to just reuse the `map` function for lists to handle the `branches`–constructor.

```
Fail Fixpoint mapRose A B (f : A -> B) (r : rose A) : rose B :=
 match r with
 | leaf x      => leaf (f x)
 | branches rs => branches (map (fun x => mapRose f x) rs)
 end.
```

However, Coq once again does not recognise that this definition of `mapRose` terminates. It raises the following error message.

> *Recursive call to mapRose has principal argument equal to "x" instead of "rs".*

Due to the usage of `map`, Coq's termination checker fails to recognise that the recursive call of `mapRose` will eventually terminate. Fortunately, there is a simple scheme to convince Coq that the function `map` is inductively recursive over its list argument and, more importantly, the supplied function does not determine the termination. That is, `map` terminates independently of its functional argument. Consider the following alternative definition of `map`.

```
Definition map A B (f : A -> B) (xs : list A) : list B :=
 let fix map' xs :=
     match xs with
     | nil       => nil
     | cons x xs => cons (f x) (map' xs)
     end
 in map' xs.
```

Instead of declaring a recursive function on top–level using the keyword `Fixpoint`, we declare a local fixpoint only. In this case, the local recursive function `map'` has only one parameter, namely the list argument, and reuses the function `f` that is passed to the top–level function `map`. Due to the explicit fixpoint definition that does not have the higher–order function as argument anymore, Coq's termination checker can realise that the termination of `map` does not depend on the function `f`.

An alternative, but equivalent definition, uses section variables to introduce the type parameters and the higher–order argument.

```
Section map.

 Variable A B : Type.
 Variable f : A -> B.

 Fixpoint map (xs : list A) : list B :=
```

```
    match xs with
    | nil       => nil
    | cons x xs => cons (f x) (map xs)
    end.

End map.
```

We can declare variables with their associated types for usage within a section. Outside of the section, the function `map` gets all arguments used in its definition as further arguments. Moreover, the code that Coq generates for the second version with section variables is exactly the first implementation given above.

Independent of which of the two implementations we chose, we can now use `map` to define `mapRose` and Coq accepts the definition as terminating.

```
Fixpoint mapRose A B (f : A -> B) (r : rose A) : rose B :=
 match r with
 | leaf x      => leaf (f x)
 | branches rs => branches (map (mapRose f) rs)
 end.
```

---

Now let us go back to our initial problem with the definition of `append'` using the higher–order, polymorphic auxiliary function `free_bind`. The problem that arose for the definition of `append'` using `free_bind` is analogous to the definition of `roseMap` using `map`. That is, we need an alternative implementation of `free_bind` that follows the scheme above: the recursion of `Free` needs to be explicitly independent of the functional argument. We then end up with the following definition of `free_bind`.

```
Definition free_bind A B (fx : Free A) (f : A -> Free B) : Free B :=
 let fix free_bind' fx :=
     match fx with
     | pure x       => f x
     | impure s pf  => impure s (fun p => free_bind' (pf p))
     end
 in free_bind' fx.
```

In order to use `free_bind` as an operator (>>=) as known from Haskell, we additionally define the following notation.

```
Notation "fx >>= f" := (free_bind fx f) (at level 40, left associativity).
```

Last but not least, we are finally equipped to define the auxiliary function `append'` as well as the final `append`–function that is the generically lifted version of the corresponding Haskell function.

```
Fixpoint append' A (xs : List A) (fys : Free (List A)) :=
  match xs with
  | nil => fys
  | cons fz fzs => pure (cons fz (fzs >>= fun zs => append' zs fys))
  end.
```

```
Definition append A (fxs fys : Free (List A)) : Free (List A) :=
  fxs >>= fun xs => append' xs fys.
```

## 5.2 Proving Properties About Haskell Programs

While we discussed the overall infrastructure and necessary preliminaries for a general framework to model non–strict effectful programs, we will focus on two effects occurring in Haskell: totality and partiality. Both instantiations are of interest to motivate our effect–generic reasoning. That is, we follow the idea of Abel et al. (2005) and prove general propositions about Haskell's concatenation functions in three steps: first, we show that the proposition holds in the total setting and then show that it also holds in the partial setting. On top of these both proofs, the last step makes the case for our generic approach: we prove the property once and for all for a whole class of effects.

### 5.2.1 Three Proofs for the Associativity of Append

Since it took us a whole section to define the function append for an effect–generic List data type, it seems reasonable to prove a property for that function. Thus, in this subsection we will focus on the associativity of append and prove this proposition in the total, partial and generic setting. Since we encounter proofs for lifted data types and functions for the first time, we will present the proofs step–by–step. We will, however, postpone technical details to the end of this subsection. Thus, we explain the details of the proof from a more abstract perspective rather than diving deep into the technical details of the applied tactics.

**Totality**

In order to refresh the memory, the definition of the container to represent total programs shown in Section 5.1.3 looks as follows.

```
Definition Zero_S := Empty.
Definition Zero_P (s : Zero_S) := Empty.
```

That is, the container representing totality has neither a position nor a shape, thus, programs modelled as `Free Zero_S Zero_P` do not have any impure values by construction. Whenever we encounter an impure–constructor, we know that the accompanying shape `s : Zero_S` cannot exist. Let us now take a look at the associativity of append in the total setting and see how we use this observation about the corresponding container when proving a proposition.

```
Lemma append_assoc_total :
 ∀ (A : Type) (fxs fys fzs : Free Zero_S Zero_P (List Zero_S Zero_P A)),
   append fxs (append fys fzs) = append (append fxs fys) fzs.
Proof.
 intros A fxs fys fzs.
 destruct fxs as [ xs | s pf ]; simpl.
```

In order to evaluate append, we make a case distinction on the first list argument `fxs`. In case of a pure value, we have the following goal and assumptions.

```
A : Type
xs : List Zero_S Zero_P A
fys, fzs : Free Zero_S Zero_P (List Zero_S Zero_P A)
============================
append' xs (append fys fzs) = append (append' xs fys) fzs
```

Now we can proceed as if we were working with ordinary lists: we proceed by induction on the `List` argument `xs`.

```
induction xs as [ | fx fxs IH ]; simpl.
```

```
            ...
            fys, fzs : Free Zero_S Zero_P (List Zero_S Zero_P A)
            ============================
            append fys fzs = append fys fzs

            ...
            IH : ForFree (fun xs => append' xs (append fys fzs)
                                  = append (append' xs fys) fzs) fxs
            ============================
            Cons fx (fxs >>= (fun zs => append' zs (append fys fzs))) =
            Cons fx (fxs >>= append (fun zs => append' zs fys) fys)
```

We then need to prove the proposition for `nil` and `cons`. The former is trivial as both sides of the terms are already equal. For the latter case, we proceed by stripping away the `Cons`–constructor and take another look at `fxs` by destructing the induction hypothesis `IH`, which produces two cases: one for `pure` and one for `impure`. Note that we will discuss the induction hypotheses `IH` that makes use of a custom proposition `ForFree` in detail later.

```
do 2 f_equal. destruct IH as [ xs H | ]; simpl.
```

```
            ...
            H : append' xs (append fys fzs) = append (append' xs fys) fzs
            ============================
            append' xs (append fys fzs) = append (append' xs fys) fzs

            ...
            s : Zero_S
            ============================
            ...
```

In the former case, the induction hypothesis brings our current goal in the right form such that we can finish the subproof easily by rewriting the hypothesis `H` and applying **reflexivity**. For the latter case we only need to focus on one of the assumptions, namely, `s : Zero_S`. Since there cannot exist any value of type `Zero_S`, the subgoal is trivially true as one of the assumptions is a contradiction. After destructing on the value `s`, there are no more subgoals and we have successfully proven the current goal.

The same reasoning applies for the last remaining goal. We started with a case distinction on the original first list arguments `fxs` and are now in the `impure`–case.

```
    ...
    s : Zero_S
```

```
============================
impure s (fun p : Zero_P s => append (pf p) (append fys fzs)) =
impure s (fun p : Zero_P s => append (append (pf p) fys) fzs)
```

The assumption s : Zero_S is once again a bogus assumption, so we clear the last remaining goal by using **destruct** s again.

Now that we have successfully proven this statement, we will take a second look at the complete proof script, without the additional remarks and assumptions. We improved the scripts a little bit and moved the **destruct** tactic directly after destructing the Free–values, to finish the second subgoal immediately. Alternatively, we could use a nested introduction pattern for s as well and destruct it directly using the pattern []. Then, the additional cases for impure would not be generated at all.

```
Lemma append_assoc_total :
 ∀ (A : Type) (fxs fys fzs : Free Zero_S Zero_P (List Zero_S Zero_P A)),
   append fxs (append fys fzs) = append (append fxs fys) fzs.
Proof.
 intros A fxs fys fzs.
 destruct fxs as [ xs | s pf ]; try destruct s; simpl.
 - induction xs as [ | fx fxs IH ]; simpl.
   + reflexivity.
   + do 2 f_equal. destruct IH as [ xs H | s pf ]; try destruct s; simpl;
     rewrite H. reflexivity.
Qed.
```

The main difference between the above proof and a proof for associativity of the append function for ordinary lists is the unboxing of the additional Free–layer of the list argument and tweaking the induction hypothesis to fit our needs.

**Partiality**

As next step, we consider an actual effect: partiality. Again, we recall the definition of the container to represent partiality when using Free as introduced in Section 5.1.3.

```
Definition One_S := unit.
Definition One_P (s : One_S) := Empty.
```

This time, the corresponding container has one possible shape, but no positions. The additional effect of representing undefined values does not contain any recursive occurence of the Free values, thus, the effect does not have any polymorphic components in its constructors.

Let us take a look at the proof for the associativity of append; we focus on the differences, as most of the proof is actually the same. Hence, this time we show the entire proof script first and take a closer look at some subgoals afterwards.

```
Lemma append_assoc_partial :
 ∀ (A : Type) (fxs fys fzs : Free One_S One_P (List One_S One_P A)),
   append fxs (append fys fzs) = append (append fxs fys) fzs.
Proof.
 intros A fxs fys fzs.
```

```
  destruct fxs as [ xs | s pf ]; simpl.
  - (* fxs = pure xs *) induction xs as [ | fx fxs IH ]; simpl.
    + (* xs = nil *) reflexivity.
    + (* xs = cons fx fxs; induction hypothesis IH *)
      do 2 f_equal. destruct IH as [ xs H | s' pf' ]; simpl.
      * (* fxs = pure xs *) rewrite H. reflexivity.
      * (* fxs = impure s' pf' *)
        do 2 f_equal. extensionality p. destruct p.
  - (* fxs = impure s pf *)
    do 2 f_equal. extensionality p. destruct p
Qed.
```

We observe that for both cases of impure–values, we do not have a contradiction in place as we had in the total setting. The contradiction in the total setting was introduced by the shape of the container that did not have any constructors. Now, we are working with $One_S$, which is a type synonym for the `unit` type and has exactly one constructor, namely `tt`. We can, however, prove the impure–cases using **destruct** again, but we need to have a position as an assumption. The position is introduced as assumption when we strip away the prefix of `impure s' pf'` on both sides of the equation such that only the position function `pf` remain to be proven equivalent.

The initial situation for the first impure–case looks as follows.

```
...
s' : One_S
===========================
impure s' (fun p => ...) =
impure s' (fun p => ...)
```

As mentioned above, after stripping away the prefix both expressions have in common, we end up with two functions that we need to prove equal. We introduce the function argument `p` using `extensionality p` and end up with `p : One_P s'` added to the assumptions and a changed goal as follows.

```
s' : One_S
p  : One_P s'
===========================
pf' p >>= (fun zs => append' zs (append fys fzs)) =
pf' p >>= (fun zs => append (append' zs fys) fzs)
```

Since $One_P$ `s'` is a type synonym for `Empty`, the position `p` has no constructors and we can finish the proof using **destruct** `p`, again. We follow the same approach to prove the second impure–case.

**Generic Effect**

Last but not least, we take a look at a generic proof for the associativity of append. Instead of introducing components of a concrete container, the shape `Sh` and position function `Pos` are universally quantified.

```
Lemma append_assoc_generic :
  ∀ (Sh : Type) (Pos : Sh -> Type) (A : Type)
    (fxs fys fzs : Free Sh Pos (List Sh Pos A)),
    append fxs (append fys fzs) = append (append fxs fys) fzs.
Proof.
intros Sh Pos A fxs fys fzs.
```

The main difference to the concrete proofs above is that in case of a generic effect, we cannot hope to have a false assumption, like a value of type `Empty`, in place to easily prove the `impure`–cases. Considering a generic effect has the consequence that we need to consider all possible instantiations. In particular, using the effect in combination with `Free` might lead to a recursive occurrence, that is, the effect might have a polymorphic argument.

By means of proving the proposition, this observation leads to the consequence that a simple case distinction over the `Free`–value `fxs` is not enough, we need to proceed by induction over `fxs`.

```
induction fxs as [ xs | s pf IH ]; simpl.
- (* fxs = pure xs *) induction xs as [ | fx fxs IH ]; simpl.
  + (* xs = nil *) reflexivity.
  + (* xs = cons fx fxs *) do 2 f_equal.
    induction IH as [ xs H | s pf _ IH' ]; simpl.
    * (* fxs = pure xs *) rewrite H. reflexivity.
    * (* fxs = impure s pf *) do 2 f_equal. extensionality p.
      apply IH'.
```

The `pure`–case stays basically the same, the difference, once again, lies in the nested `impure`–case.

```
...
s : Sh
IH' : ∀ p : Pos s,
      pf p >>= (fun zs => append' zs (append fys fzs)) =
      pf p >>= (fun zs => append' zs fys) >>= (fun zs => append' zs fzs)
p : Pos s
============================
pf p >>= (fun zs => append' zs (append fys fzs)) =
pf p >>= (fun zs => append' zs fys) >>= (fun zs => append' zs fzs)
```

Here, we cannot prove the subgoal by contradiction but need to actually use the induction hypothesis `IH'` generated by `induction IH`.

A similar situation appears for the second `impure`–case that originated from the initial induction on `fxs`.

```
IH : ∀ p : Pos s,
     append (pf p) (append fys fzs) = append (append (pf p) fys) fzs
============================
impure s (fun p : Pos s => append (pf p) (append fys fzs)) =
impure s (fun p : Pos s => append (append (pf p) fys) fzs)
```

In this case, the induction hypothesis is directly applicable when proving the functions equal. This last subgoal then finishes the proof.

```
  - (* fxs = impure s pf *) do 2 f_equal. extensionality p.
    apply IH.
Qed.
```

**The Generic Proof Owns them All**

Here comes the best part about the generic proof: the generic proof is applicable to the total and partial setting as well. That is, instead of proving the proposition for all concrete effects again and again, as necessary in the approach introduced by Abel et al. (2005), we prove the proposition once and for all.

Of course, if we want to prove a property in a concrete setting, we can still reuse the generic propositions as the following alternative proofs for the concrete settings of totality and partiality illustrate.

```
Lemma append_assoc_total' :
 ∀ (A : Type) (fxs fys fzs : Free Zero_S Zero_P (List Zero_S Zero_P A)),
   append fxs (append fys fzs) = append (append fxs fys) fzs.
Proof.
 apply append_assoc_generic.
Qed.

Lemma append_assoc_partial' :
 ∀ (A : Type) (fxs fys fzs : Free One_S One_P (List One_S One_P A)),
   append fxs (append fys fzs) = append (append fxs fys) fzs.
Proof.
 apply append_assoc_generic.
Qed.
```

### 5.2.2 Induction Principle for Free and List

Recall that the `induction` tactic we use in proofs makes use of the associated induction principle for the corresponding type. We discussed that an induction principle is an ordinary function of type `Prop` using the example of Coq's predefined list data type in Section 2.3.2. Since our proofs depend on induction on values of type `Free`, we now consider the corresponding induction principle that Coq generates.

```
Free_ind : ∀ (Sh : Type) (Pos : Sh -> Type) (A : Type) (P : Free A -> Prop),
   (∀ (x : A), P (pure x)) ->
   (∀ (s : Sh) (pf : Pos s -> Free A),
     (∀ p : Pos s, P (pf p)) -> P (impure s pf)) ->
   ∀ (fx : Free A), P fx
```

For the base case we need to show that a given predicate `P` holds for `pure x` for all `x` of appropriate type. The `impure` case is a bit more interesting: given a shape `s` and a position function `pf`, we need to prove that the predicate holds for `impure s pf` with an additional induction hypothesis in place. The hypothesis states that, for all possible positions `p`, the predicate already holds for all recursive values that we can access using the position function `pf`. Note that these arguments correspond to the variables introduced in the patterns when using the `induction` tactic. For example, the code above used the tactic using an introduction pattern as follows.

```
induction fxs as [ xs | s pf IH ].
```

The variable `xs` in the left branch of the introduction pattern corresponds to the pure value, while `s pf` are the arguments of the `impure` constructor and `IH` is the generated induction hypothesis.

In case of data types that contain arguments of type `Free` due to the lifting of the constructors, the induction principle generated by Coq is not strong enough to fulfil its purpose. Consider, for example, the induction principle Coq generates for the `List` data type defined in Section 5.1.3.

```
List_ind : ∀ (Sh : Type) (Pos : Sh -> Type) (A : Type)
       (P : List A -> Prop),
       P nil ->
       (∀ (fx : Free A) (fxs : Free (List A)), P (cons fx fxs)) ->
       ∀ l : List Pos A, P l
```

While the base case for `nil` looks fine, the generated function for the `cons` constructor looks a bit odd. More precisely, the second function misses an induction hypothesis for the list `fxs`. Without such a hypothesis at hand, there are not many propositions that we would be able to prove. Fortunately, we can define a custom induction principle for `List` and other data types that we might define. Since an induction principle is an ordinary proposition, that is, a function of type `Prop`, we just need to implement a function of the right type. By enabling an option,[3] we deactivate the automatic generation and define the custom induction principle named `List_ind`. That way, we can use **induction** as we are used to.

First of all, we declare the types and the functions we want to use in the induction principle. The keyword **Hypothesis** is commonly used for proposition functions in custom induction principles instead of using the keyword **Variable**; both keywords fulfil the same purpose though.

```
Variable A : Type.
Variable P : List A -> Prop.

Hypothesis nilP : P nil.
Hypothesis consP : ∀ fx fxs, ?IH -> P (cons fx fxs).
```

In order to specify the missing induction hypothesis in case of `cons` (that we currently named `?IH`), we need to work out how this hypothesis looks like in the lifted case. When we consider ordinary lists, we require that the predicate holds for a list `fxs` in order to prove that it also holds for the list `cons fx fxs` for an arbitrary `fx` of appropriate type as well. In case of a lifted list, the predicate is supposed to hold for all pure and effectful lists. The needed property is similar to stating that a predicate of type `A -> Prop` should hold for all elements of a list–like structure. In our case the list–like structure is the `Free` wrapper with its two constructors `pure` and `impure`. We define an auxiliary proposition `ForFree` that states if a predicate of type `A -> Prop` holds for all elements occurring in a value wrapped using `Free`. Remember that propositions in Coq are data type declarations that yield a value of type `Prop` instead of `Type`.

---

[3] `Unset Elimination Schemes.`

```
Inductive ForFree (P : A -> Prop) : Free A -> Prop :=
| forPure   : ∀ x, P x -> ForFree P (pure x)
| forImpure : ∀ s pf, (∀ p, ForFree P (pf p)) -> ForFree P (impure s pf).
```

In case of `pure x` we can directly apply the predicate to the element `x`. The `impure` constructor, on the other hand, embeds values of type `Free A`. Thus, we can apply the proposition `ForFree` recursively. We access the values of type `Free A` by applying the position function to all valid positions with respect to the used container. That is, the predicate holds for `impure s pf` if `ForFree P` already holds for all elements that can be accessed using the position function.

Now we use the proposition `ForFree` to redeclare the necessary induction hypothesis in case of the `cons` constructor.

```
Hypothesis consP : ∀ fx fxs, ForFree P fxs -> P (cons fx fxs).
```

Note that the predicate `P` is of type `List A -> Prop`, that is, it is a predicate on lists. Based on these hypotheses, we define our custom induction principle by implementing a function with the following type signature.

```
Fixpoint List_ind (xs : List A) : P xs.
```

The above declaration corresponds to a **Lemma**–statement, that is, we can define the function like proofs using tactics. Due to the declared hypothesis, the goal and associated assumptions look as follows.

```
A : Type
P : List A -> Prop
nilP : P nil
consP : ∀ fx fxs, ForFree P fxs -> P (cons fx fxs)
List_ind : ∀ xs : List A, P xs
xs : List A
============================
P xs
```

Since the function is recursive, `List_ind` itself appears as an assumption as well.

In order to show that the predicate `P` holds for the list `xs`, we make a case distinction that leads to the following two subgoals.

```
destruct xs as [ | fy fys ].
```

```
                    ...
                    nilP : P nil
                    ============================
                    P nil

                    ...
                    consP : ∀ fx fxs, ForFree P fxs -> P (cons fx fxs)
                    List_ind : ∀ xs : List A, P xs
                    fy : Free A
                    fys : Free  (List A)
                    ============================
                    P (cons fy fys)
```

We finish the first subgoal by applying the hypothesis `nilP`; for the second goal we proceed by applying the hypothesis `consP`. The new subgoal is the precondition of the hypothesis `consP` instantiated with `fy` and `fys` for the quantified variables `fx` and `fxs`, namely `ForFree P fys`. We proceed by induction on `fys` since the proposition `ForFree` is an inductively defined proposition with constructors in case of `pure` and `impure` values.

```
induction fys as [ ys | s pf IH ].     ...
                                       List_ind : ∀ xs : List A, P xs
                                       ys : List A
                                       ============================
                                       ForFree P (pure ys)

                                       ...
                                       IH : ∀ (p : Pos s), ForFree P (pf p)
                                       ============================
                                       ForFree P (impure s pf)
```

In both cases we apply the corresponding constructor of the proposition `ForFree`, namely, `forPure` and `forImpure`.

```
...
List_ind : ∀ xs : List A, P xs
============================
P ys

...
IH : ∀ (p : Pos s), ForFree P (pf p)
============================
∀ p : Pos s, ForFree P (pf p)
```

We finish both subgoals by simply applying the appropriate assumption. In the first case we apply the recursive function `List_ind` that we are currently defining. For the `impure`–case we use the induction hypothesis generated for `Free`. The complete proof script is finalised by using `Defined` and looks as follows.

```
Fixpoint List_ind (xs : List Sh Pos A) : P xs.
 destruct xs as [ | fy fys ].
 - apply nilP.
 - apply consP.
   induction fys as [ ys | s pf IH ].
   + apply forPure.
     apply List_ind.
   + apply forImpure.
     apply IH.
Defined.
```

Note that the difference between `Defined` and `Qed` to finalise a proof influences the visibility (in context of Coq often called *opacity*) of the associated function. The difference is rather technical; for here, it is sufficient to know that Coq can unfold definitions that are marked using `Defined`, while a definition marked with `Qed` prevents unfolding.

As alternative to the proof–mode version, we can, of course, implement the induction principle as function definition directly. In that case, instead of using induction on `Free`, we need a recursive helper function `free_ind` to work as induction hypothesis in case of `impure` values.

```
Fixpoint List_ind (xs : List A) : P xs :=
 match xs with
 | nil        => nilP
 | cons fy fys =>
   consP fy (let fix free_ind (fxs : Free (List A)) : ForFree P fxs :=
               match fxs with
               | pure xs => forPure P xs (List_ind xs)
               | impure s pf => forImpure (fun p => free_ind (pf p))
               end in free_ind fys)
 end.
```

Note that the preceding sections used this definition of `List_ind` whenever we used the tactic `induction` on list values. Hence, the arguments in an introduction pattern like `[ | fy fys IH ]` correspond to the arguments of the `consP` constructor and the induction hypothesis of type `ForFree`.

## 5.3 First Ideas to Model Curry Programs

In this section we want to take a look at how to model non–strict non–determinism, motivated by the functional logic programming language Curry. As we have seen in previous chapters, Curry combines the functional programming paradigm with concepts known from logic programming. More precisely, Curry combines non–determinism and laziness; the latter is introduced by shared variables via let–bindings. This combination leads to a call–time choice semantics. The approach we illustrated in the first sections enables reasoning about non–strict functional programs. The difference that we want to illuminate in this section is that sharing expressions via let–bindings in a model of Haskell with `undefined` values is an optimisation only but cannot be observed by the programmer. In Curry, on the other side, we can observe the difference between sharing a non–deterministic computation that is evaluated once and duplicating non–deterministic computations due to Curry's call–time choice semantics.

Before we dive deep into the problems when modelling call–time choice, we first implement Curry's non–determinism effect using `Free`. Based on this implementation, we discuss why the model cannot express call–time choice and give an outlook on ideas to implement the intended semantics. There are two aspects we want to emphasise here.

First, the outlook we give on modelling call–time choice can be used to incorporate sharing for other effects as well. For instance, a common extension of Haskell is the function `trace` that enables to track information of type `String` while evaluating a program. The effect of tracing results in an observable difference between sharing and duplicating an effectful expression. Hence, we need a similar technique to model call–time choice semantics as well as for more involved effects than we have seen so far.

Second, the ideas about how to model sharing are still work–in–progress. That is, we illustrate the motivation only and then reference related work that tackles similar problems and that we think is applicable for our problem.

## 5.3.1 Non–Determinism as Effect

We discussed how to model totality and partiality using `Free` in Section 5.1.3 (a) and Section 5.1.3 (b), respectively. In the same manner, we now take a look at how to model non–determinism. That is, we define a container representation and prove that this representation is isomorphic to a data structure that combines a pair–structure with a nullary constructor. With these definitions at hand, we define simple non–deterministic functions and prove exemplary properties.

**Representing Non–determinism using Containers**

In Curry the built–in non–determinism effect comes with two primitives: a polymorphic value `failed :: a` to represent finite failures as known from logic programming as well as the operator `(?) :: a -> a -> a` to construct a non–deterministic choice between two expressions. That is, we define the functor representing the non–determinism effect using two constructors as follows.

```
Inductive ND (A : Type) :=
| choice : A -> A -> ND A
| failed : ND A.
```

In order to transfer this functor to its isomorphic container representation, we need to define the shape and position type function involved. As we have two constructors to represent, we could simply use `bool` as the type for shapes. In order to increase the readability of the resulting code, we use a custom data type with two constructors instead.

```
Inductive ND_S :=
| ch : ND_S
| fd : ND_S.
```

In contrast to definitions of position type functions we have seen before, due to the two different shapes, we have to take the shape into account to determine the final position type. That is, for the first time the dependently typed position type function actually depends on the given shape. There are no positions to access in case of the `failed` constructor and two possible positions — the left or right argument — in case of `choice`. Hence, we use types with zero and two values as position types: `Empty` and `bool`, respectively.

```
Definition ND_P (s : ND_S) :=
 match s with
 | fl => Empty
 | ch => bool
 end.
```

As preparation to show that the constructed container is indeed isomorphic to the introduced functor, we define functions to transform values from one type to the other and vice versa.

```
Definition from_ND A (nd : ND A) : Ext ND_S ND_P A :=
 match nd with
 | choice x y => ext ch (fun (p : ND_P ch) => if p then x else y)
 | failed     => ext fl (fun (p : ND_P fl) => match p with end)
```

```
end.

Definition to_ND A (e : Ext ND_S ND_P A) : ND A :=
 match e with
 | ext ch pf => choice (pf true) (pf false)
 | ext fl pf => failed
 end.
```

Since the code corresponding to `failed` resembles the implementation we have seen for `One` before, the interesting part in both definitions is the case of `choice`. In the definition of `from_ND` we decide between the left argument `x` and right argument `y` depending on the given position `p` that is of type `bool`. We then implement `to_ND` such that it complies with its counterpart `from_ND`, that is, we access the left argument using `true` and the right argument using `false`.

These definitions indeed form an isomorphism; since we have already seen quite a few of these proofs, we refrain from showing the whole proof script here but provide the definition in Appendix A.5.

```
Lemma from_to_ND : ∀ (A : Type) (e : Ext ND_S ND_P A),
   from_ND (to_ND e) = e.

Lemma to_from_ND : ∀ (A : Type) (nd : ND A),
   to_ND (from_ND nd) = nd.
```

**Proving Properties About Non–deterministic Functions**

In the following sections the suffix `ND` for types is used as synonym for the concrete instantiation of data types with the non–determinism effect, e.g., `FreeND = Free ND_S ND_P` and `ListND = List ND_S ND_P`.

We define the following primitives `Failed` and `Choice` to construct values introduced by the non–determinism effect; we use the operator `?` as notation for the latter primitive. Furthermore, most of the following lemmas are stated without the corresponding proof script. We provide these missing proofs scripts in Appendix A.5.

```
Definition Failed (A : Type) : FreeND A :=
  impure fl (fun (p : ND_P fl) => match p with end).
Definition Choice (A : Type) (fx : FreeND A) (fy : FreeND A) :=
  impure ch (fun (p : ND_P ch) => if p then fx else fy).
```

**First Example**  For our first example we work on a non–deterministic (positive) number that is the choice between 1 and 2. Note that we work with (positive) numbers as representatives for `Int` values in Curry, that is, these numbers are primitive and only need to be lifted at the top–level constructor. Thus, we reuse Coq's type for Peano numbers `nat` to model positive numbers.

```
Definition oneOrTwo : FreeND nat :=
  pure 1 ? pure 2.
```

We define the following function `even` for arbitrary effects based on predefined Coq functions for `nat` values.

```coq
Definition even (fn : Free nat) : Free bool :=
  liftM1 Nat.even fn.

Definition liftM1 A R (f : A -> R) (x : Free A) : Free R :=
  x >>= fun x' => pure (f x').
```

Here, the auxiliary function `liftM1` is used to lift primitive functions — like the `even` function on Peano numbers. We will additionally use the notations `TTrue` and `FFalse`, respectively, for lifted Boolean values to distinguish them more comfortably from Coq's predefined Boolean values.

Now we are ready for some simple propositions about the defined functions: we prove that doubling the non–deterministic computation `oneOrTwo` non–deterministically and checking if the value is even yields `TTrue` twice. That is, doubling the value `oneOrTwo` is always even.

```coq
Lemma even_oneOrTwo : even (liftM2 mult oneOrTwo (pure 2)) = TTrue ? TTrue.
```

As an alternative formulation of this proposition, we can reuse `ForFree` to define the proposition `AllND` to check a given predicate for all values of a non–deterministic computation.

```coq
Definition AllND (A : Type) (P : FreeND A -> Prop) (fx : FreeND A) : Prop :=
  ForFree (fun x => P (pure x)) fx.
```

Given the definition `AllND`, we can simplify the statement above using a suitable proposition instead of reconstructing all non–deterministic choices on the right–hand side manually.

```coq
Lemma even_oneOrTwo_allND :
  AllND (fun fb => fb = TTrue) (even (liftM2 mult oneOrTwo (pure 2))).
```

Note that this alternative definition is more general than the one above: here we say that all results are even, whereas the lemma `even_oneOrTwo` states that there are exactly two results, which are both even.

**Second Example**  For the second example we consider the function `inc`.

```coq
Definition inc (fn : Free nat) : Free nat :=
  liftM1 (fun n => S n) fn.
```

We prove that applying the function `inc` to a non-deterministic choice between two values is equivalent to applying the function to both sides of the choice.

```coq
Lemma pulltab_inc : ∀ (fx fy : FreeND nat),
  inc (fx ? fy) = inc fx ? inc fy.
```

We can generalise this property of Curry's evaluation, often referred to as *pull–tabbing*, for any strict function.[4] The strictness property states that applying the function to an argument is equivalent to evaluating its argument to head normal form first and applying the function to the pure value.

```coq
Lemma pulltab_if_strict : ∀ (A B : Type)
  (f : FreeND A -> FreeND B) (fx fy : FreeND A)
  (Hstrict : ∀ fz, f fz = fz >>= fun z => f (pure z)),
  f (fx ? fy) = f fx ? f fy.
```

---

[4]In a setting with arbitrary effects, Filinski and Støvring (2007) call Curry's strictness property we describe here rigid.

**Third Example**   The third example considers not only a non–deterministic computation but a non–deterministic function. We reimplement the function `insert`, which non–deterministically inserts a value at each position of a list, that we introduced in Figure 2.2.1.

```
Fixpoint ndInsert' (A : Type) (fx : FreeND A) (xs : ListND A)
 : FreeND (ListND A) :=
 match xs with
 | nil         => Cons fx Nil
 | cons fy fys => Cons fx (Cons fy fys)
                  ? Cons fy (fys >>= fun ys => ndInsert' fx ys)
 end.

Definition ndInsert (A : Type) (fx : FreeND A) (fxs : FreeND (ListND A))
 : FreeND (ListND A) :=
 fxs >>= fun xs => ndInsert' fx xs.
```

Note, that due to the recursive structure of the function, we need to split the definition into two parts as already described in Section 5.1.1 and Section 5.1.4.

The property we want to define about `ndInsert` states that for each resulting list, the length of the list increases by one in contrast to the original list. The definition for `length` is defined as follows for an arbitrary effect.

```
Fixpoint length' (A : Type) (xs : List' A) : Free nat :=
  match xs with
  | nil       => pure 0
  | cons _ fxs => inc (fxs >>= fun xs => length' xs)
  end.

Definition length (A : Type) (fxs : Free (List A)) : Free nat :=
  fxs >>= fun xs => length' xs.
```

More precisely, the property we define is more specific, as we do not universally quantify over the list argument, but show it for a concrete list only.

```
Lemma ndInsert_inc :
  AllND (fun fxs => length fxs = pure 3)
        (ndInsert (pure 1) (Cons (pure 2) (Cons (pure 3) Nil))).
```

Let us take a more detailed look at this example, as a first naive try to prove the property fails. After simplifying the expression, that is, evaluating the expression

```
ndInsert (pure 1) (Cons (pure 2) (Cons (pure 3) Nil))
```

the goal looks as follows.

```
simpl.    ===========================
          AllND (fun fxs => length fxs = pure 3)
                (Cons (pure 1) (Cons (pure 2) (Cons (pure 3) Nil))
                 ? Cons (pure 2) (Cons (pure 1) (Cons (pure 3) Nil)
                                  ? Cons (pure 3) (Cons (pure 1) Nil)))
```

As `AllND` is defined via `ForFree`, we can use the second rule of the proposition to process the goal. Based on the decision for the non–deterministic choice, we get the first resulting list for the left branch and the recursive call for the right branch. We make a case distinction on that decision to proceed, which results in two subgoals.

```
apply forImpure. intros p; destruct p.
            ============================
            ForFree (fun x => length' x = pure 3)
                    (Cons (pure 1) (Cons (pure 2) (Cons (pure 3) Nil)))

            ForFree (fun x => length' x = pure 3)
                    (Cons (pure 2) (Cons (pure 1) (Cons (pure 3) Nil)
                                   ? Cons (pure 3) (Cons (pure 1) Nil)))
```

For the first subgoal, we use the first rule of the `ForFree`–proposition to handle the pure list constructor and then only need to show that the list has indeed `pure 3` elements.

```
- apply forPure. reflexivity.
```

In the second subgal the outermost constructor is the pure list constructor `Cons`, that is, we use the first rule of the `ForFree`–proposition again to process the goal.

```
- apply forPure.
            ============================
            length (Cons (pure 2) (Cons (pure 1) (Cons (pure 3) Nil)
                                   ? Cons (pure 3) (Cons (pure 1) Nil)))
            = pure 3
```

When restructuring the left–hand side of the goal, we can unroll the definition of `length` once,[5] yielding the following expression.

```
liftM1 S (length (Cons (pure 1) (Cons (pure 3) Nil)
                 ? Cons (pure 3) (Cons (pure 1) Nil)))
= pure 3
```

Fortunately, we know from the preceding examples that a strict function applied to a non–deterministic value performs a pull–tab step. That is, we can rewrite the above equation as follows. Note that we prove the hypothesis about the strictness of the applied function `liftM1 (fun n => S n) . length` by induction on the argument of type `FreeND (ListND nat)`.

```
rewrite pulltab_f_strict with (f := fun fxs => inc (length fxs)); simpl.
                                        ============================
                                        pure 3 ? pure 3 = pure 3
```

We finally realise that we have a non–deterministic choice on the left–hand side, while we have a pure value on the right–hand side. That is, we cannot prove this property. The problem with the original property is that we disregarded that the non–determinism introduced by the function `ndInsert` occurs nested in the list constructors.

---

[5]This unrolling is a simplified illustration for the reader and not exactly the expression that Coq yields when using the tactic `simpl` here. We can, however, define a lemma proving that this simplification is valid.

We regain the correctness of the property we initially wanted to state by evaluating the resulting list to normal form, that is, pulling all nested occurrences of non–deterministic choices to the top–level. Remember that in our setting the usage of the operator >>= mimics the evaluation to head normal form, that is, we can define the evaluation to normal form by traversing a given data structure and evaluating all constructor arguments to head normal form. Here, we only present the definition for `ListND` **nat**, but the nature of the functions is more generic: we can use the same definition for arbitrary effects and arbitrary types.

```
Fixpoint nfList' (ns : ListND nat) : FreeND (ListND nat) :=
 match ns with
 | nil        => Nil
 | cons fn fns => fn >>= fun n' =>
                  (fns >>= nfList') >>= fun ns' =>
                  Cons (pure n') (pure ns')
 end.

Definition nfList (fns : FreeND (ListND nat)) : FreeND (ListND nat) :=
 fns >>= nfList'.
```

Taking this idea one step further, we could define an overloaded function `nf` using a type class and implement instances for all translated data types. Using the function `nfList`, we redefine the property by evaluating the list argument to normal form and finally succeed with the proof.

```
Lemma ndInsert_inc :
  AllND (fun fxs => length fxs = pure 3)
        (nfList (ndInsert (pure 1) (Cons (pure 1) (Cons (pure 2) Nil)))).
```

### 5.3.2 Sharing as Effect

The preceding examples illustrate how to model simple non–deterministic values and functions as well as how to prove properties about them. The model we use so far, however, is not a suitable model for Curry's call–time choice semantics. The difference between Curry's call–time choice semantics and the non–strict non–determinism we are currently able to model can be generalised to the difference between call–by–name and call–by–need. In the following, we first compare our example definition of doubling a number with an alternative version to exhibit the call–by–name behaviour of our model. We discuss an ad–hoc solution to mimic call–by–need for concrete examples and give an example that shows the limitations of that idea. Secondly, we propose a combination of the underlying mechanism of the *KiCS2* compiler with related work on algebraic effects with scoped operations to implement Curry's call–time choice semantics.

We only provide proof scripts of the stated propositions if we highlight a specific problem and discuss a peculiarity, respectively. Otherwise we refrain from showing the proof scripts here but provide the missing code in Appendix A.5.

**Call–by–name vs. Call–by–need**

Consider the following two implementations for doubling a number: the first version multiplies the given value by two and the second version uses addition instead.

```
Definition doubleMult (fx : Free nat) : Free nat :=
  liftM2 mult fx (pure 2).

Definition doublePlus (fx : Free nat) : Free nat :=
  liftM2 plus fx fx.
```

As we know from the proof in Section 5.3.1, the check on evenness for doubling the non–deterministic computation `oneOrTwo` yields `True` for both non–deterministic branches.

```
Lemma even_doubleMult_Choice :
  even (doubleMult oneOrTwo) = TTrue ? TTrue.
```

We can, however, not prove the same property for `doublePlus` as the expression yields four instead of two results.

```
Lemma even_doublePlus_Choice :
 even (doublePlus oneOrTwo) = TTrue ? TTrue.
Proof.
  simpl. f_equal. extensionality p.
  destruct p.
  - (* even (pure 1 + oneOrTwo) = TTrue *)
    simpl.
    (* even (pure 2) ? even (pure 3) = TTrue *)
    admit.
  - (* even (pure 2 + oneOrTwo) = TTrue *)
    simpl.
    (* even (pure 3) ? even (pure 4) = TTrue *)
    admit.
Abort.
```

Using the tactic `admit` we give up a try to prove a subgoal and the keyword `Abort` marks that we give up on the whole proof of the lemma. The comments highlight that the argument `oneOrTwo` is evaluated two times: once as first argument of the addition and once as the second one. The two occurrences of the variable `fx` in the definition of `doublePlus` result in two individual evaluations. That is, the expression is equivalent to the inlined version using addition directly on two individual calls to the non–deterministic computation `oneOrTwo`.

```
Lemma doublePlus_inline :
  doublePlus oneOrTwo = liftM2 plus oneOrTwo oneOrTwo.
```

Applying the predicate `even` to this computation then yields four results, which we illustrate with the second proof.

```
Lemma even_doublePlus_Choice :
  even (doublePlus oneOrTwo) = (TTrue ? FFalse) ? (FFalse ? TTrue).
```

Note, that the expression `doubleMult oneOrTwo` introduces just one level of non–determinism and is, thus, equivalent to `pure 2 ? pure 4`.

```
Lemma doubleMult_inline :
  doubleMult oneOrTwo = pure 2 ? pure 4.
```

Hence, the unwanted behaviour origins from duplicating non–deterministic expressions: in the definition of `doublePlus`, we use the argument `fx` two times on the right–hand side of the definition. Note that the same unwanted behaviour when using `doublePlus` cannot be observed when applying the function to the value `Failed`.

```
Lemma even_doublePlus_Failed :
  even (doubleMult Failed) = Failed.
```

We observe here that the difference between call–by–name and call–by–need — that is, between sharing a value or duplicating its evaluation — cannot be observed for effects like partiality — as mimicked using `Failed` — but becomes observable for an effect like non–determinism.

**An Ad–hoc Solution**

In order to regain the property that we want to prove, we need to evaluate a shared non–deterministic computation once and use the pure value of that computation for all occurrences in the expression.

We can mimic this behaviour with the following redefinition of `doublePlus`.

```
Definition doubleSharePos (fn : FreeND nat) : FreeND nat :=
  match fn with
  | pure _     => liftM2 plus fn fn
  | impure s pf => impure s (fun p => doublePlus (pf p))
  end.
```

If the argument `fn` is already a pure number, we apply the function `plus` directly. In case of an impure value, we reconstruct the constructor of the observed effect — by using the same shape `s` — and then apply `doublePlus` on all values the position function `pf` yields for the given position `p`. Using this alternative definition with explicit sharing of the argument `fn`, we can prove the wanted property.

```
Lemma even_doubleSharePos_Choice :
  even (doubleSharePos oneOrTwo) = TTrue ? TTrue.
```

Since it is inconvenient to handle sharing of the argument `fn` by using explicit pattern matching on the `Free`–structure, we introduce an auxiliary function for explicit sharing.

```
Definition shareStrict (A : Type) (fx : Free A) : Free (Free A) :=
  match fx with
  | pure x     => pure (pure x)
  | impure s pf => impure s (fun p => pure (pf p))
  end.
```

Note that we define this explicit sharing function for an arbitrary effect, we do not need to know about the possible shapes that we are reconstructing on the right–hand side. In order to define this functionality in a separate function, we yield a nested value, that is, a value of type `Free (Free A)`. The inner layer of `Free` represents the shared computation while the outer layer is a reconstruction of the effect occurring in the argument `fx` that we want to share.

We implement an additional version of `doublePlus` using the explicit sharing function and prove that our running example for the property holds as well.

```
Definition doubleShare (fn : FreeND nat) : FreeND nat :=
  shareStrict fn >>= fun fn' => doublePlus fn'.

Lemma even_doubleShare_Choice :
  even (doubleShare oneOrTwo) = TTrue ? TTrue.
```

Note that the naming of the explicit sharing function was not picked by chance: the above usage for the definition of `doublePlus` only works as expected, because `doublePlus` is strict in its argument. That is, we can only use `shareStrict` if we are sure that the computation that we are sharing needs to be evaluated anyway. Otherwise we might trigger a non–deterministic computation or a partial value that was not needed in the first place. In order to construct an example that shows the limitations of the explicit sharing function we consider a nested application of the effect–generic function `const` that takes two arguments, ignores its latter one and yields the former.

```
Definition const (A B : Type) (fx : Free A) (fy : Free B) : Free A :=
  fx.
```

The following lemma shows a nested usage of the function `const` on an arbitrary expression as first argument and passing a shared non–deterministic computation as second and third argument. The expected property is that this nested application is equivalent to yielding just the first argument.

```
Lemma share_with_const : ∀ (A : Type) (fx : FreeND A),
    shareStrict oneOrTwo >>= fun fy' => const (const fx fy') fy' = fx.
Proof.
  intros A fx.
  simpl.
  (*   const (const fx (pure 1)) (pure 1)
     ? const (const fx (pure 2)) (pure 2) = fx *)
  admit.
Abort.
```

However, when using `shareStrict` the left–hand side of the equality is — as the name of the function already suggests — too strict. When we use `shareStrict` on an expression that does not need to be evaluated — here `const` yields its first argument and ignores the second one — the usage results in an unwanted behaviour in case of an effectful computation. That is, the above example application triggers the non–deterministic computation `oneOrTwo` and, thus, yields its first argument `fx` twice.

```
Lemma share_with_const : ∀ (A : Type) (fx : FreeND A),
  shareStrict oneOrTwo >>= fun fy' => const (const fx fy') fy' = fx ? fx.
```

At the end, although we can use the explicit sharing function `shareStrict` to mimic the sharing behaviour of Curry for some concrete examples, the usage is limited to situations that demand the evaluation of the shared expression anyway. In order to enable a more generic version of the explicit sharing function that does not rely on the demand of the shared expression, a more sophisticated underlying model is required.

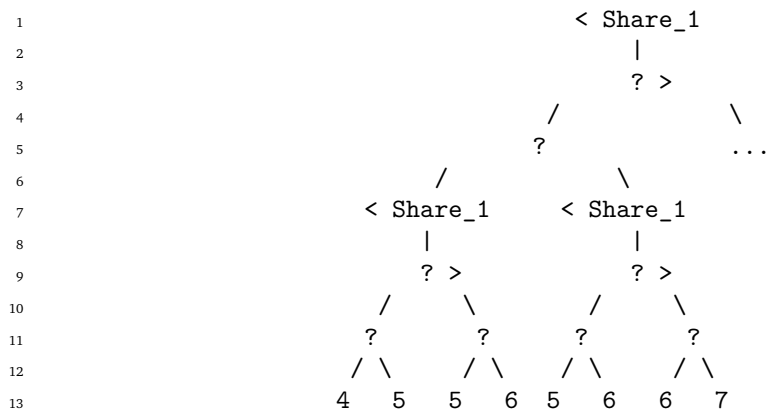

**Figure 5.1.:** The decision tree representation of the expression `example` in the simple Coq model with sharing nodes adapted with explicit begin (`<`) and end (`>`) markers

**Free Monads with Scope**

A promising approach to model Curry's call–time choice semantics is to use scoped operations as discussed by Wu et al. (2014) and Piróg et al. (2018). More precisely, Wu et al. present two approaches for implementations in Haskell: (a) using explicit primitives to mark the beginning and the end of a scoped expression; (b) adapting the free monad construction by replacing the underlying functor with a higher–order functor. The higher–order functor is a type constructor that takes a functor as argument, i.e., is of kind (* -> *) -> * -> *. The former approach leads to an administrative overhead to keep track of well–balanced begin and end tags and the problem that the construction of an unbalanced program is possible in the first place. The usage is more convenient in the latter approach since the sharing primitive explicitly takes the scoped expression explicitly as its argument. On the other hand, Piróg et al. derive a representation of scoped effects using category theory. They also give a Haskell implementation for their representation of the free monad with scoped effects by adding an additional constructor `Scope`.

```
data Prog f g a = Return a
                | Op (f (Prog f g a))
                | Scope (g (Prog f g (Prog f g a)))
```

Note that the resulting data type `Prog` is parametrised over an additional type constructor g. This type constructor is the functor for the effect with scope — the type constructor `f` is still used for the effects, like partiality or non–determinism, that we have seen so far. The nested structure of the `Scope` constructor enables the separation of *ordinary* and scoped effects: the inner `Prog` layer is not affected when using (>>=) to transform the structure.

In order to illustrate how a scoped effect can help in case of Curry's call–time choice semantics, Figure 5.1 shows an adapted version of the decision tree representation for the following Curry expression.

```
example :: Int
example = let x = 1 ? 2 in (x + (1 ? 2)) + (x + (1 ? 2))
```

More precisely, the illustration enhances the decision tree representation using explicit markers. Due to the explicit markers, we can number the `?`–edges corresponding to the given label of the `Share`–edges. That is, the choices in line number 5 and 11 are not affected by the sharing effect and do not get any number. Note that a numbering that affects only the choices within the markers corresponds to the tree representation when evaluating the example expression in Curry.

## 5.4 Related Work

In this section we present and discuss related and future work concerning formal reasoning about functional programs with effects. We classify the related work using three categories: work on modelling algebraic effects in general, but especially using free monads; modelling and reasoning techniques that focus on Haskell programs; and approaches that model Curry's call–time choice semantics. If applicable, we state our interests in future work for specific topics that we discuss as related work directly.

### 5.4.1 Modelling Algebraic Effects

In his attempt to combine several functional programming idioms to describe a technique that established reusable components, Swierstra (2008) manifests free monads and their usage for modelling effects in the functional programming community.

Plotkin and Pretnar (2009) give a more theoretical overview of algebraic effects and the corresponding handlers. They were motivated by finding models for exception handlers, but also present algebraic theories for effects like non–determinism, IO and state. For all these effects the underlying computation monad is the free–model of the respective theory. That is, they derive the representation of free monads to model effects from an algebraic theory. Starting with their work on algebraic effects, by now there are implementations in Idris (Brady, 2013), Scala (Brachthäuser and Schuster, 2017), OCaml (Kiselyov and Sivaramakrishnan, 2016), and Haskell (Kiselyov and Ishii, 2015; Kammar et al., 2013) as well as new languages with primitives for effects and handlers like Eff (Pretnar, 2015), Koka (Leijen, 2016), Frank (Lindley et al., 2017), and F* (Swamy et al., 2016).

Concerning modelling effects in dependently typed languages, McBride (2015) introduces the `General` monad to model recursion in Agda.

```
data General (S : Set) (T : S -> Set) (X : Set) : Set where
  !!   : X -> General S T X
  _??_ : (s : S) -> (T s -> General S T X) -> General S T X
```

He describes `General` as a request–response tree with values of type `X`. The definition is basically the implementation of the free monad we have presented and used here modulo renaming: `!!` embeds value as we do using `pure` and `impure` corresponds to `??`.

The formalism FreeSpec (Letan et al., 2018) allows to model components as programs with algebraic effects in a modular fashion and verify properties of these components. Letan et al. implement the framework in Coq based on the `Program` monad defined in the `operational` package[6] in Haskell. In contrast to the free monad we use in this chapter, their `Program` monad adds a third constructor to explicitly model processing a program using a continuation via the additional constructor `Bind`.

```
Inductive Program (I : Type -> Type) (A : Type) :=
| Pure    : A -> Program I A
```

---
[6]https://hackage.haskell.org/package/operational

```
| Bind    : ∀ (B : Type), Program I B -> (B -> Program I A) -> Program I A
| Request : I A -> Program I A.
```

In a newer version of the framework the constructors `Bind` and `Request` were merged into one constructor, hence, resemble the definition of freer monads as introduced by Kiselyov and Ishii (2015) more closely.

```
Inductive Program (I : Type -> Type) (A : Type) :=
| Pure    : A -> Program I A
| Request : ∀ (B : Type), I B -> (B -> Program I A) -> Program I A.
```

In contrast to the usual free monad definition using a functor, the freer representation removes the functor constraint by using a continuation `B -> Program I A` instead of a direct description of how to process the instruction `I B`.

The work of Koh et al. (2019) on *interaction trees* adapts the freer monad in Coq as well. The associated DeepSpec project on verifying networked servers implemented in C uses interaction trees as general structure to represent reactive components. Again, a more recent version of the framework uses a slightly modified structure. They currently implement a coinductive variant of the free monad to model diverging computations (Xia et al., 2019).

```
CoInductive itree (E: Type -> Type) (R: Type) :=
| Ret : R -> itree E R
| Tau : itree E R -> itree E R
| Vis : ∀ (A : Type), E A -> (A -> itree E R) -> itree E R.
```

It would be interesting to investigate if a coinductive definition of `Free` might lead to a model for non–termination in our setting as well.

Concerning other ideas that are worth further investigation, Atkey and Johann (2015) examine how the reasoning about inductive data types can be eased using the categorical principle of initial algebras. They are particularly interested in data types that are interleaved with effects — similar to the monadic liftings of the data types that we have discussed. Their approach does not only parametrise over the effects but also over the data structures. That is, they describe an abstraction using *initial f–and–m–algebras*, where *m* describes the interleaved effects as monads and *f* is a functor that describes the pure part of the respective data type. The part we like to investigate further is the approachability of the custom proof principle they use to prove properties about programs modelled as f–and–m–algebras for our setting.

### 5.4.2 Reasoning About Haskell Programs

In addition to related work that model effects using monads, there is also a variety of work that focuses on modelling and reasoning about Haskell programs. Most prominently represented in this thesis is the work by Abel et al. (2005). We based our approach on their original idea to model Haskell programs in the dependently typed language Agda using a monadic lifting for data types and functions. The monadic lifting models two aspects of Haskell. First, Haskell is a non–strict language, such that the monadic >>=–operator comes in handy when defining functions that need to pattern match on their arguments. Second, since Abel et al. want to model total as well as partial Haskell programs, the monadic approach enables the reuse of generic data type and function definitions. By instantiating the monadic parameter with a concrete instance the model can be used for either one of the wanted settings.

The interactive proof assistant Sparkle developed by de Mol et al. (2002) is integrated in the programming language Clean (Brus et al., 1987). Since Clean is a non–strict functional language like Haskell, reasoning about Clean is similar to reasoning about Haskell programs. Furthermore, Sparkle considers the partiality of Clean as default and, thus, explicitly models undefined values. As an especially interesting use–case of this explicit model, van Eekelen and de Mol (2006) reason about the strictness of functions. As future work, it would be interesting to model such strictness properties in the framework used in this thesis in the same manner as they do in Sparkle.

A conceptually similar approach that targets Haskell instead of Clean is HERMIT (Farmer et al., 2015). HERMIT is a toolkit for reasoning about Haskell programs that integrates directly into the GHC pipeline. The tool was successfully used by Farmer et al. to formalise and prove properties about various type classes and their corresponding laws as well as pen–and–paper proofs taken from *Pearls of Functional Algorithm Design* (Bird, 2010).

More recently, the translation of Haskell to Coq is a line of research that is of growing interest. In the process, the goal is to have generated Coq code as close to the original structure of the Haskell code as possible. The experience report by Dijkstra (2012) describes such an approach for an automatical translation. Dijkstra models partial functions using a method presented by Bove and Capretta (2007) that extends all functions with additional proof arguments. These proof arguments pose certain invariants about all function applications. For example, the additional argument for the partial function *head* on lists asserts that the applied list argument is not empty. In contrast to this explicit modelling of Haskell's partiality, Spector-Zabusky et al. (2018) present a translation from Haskell to Coq that focuses on total Haskell programs. The advantage of focussing on the total subset of Haskell is that they can translate functions one–to–one to Coq. In case of a partial function, like *head*, their translation scheme uses, however, an additional axiom that can be understood as a representation of the polymorphic value *undefined*. As they state, this axiom makes their model unsound and is only meant as a temporarily solution. That is, the common workflow suggests that the user can start proving properties about partial functions right away and totalise these functions later. Most recently, Breitner et al. (2018) present a more rigorous case study of this approach that proves properties about real world Haskell libraries. For this publication the approach concerning partial functions was slightly adapted. They still use an additional axiom for partial functions, but define it in such a way that prevents Coq from unfolding the definition. This approach enables a more practical way to reason about some simple properties involving partial functions.

### 5.4.3 Modelling Curry Programs

Besides the generic view on modelling effects and the focus on modelling or reasoning about Haskell, we also take a look at related work on proving properties about functional logic programs to find connections and inspirations for our findings about modelling Curry programs. Cleva et al. (2004) discuss that common equational reasoning techniques are not valid in functional logic languages that combine non–deterministic functions and call–time choice semantics. They suggest a proof calculus based on *CRLW* (constructor–based rewriting logic) (González-Moreno et al., 1996) to prove properties about first–order functional logic programs. On top of that, they encode this approach in various proof assistants for simple exemplary programs and corresponding properties.

Instead of using term rewriting systems like *CRLW* as basis, Fischer et al. (2009) present an embedding of Curry's functional logic features, that is, the combination of non–strictness, sharing and non–determinism, in the functional language Haskell. That is, they are more interested in a denotational, executable model of Curry. Their approach follows the same idea as Abel et al. (2005) and uses a monadic lifting of data types and functions to incorporate non–determinism in combination with non–strictness. Due to the combination with non–determinism, the examples they are most interested in are monadic components that support the type class `MonadPlus`.

```
class MonadPlus m where
  mplus :: m a -> m a -> ma
  mzero :: m a
```

Recall that the type constructor class `MonadPlus` comes with two functions `mplus` and `mzero` that represent operations for non–deterministic choices and failure values, respectively. As we have already discussed in Section 5.3.2, such a monadic representation models call–by–name rather than the concept of call–by–need. In order to overcome this issue, they introduce an explicit sharing function `share :: MonadPlus m => m a -> m (m a)`. We used the same interface for the sharing operator in our approach as discussed in Section 5.3.2. These two layers for the resulting value of the sharing function is necessary to retain the non–strictness property. The first monadic layer additionally performs bookkeeping when sharing non–deterministic computations. In order to track the decisions of shared non–deterministic computations during evaluation, Fischer et al. use an untyped heap that stores computations corresponding to variable references. That is, their implementation is a realisation of the heap used in the operational call–by–need semantics as introduced by Launchbury (1993). Petricek (2012) presents a generalisation of the `share` operator namend `malias :: Monad m => m a -> m (m a)` that can be used to model call-by-value and call-by-name semantics for any monad `m` as well as to model call-by-need semantics for certain kind of monads. The main contribution of an implementation for Curry's call–time choice in Coq using `Free` in contrast to the implementation of Fischer et al. is to avoid impure features. This avoidance originates in our usage of Coq, that does not allow such features, but is also of interest for an implementation in Haskell as well. As future work, we plan to investigate if a Haskell implementation that tracks the decisions of choices instead of references to computations can compete with respect to performance.

An approach by Antoy et al. (2017) aims at modelling Curry programs in Agda. In contrast to our as well as the goal of Fischer et al. to implement a generic model that combines non–strictness, sharing and effects like non–determinism, Antoy et al. explicitly tackle Curry's choice operator only. They present two different representations of Curry's non–determinism called *planned choices*, and *set of values* respectively. The translation for the first approach adds an extra argument to the translated function. This argument represents the choice to make for non–deterministic choices during the computation of the translated function and, hence, makes the function deterministic. For us, the second approach is more interesting because of its high resemblance to a monadic interface. The second approach translates a non–deterministic Curry function into a deterministic function in Agda that explicitly models the potential non–determinism. The following data type represents the explicit model used in Agda.

```
data ND (A : Set) : Set where
  Val  : A -> ND A
```

```
_??_ : ND A -> ND A -> ND A
```

Here, it becomes apparent that their model is a specialisation of the generic monadic model; in case of our model we can instantiate `Free` with a variant of the non–determinism effect `ND` without `failed` results[7] in a representation isomorphic to their data type `ND`. The provided operators `$*` and `*$*` underline the correspondance to a monadic interface: The former operator corresponds to the functor method `fmap` and the latter to the monadic bind–operation `>>=`.

Lastly, we like to mention the promising prototype developed by Bunkenburg (2019) that was developed during his master's thesis. Bunkenburg implements Curry's call–time choice semantics in Coq using the `Free` model in combination with the results of Wu et al. (2014) about scoped effects. Wu et al. describe two representations for scoped effects: explicit syntax with begin–and end–tags to delimit the scope and higher–order syntax that uses a higher–order functor to describe effects. The first approach leads to cumbersome implementations for handlers of scoped effects when also considering possible mismatches of the tags. The higher–order syntax, on the other hand, takes the scoped program directly as argument and, thus, avoids to mark the begin and end of the scope manually. Concerning the implementations in Coq, the higher–order syntax, however, opens a new can of worms when trying to circumvent Coq's strict positivity restriction for data types, especially for the definition of the container representation of higher–order functors. As future work we like to try out this prototype using several case studies to detect the limits of the approach or the implementation in particular. With respect to the difficulty of defining the container representation for the higher–order approach in Coq, we hope to circumvent the problem by finding an appropriate implementation that is probably less general than the implementation in Haskell but adheres to Coq's restrictions. First experiments revealed that Agda accepts a slightly adjusted definition of the data type Bunkenburg tried to implement in Coq. Hence, it might be worthwhile to reimplement the model for call–time choice in Agda to try out first proofs about Curry programs for the higher–order approach. A fairly long–term goal is to show that our model using `Free` correctly models Curry's call–time choice. Thanks to the implementation in Coq, we can use the model to prove the equivalence with an operational or denotational model of Curry. The implementation of the operational or denotational model of Curry is an interesting line of work itself that is useful for a wide range of questions and issues.

## 5.5 Future Work

Besides the special case of modelling Curry programs, we are interested in adapting our approach to be applicable to call–by–need semantics in general. In Haskell, for example, we can also observe the difference between call–by–name and call–by–need when considering tracing as effect. Consider the following exemplary Haskell program and its evaluation assuming a call–by–need and call–by–name semantics, respectively.

```
> let xNeed = trace "example" 42 in xNeed + xNeed
example84

> let xName = trace "example" 42 in xName + xName
```

---

[7]As already noted, the approach by Antoy et al. does not consider *failure* values but focuses on a non–deterministic choice operator instead.

```
exampleexample84
```

The function `trace :: String -> a -> a` takes a message as first argument that is printed on the console and yields its second argument. Note that using call–by–name semantics allows us to inline the let–binding, yielding the following expression that triggers the tracing–effect twice.

```
> trace "example" 42 + trace "example" 42
exampleexample84
```

We plan to apply the first insights we gained from modelling Curry programs to model Haskell's tracing effect with call–by–need semantics. In a first attempt to investigate the current approach (Christiansen et al., 2019), we took a more detailed look at the effects occurring in Haskell and defined a category of effects that behave the same under call–by–name and call–by–need semantics using a syntactic criteria for the effects involved. In that paper, we have shown that a variety of propositions can be successfully modelled using our approach, assuming that the functions involved do not share computations.

When thinking about non–deterministic functions, it becomes apparent that we cannot model all possible functions with our current approach. Consider for example the following two function definitions of the same type.

```
fun1 :: Int -> Int
fun1 x = (x + 1) ? (x * 2)

fun2 :: Int -> Int
fun2 = (+1) ? (*2)
```

Both functions increment the input value by one as well as multiply it by two by using non–determinism. The crucial difference is that the first function non–deterministically yields a value whereas the second one yields a non–deterministic function. When using these functions as argument to `map`, for example, we can observe that these implementation behave differently.

```
> map fun1 [1,1]
[2,2]
[2,2]
[2,2]
[2,2]

> map fun2 [1,1]
[2,2]
[2,2]
```

These examples are also used by Mehner et al. (2014) to show that eta–equivalence does not hold for Curry. Translating the second function into our framework is, however, not possible. Up to now, we have translated all function types by lifting each individual argument. In case of `fun2`, we need to implement the non–deterministic choice between two functions, thus, the final type is a function type that must be lifted. More precisely, let us take a look at the definition of `fun1` and the try to define `fun2` with the usual pattern.

```
Definition fun1 (A : Type) (fn : Free nat) : Free nat :=
  (liftM2 plus fn (pure 1)) ? (liftM2 mult fn (pure 2)).

Definition fun2 (A : Type) : Free nat -> Free nat :=
  (fun fn => liftM2 plus fn (pure 1)) ? (fun fn => liftM2 mult fn (pure 2)).
```

The second definition will not typecheck, because both arguments of ? are functions, not the resulting expression. The resulting expression is of type `Free (Free nat -> Free nat)`. When thinking about the translation of types in more detail, it indeed makes sense to translate a Curry (or Haskell) function of type `a -> a` to `Free (Free a -> Free a)` in our monadic lifting. Currently, we have corresponding type definitions that lift all arguments of the corresponding constructors. For example, a type for pairs has a constructor `pair` of the following type.

```
Π> Check pair.
  pair : Free A -> Free B -> Pair A B.
```

Applying this rule to a function type, we need a constructor for lifted arguments as well.

```
Inductive Arrow (A B : Type) :=
| arrow : (Free A -> Free B) -> Arrow A B.
```

Using the definition of `Arrow` as translation for function types, we can then define the two functions above as follows.

```
Definition fun1 (A : Type) :  Free (Arrow (Free nat) (Free nat)) :=
  (pure (arrow (fun fn => (liftM2 plus fn (pure 1))))) ?
  (pure (arrow (fun fn => (liftM2 mult fn (pure 2))))).

Definition fun2 (A : Type) : Free (Arrow (Free nat) (Free nat)) :=
  pure (arrow (fun fn => (liftM2 plus fn (pure 1))
                         ? (liftM2 mult fn (pure 2)))).
```

Although both functions have the same type, we can now see the difference of the implementation more explicitly: the first definition produces a non–deterministic result that consists of two functions first, while the second definition yields a pure function with non–deterministic values as argument for the `arrow` constructor. Note that a similar construction is also possible in the partiality setting we discussed for Haskell. The difference between such two function definitions can, however, only be observed in the presence of the function `seq :: a -> b -> b` that evaluates its first argument to head normal form and yields its second argument. The example consisting of two functions in case of partiality in Haskell look as follows.

```
fun1 :: a -> b
fun1 x = undefined

fun2 :: a -> b
fun2 = undefined
```

In order to make the framework approachable for users, we like to investigate if all functions should be translated using `Arrow` or only functions that cannot be constructed otherwise. For both variants it makes sense to conduct a case study to check how these decisions affect the usage of the framework when writing proofs.

As a different direction for future work, it would be interesting to consider inequational propositions as well. Using inequational propositions, we can, for example, compare the strictness of two implementations of the same function. Consider the following two variants to define multiplication on Peano numbers.

```
mult :: Peano -> Peano -> Peano
mult Zero     _ = Zero
mult (Succ m) n = add n (mult m n)

mult2 :: Peano -> Peano -> Peano
mult2 Zero      _    = Zero
mult2 (Succ _) Zero = Zero
mult2 (Succ m) n    = add n (mult2 m n)
```

We refer to the former definition as the default implementation, whereas the latter definition is a more advanced variant that is less–strict. Assuming a less–strict–relation $\sqsubseteq$, we are then interested in a formal specification that `mult2` is less strict than `mult`.

$$\forall\ p\ q : \text{Peano.\ mult}\ p\ q \sqsubseteq \text{mult2}\ p\ q \land \exists\ p\ q : \text{Peano.\ mult2}\ p\ q \not\sqsubseteq \text{mult}\ p\ q$$

A first idea to model the relation in the setting of partiality using the `Free`–approach looks as follows.

```
Inductive le_Free_Partial A (leA : A -> A -> Prop)
  : Free A -> Free A -> Prop :=
| le_pure         : ∀ x y, leA x y -> le_Free_Partial leA (pure x) (pure y)
| le_impure_pure : ∀ e x, le_Free_Partial leA (impure e) (pure x)
| le_impure       : ∀ e1 e2, le_Free_Partial leA (impure e1) (impure e2).
```

In a similar manner, we can define a specialised version for an effect like `error`. More interesting, however, is the objective to generalise the relation to arbitrary effects. For example, an attempt to generalise the definition for the partiality and error setting passes an additional function to compare the shape of the effects involved.

```
Inductive le_Free_Hoare (A : Type) (leS : Sh -> Sh -> Prop)
    (leA : A -> A -> Prop) : Free C A -> Free C A -> Prop :=
| le_pure_Hoare        : ∀ x y,
    leA x y -> le_Free_Hoare leS leA (pure x) (pure y)
| le_impure_pure_Hoare : ∀ s pf x,
    (∀ p, le_Free_Hoare leS leA (pf p) (pure x)) ->
      le_Free_Hoare leS leA (impure (ext s pf)) (pure x)
| le_impure_Hoare      : ∀ s1 pf1 s2 pf2,
    (∀ p1, exists p2, le_Free_Hoare leS leA (pf1 p1) (pf2 p2)) ->
    leS s1 s2 ->
    le_Free_Hoare leS leA (impure (ext s1 pf1)) (impure (ext s2 pf2))
| le_pure_impure_Hoare : ∀ x s pf,
    (exists p : Pos s, le_Free_Hoare leS leA (pure x) (pf p)) ->
      le_Free_Hoare leS leA (pure x) (impure (ext s pf)).
```

In contrast to the version of partiality, the constructors `le_impure_pure_Hoare` and `le_impure_Hoare` have additional hypotheses that take care of possible recursive effects. Moreover, the constructor `le_pure_impure_Hoare` needs to be added for effects like non–determinism that combine several values when using its primitive `choice`–operator. The definition above is not the only relational interpretation that is reasonable. Possible alternatives are constructions corresponding to Smith or Plotkin powerdomains (Abramsky and Jung, 1994). We have presented these ideas in Christiansen and Dylus (2019) and plan to continue this approach in the future.

## 5.6 Conclusion

In this chapter we discussed an approach to model and reason about non–strict effectful programs in a proof assistant like Coq. Here, partiality as most notably occurring in Haskell as well as Curry's non–determinism were effects of special interest. We started with a detailed motivation that resembles the obstacles we stumbled upon and the corresponding ideas and final solutions we came up with to tackle the problems. After establishing the preliminary concepts underlying the framework, we define first Haskell functions and associated properties. One of the advantages of the framework is its generality: although we are especially interested in an effect like partiality that can occur in Haskell, we emphasise the definition of effect–generic functions and proofs. The former is a common idiom in functional programming, we generalise functions using, for example, a monadic abstraction. In case of our framework, the latter generalisation is even more interesting: if a proposition holds for arbitrary effects, we can reuse this proposition in other generic proofs as well as in proofs about more specialised settings like partiality. That is, we can even reuse a variety of functions and properties we define to model Haskell programs when reasoning about Curry programs. Of course, for Curry programs we need a non–deterministic choice between two computations as an additional primitive, nonetheless, effect–generic functions and propositions can be reused. In the section about Curry, we additionally observe that the current model we use for Haskell as well as Curry follows call–by–name semantics. In both cases, call–by–need semantics is what we are actually aiming for. In case of Curry, however, we more naturally stumble upon this problem, because in case of non–determinism we can observe the difference between call–by–name and call–by–need whereas we cannot observe this difference in case of partiality only. For an effect like tracing, on the other hand, that is common for Haskell programmers as well, we can again observe the difference. Thus, this observation leads us to the goal to define a model that enables us to reason about call–by–need semantics. We discussed a first naive approach to reproduce the wanted semantics for simple programs, but could not use this approach in a more general setting. In order to overcome this obstacle, we refer to a prototypical implementation using scoped operations to model Curry's call–time choice semantics. We hope to apply the ideas on scoped operations to reason about Haskell's tracing effects and other effects that rely on call–by–need semantics.

## 5.7 Final Remarks

The framework presented in this chapter as well as additional content has been previously published. We published the general approach to represent Haskell programs as monadic Coq programs using free monads and container representations for functors in The Art, Science, and Engineering of Programming, Volume 3 (Dylus et al., 2019). In the publication we additionally present a case study on queues that we did not include here. Furthermore, the publication has a fairy tale theme that allowed us to present the obstacles and their solution in a more tutorial–like manner. Note that we decided here to define `Free` with an inlined version of `Ext` instead of using the data type to represent the container extension explicitly. The additional content about Haskell's tracing effect and the observations concerning modelling call–by–name and call–by–need were published recently in the Proceedings of the 12th International Symposium on Haskell (Christiansen et al., 2019). The publication also covers the translation of function types using an additional layer of monadic liftings as we discuss properties about Haskell's primitive function `seq`.

# Conclusion  6

This thesis investigated effectful declarative programming by means of two practical applications as well as formal reasoning. In case of the applications, we set the focus on the effect of non–determinism, especially, the combination of non–determinism and non–strictness. We implemented both applications using the functional logic programming language Curry. Furthermore, we investigated how to formally reason about effectful non–strict programs using a proof assistant like Coq. Here, the effects of main interest were partiality — as it occurs in Haskell — and, again, non–determinism as known from Curry. The results for Haskell–like partiality are mature, whereas modelling Curry's call–time choice semantics remains a challenge. First ideas to tackle this challenge are, however, promising.

## 6.1 Combination of Non–determinism and Non–strictness

In Chapter 3 we implemented a variety of sorting functions parametrised over a comparison function and applied these sorting functions to a non–deterministic comparison function. The resulting functions enumerate permutations of the input list. We implemented this approach in Curry using the built–in non–determinism as well as in Haskell using lists as representation for non–deterministic computations. The Haskell implementation uses a generic, monadic lifting of the ordinary, pure sorting function in order to use a non–deterministic comparison function and compute non–deterministic results. One particularly interesting observation was that the Haskell version of selection sort computes duplicated results whereas the Curry version computes exactly all permutations. This difference was the main reason we investigated the difference of both implementations in the first place. Furthermore, we observed that the Curry version of these implementations can exploit non–strictness better than their Haskell counterparts. We investigated this property for all sorting functions that computed exactly all permutations in Curry as well as in Haskell. In order to check for this property, we computed only the head elements of the permutations and counted the number of non–deterministic choices that were demanded to compute the result. The most impressive sorting functions for this example are selection sort and bubble sort implemented in Curry as they only demanded $n$ non–deterministic choices for a list of length $n$. On top of that, none of the Curry implementations need to demand all $n!$ non–deterministic computations for a list of length $n$, whereas the Haskell implementations demand at least $n!$ computations. This comparison shows a clear advantage of the Curry implementation: $n!$ non–deterministic computations in Haskell corresponds to evaluating all non–deterministic computations that occur for an implementation that yields exactly all permutations. On the other hand, selecting only the head element of the permutations has no effect on the non–determinism that needs to be demanded for these examples in Curry.

In Chapter 4 we presented an implementation for probabilistic programming in Curry. Such a library proves to be a good fit for an implementation using a functional logic language, because both paradigms share similar features. While other libraries need to reimplement features specific to probabilistic programming, we solely rely on core features of functional logic languages. The key idea of the library is to use non–determinism to model distributions, which consist of pairs of an event and the corresponding probability. We discussed design choices as well as the disadvantages and advantages that result from this approach. Besides modelling distributions, users of probabilistic programming are interested in asking queries about their models. The presented implementation provides non–strict probabilistic combinators in order to avoid spawning unnecessary non–deterministic computations when modelling distributions and performing queries. On the one hand, these non–strict combinators have benefits in terms of performance due to early pruning when performing queries. Using combinators that are too strict, on the other hand, leads to a loss of these performance benefits. Fortunately, the user does not have to worry about strictness as long as they only use the provided combinators. On top of that, we showed that the library operations that correspond to a monadic interface obey the expected monad laws, if the user meets two restrictions concerning their usage.

## 6.2 Modelling Effectful Programs in Coq

In Chapter 5 we discussed an approach to model and reason about non–strict effectful programs in the proof assistant Coq. One of the advantages of the framework is its generality: although we are mainly interested in an effect like partiality that can occur in Haskell, we emphasise the definition of effect–generic functions and proofs. In case of our framework the generalisation of proofs is especially beneficial: if a proposition holds for arbitrary effects, we can reuse this proposition in other generic proofs as well as in proofs about more specialised settings like partiality. That is, we can even reuse a variety of functions and properties we define to model Haskell programs when reasoning about Curry programs.

We observed that the current model we use for Haskell as well as Curry follows call–by–name semantics instead of the wanted call–by–need semantics. We stumbled upon this problem in case of Curry, because we can observe the difference between call–by–name and call–by–need in case of non–determinism. In case of Haskell, we cannot observe this difference when modelling partiality only. In Haskell we can, however, again observe the difference for an effect like tracing. Thus, this observation leads us to the goal to define a model that enables us to reason about call–by–need semantics. We hope to apply ideas of a prototypical implementation using scoped operations to model call–time choice to reason about Haskell's tracing effects and other effects that rely on call–by–need semantics as well.

# Bibliography

Michael Abbott, Thorsten Altenkirch, and Neil Ghani. 2003. Categories of Containers. In *Foundations of Software Science and Computation Structures* (Lecture Notes in Computer Science). Volume 2620. Springer, 23–38 (cited on pages 24, 102).

Andreas Abel, Marcin Benke, Ana Bove, John Hughes, and Ulf Norell. 2005. Verifying Haskell Programs Using Constructive Type Theory. In *Proceedings of the 2005 ACM SIGPLAN Workshop on Haskell* (Haskell '05). ACM, 62–73 (cited on pages 98, 104, 111, 116, 132, 134).

Samson Abramsky and Achim Jung. 1994. Domain theory. In *Handbook of Logic in Computer Science*. Clarendon Press, 1–168 (cited on page 139).

Abdulla Alqaddoumi, Sergio Antoy, Sebastian Fischer, and Fabian Reck. 2010. The pull-tab transformation. In *Proceedings of the Third International Workshop on Graph Computation Models*, 127–132 (cited on page 12).

Thorsten Altenkirch and Conor McBride. 2003. Generic programming within dependently typed programming. In *Generic Programming*. IFIP — The International Federation for Information Processing. Volume 115. Springer, 1–20 (cited on page 24).

Thorsten Altenkirch and Gun Pinyo. 2017. Monadic Containers and Universes (Abstract). In *Abstracts of 23rd International Conference on Types for Proofs and Programs* (Lecture Notes in Computer Science). Volume 10608. Springer, 20–21 (cited on page 102).

Sergio Antoy. 2005. Evaluation Strategies for Functional Logic Programming. *Journal of Symbolic Computation*, 40, 1, 875–903 (cited on page 63).

Sergio Antoy, Michael Hanus, and Steven Libby. 2017. Proving Non-Deterministic Computations in Agda. In *Proceedings of the 24th International Workshop on Functional and (Constraint) Logic Programming* (Electronic Proceedings in Theoretical Computer Science). Volume 234. Open Publishing Association, 180–195 (cited on pages 134, 135).

Zena M. Ariola and Matthias Felleisen. 1997. The call-by-need lambda calculus. *Journal of Functional Programming*, 7, 3, 265–301 (cited on page 4).

Ken Arnold, James Gosling, and David Holmes. 2005. *The Java Programming Language*. Addison Wesley Professional (cited on page 1).

Robert Atkey and Patricia Johann. 2015. Interleaving data and effects. *Journal of Functional Programming*, 25, E20 (cited on page 132).

Bruno Barras, Samuel Boutin, Cristina Cornes, et al. 1997. The Coq Proof Assistant Reference Manual: Version 6.1. Technical Report. INRIA, 214 (cited on page 3).

Richard Bird. 2010. *Pearls of Functional Algorithm Design*. Cambridge University Press (cited on page 133).

Frédéric Blanqui, Jean-Pierre Jouannaud, and Mitsuhiro Okada. 2002. Inductive-data-type systems. *Theories of Types and Proofs 1997*, 272, 1, 41–68 (cited on page 99).

Ana Bove and Venanzio Capretta. 2007. Computation by Prophecy. In *Typed Lambda Calculi and Applications*. Lecture Notes in Computer Science. Volume 4583. Springer, 70–83 (cited on page 133).

Jonathan Immanuel Brachthäuser and Philipp Schuster. 2017. Effekt: Extensible algebraic effects in Scala (short paper). In *Proceedings of the 8th ACM SIGPLAN International Symposium on Scala* (SCALA 2017). ACM, 67–72 (cited on page 131).

Edwin Brady. 2013. Programming and Reasoning with Algebraic Effects and Dependent Types. In *Proceedings of the 18th ACM SIGPLAN International Conference on Functional Programming* (ICFP '13). ACM, 133–144 (cited on page 131).

Bernd Braßel. 2009. A Technique to Build Debugging Tools for Lazy Functional Logic Languages. *Proceedings of the 17th International Workshop on Functional and (Constraint) Logic Programming (WFLP 2008)*. Electronic Notes in Theoretical Computer Science 246, 39–53 (cited on page 90).

Bernd Braßel, Michael Hanus, and Frank Huch. 2004. Encapsulating Non-Determinism in Functional Logic Computations. *Journal of Functional and Logic Programming*, 2004, 6 (cited on pages 17, 60).

Joachim Breitner, Antal Spector-Zabusky, Yao Li, et al. 2018. Ready, Set, Verify! Applying Hs-to-coq to Real-world Haskell Code (Experience Report). *Proceedings of the ACM on Programming Languages*, 2, ICFP (cited on page 133).

TH Brus, Marko CJD van Eekelen, MO Van Leer, and Marinus J Plasmeijer. 1987. Clean — a Language for Functional Graph Rewriting. In *Conference on Functional Programming Languages and Computer Architecture* (Lecture Notes in Computer Science). Volume 247. Springer, 364–384 (cited on page 133).

Niels Bunkenburg. 2019. *Modeling Call-Time Choice as Effect Using Scoped Free Monads*. Master's Thesis. Christian-Albrechts University of Kiel (cited on page 135).

Adam Chlipala. 2013. *Certified Programming with Dependent Types: A Pragmatic Introduction to the Coq Proof Assistant*. The MIT Press (cited on pages 18, 98).

Jan Christiansen and Sandra Dylus. 2019. Proving Inequational Propositions about Haskell Programs in Coq. International Conference on the Art, Science, and Engineering of Programming. Poster. (2019) (cited on page 139).

Jan Christiansen and Sebastian Fischer. 2008. EasyCheck — Test Data for Free. In *Proceedings of the International Symposium on Functional and Logic Programming* (Lecture Notes in Computer Science). Volume 4989. Springer, 322–336 (cited on pages 1, 89).

Jan Christiansen, Nikita Danilenko, and Sandra Dylus. 2016. All Sorts of Permutations (Functional Pearl). In *Proceedings of the 21st ACM SIGPLAN International Conference on Functional Programming* (ICFP 2016). ACM, 168–179 (cited on page 53).

Jan Christiansen, Daniel Seidel, and Janis Voigtländer. 2010. Free Theorems For Functional Logic Programs. In *Proceedings of the 4th ACM SIGPLAN Workshop on Programming Languages Meets Program Verification* (PLPV '10). ACM, 39–48 (cited on page 90).

Jan Christiansen, Sandra Dylus, and Niels Bunkenburg. 2019. Verifying Effectful Haskell Programs in Coq. In *Proceedings of the 12th ACM SIGPLAN International Symposium on Haskell* (Haskell 2019). ACM, 125–138 (cited on pages 136, 140).

José Miguel Cleva, Javier Leach, and Francisco J. López-Fraguas. 2004. A logic programming approach to the verification of functional-logic programs. In *Proceedings of the 6th ACM SIGPLAN International Conference on Principles and Practice of Declarative Programming* (PPDP '04). ACM, 9–19 (cited on page 133).

William Clinger. 1982. Nondeterministic call by need is neither lazy nor by name. In *Proceedings of the 1982 ACM Symposium on LISP and Functional Programming* (LFP '82). ACM, 226–234 (cited on page 62).

Thierry Coquand and Gérard Huet. 1986. *The Calculus of Constructions*. PhD Thesis. INRIA (cited on page 18).

Thierry Coquand and Christine Paulin. 1988. Inductively Defined Types. In *Proceedings of the International Conference on Computer Logic* (Lecture Notes in Computer Science). Volume 417. Springer, 50–66 (cited on page 17).

Evan Czaplicki. 2012. *Elm: Concurrent FRP for Functional GUIs*. Senior Thesis. Harvard University (cited on page 1).

Nils Anders Danielsson, John Hughes, Patrik Jansson, and Jeremy Gibbons. 2006. Fast and Loose Reasoning is Morally Correct. In *Conference Record of the 33rd ACM SIGPLAN-SIGACT Symposium on Principles of Programming Languages* (POPL '06). ACM, 206–217 (cited on page 93).

Maarten de Mol, Marko van Eekelen, and Rinus Plasmeijer. 2002. Theorem Proving for Functional Programmers. In *Implementation of Functional Languages*. Lecture Notes in Computer Science. Volume 2312. Springer, 55–71 (cited on page 133).

Luc De Raedt, Angelika Kimmig, and Hannu Toivonen. 2007. ProbLog: A Probabilistic Prolog and Its Application in Link Discovery. In *Proceedings of the 20th International Joint Conference on Artifical Intelligence* (IJCAI '07). Morgan Kaufmann Publishers Inc., 2468–2473 (cited on pages 55, 63).

Gabe Dijkstra. 2012. Experimentation Project Report: Translating Haskell Programs to Coq Programs. Experimentation Project. Utrecht University, 17 (cited on page 133).

Sandra Dylus, Jan Christiansen, and Finn Teegen. 2020. Implementing a Library for Probabilistic Programming Using Non-strict Non-determinism. *Theory and Practice of Logic Programming*, 20, 1, 147–175 (cited on page 91).

Sandra Dylus, Jan Christiansen, and Finn Teegen. 2019. One Monad to Prove Them All. *The Art, Science, and Engineering of Programming*, 3, 3-8 (cited on page 140).

Sandra Dylus, Jan Christiansen, and Finn Teegen. 2018. Probabilistic Functional Logic Programming. In *Practical Aspects of Declarative Languages* (Lecture Notes in Computer Science). Volume 10702. Springer, 3–19 (cited on page 91).

Martin Erwig and Steve Kollmansberger. 2006. Functional Pearls: Probabilistic Functional Programming in Haskell. *Journal of Functional Programming*, 16, 1, 21–34 (cited on pages 57, 59, 64, 89).

Andrew Farmer, Neil Sculthorpe, and Andy Gill. 2015. Reasoning with the HERMIT: Tool Support for Equational Reasoning on GHC Core Programs. In *Proceedings of the 8th ACM SIGPLAN Symposium on Haskell* (Haskell '15). ACM, 23–34 (cited on page 133).

Andrezej Filinski. 1996. *Controlling Effects*. PhD thesis. Carnegie-Mellon University Pittsburgh (cited on page 5).

Andrzej Filinski and Kristian Stø vring. 2007. Inductive reasoning about effectful data types. In *Proceedings of the 12th ACM SIGPLAN International Conference on Functional Programming* (ICFP '07). ACM, 97–110 (cited on page 123).

Sebastian Fischer, Oleg Kiselyov, and Chung-chieh Shan. 2009. Purely Functional Lazy Non-deterministic Programming. In *Proceedings of the 14th ACM SIGPLAN International Conference on Functional Programming* (ICFP '09). ACM, 11–22 (cited on pages 53, 55, 134).

Jeremy Gibbons and Ralf Hinze. 2011. Just do it: Simple Monadic Equational Reasoning. In *Proceedings of the 16th ACM SIGPLAN International Conference on Functional Programming* (ICFP '11). ACM, 2–14 (cited on pages 71, 93).

Juan C González-Moreno, Maria Teresa Hortalá-González, Francisco Javier López-Fraguas, and Mario Rodríguez-Artalejo. 1996. A rewriting logic for declarative programming. In *European Symposium on Programming* (Lecture Notes in Computer Science). Volume 1058. Springer, 156–172 (cited on page 133).

Noah D. Goodman and Andreas Stuhlmüller. 2014. The Design and Implementation of Probabilistic Programming Languages. http://dippl.org. (2014) (cited on pages 57, 63).

Noah D. Goodman, Vikash K. Mansinghka, Daniel M. Roy, Keith Bonawitz, and Joshua B. Tenenbaum. 2008. Church: a Language for Generative Models. In *Proceedings of the 24th Conference in Uncertainty in Artificial Intelligence* (UAI 2008). AUAI Press, 220–229 (cited on page 57).

Andrew D. Gordon, Thomas A. Henzinger, Aditya V. Nori, and Sriram K. Rajamani. 2014. Probabilistic Programming. In *Proceedings of the on Future of Software Engineering* (FOSE 2014). ACM, 167–181 (cited on page 57).

Michael Hanus. 2017. *PAKCS: The Portland Aachen Kiel Curry System* (cited on page 11).

Michael Hanus. 1994. The Integration of Functions into Logic Programming: From Theory to Practice. *Journal of Logic Programming*, 19/20, 583–628 (cited on page 32).

Michael Hanus, Herbert Kuchen, and Juan Jose Moreno-Navarro. 1995. Curry: A truly functional logic language. In *Proceedings of the ILPS '95 Workshop on Visions for the Future of Logic Programming*, 95–107 (cited on page 3).

Michael Hanus, Bernd Braß el, Björn Peemöller, and Fabian Reck. 2014. *KiCS2 - The Kiel Curry System User Manual Version 0.3.2 of 2014-07-30* (cited on page 11).

Ralf Hinze. 2000. A new approach to generic functional programming. In *Proceedings of the 27th ACM SIGPLAN-SIGACT Symposium on Principles of Programming Languages* (POPL '00). ACM, 119–132 (cited on page 24).

Ellis Horowitz. 1983. *Fundamentals of Programming Languages*. (Second edition). Springer (cited on page 3).

Paul Hudak, John Hughes, Simon Peyton Jones, and Philip Wadler. 2007. A History of Haskell: Being Lazy with Class. In *Proceedings of the Third ACM SIGPLAN Conference on History of Programming Languages* (HOPL III). ACM, 12–1–12–55 (cited on page 3).

Graham Hutton. 2016. *Programming in Haskell*. (Second edition). Cambridge University Press (cited on page 3).

Graham Hutton and Diana Fulger. 2008. Reasoning about effects: Seeing the wood through the trees. In *Proceedings of the Ninth Symposium on Trends in Functional Programming* (cited on page 93).

Johan Jeuring, Patrik Jansson, and Cláudio Amaral. 2012. Testing Type Class Laws. In *Proceedings of the 2012 Haskell Symposium* (Haskell '12) number 12. Volume 12. ACM, 49–60 (cited on page 93).

Selmer M. Johnson. 1963. Generation of Permutations by Adjacent Transposition. *Mathematics of Computation*, 17, 83, 282–285 (cited on page 52).

Ohad Kammar, Sam Lindley, and Nicolas Oury. 2013. Handlers in action. In *Proceedings of the 18th ACM SIGPLAN International Conference on Functional Programming* (ICFP '13). ACM, 145–158 (cited on page 131).

Angelika Kimmig, Bart Demoen, Luc De Raedt, Vítor Santos Costa, and Ricardo Rocha. 2011. On the Implementation of the Probabilistic Logic Programming Language ProbLog. *Theory and Practice of Logic Programming*, 11, 2-3, 235–262 (cited on page 57).

Oleg Kiselyov and Hiromi Ishii. 2015. Freer Monads, More Extensible Effects. In *Proceedings of the 2015 ACM SIGPLAN Symposium on Haskell* (Haskell '15). ACM, 94–105 (cited on pages 131, 132).

Oleg Kiselyov and Chung-chieh Shan. 2009. Embedded Probabilistic Programming. In *Domain-Specific Languages* (Lecture Notes in Computer Science). Volume 5658. Springer, 360–384 (cited on page 90).

Oleg Kiselyov and K. C. Sivaramakrishnan. 2016. Eff directly in OCaml. In *Proceedings ML Family Workshop / OCaml Users and Developers Workshops, ML/OCAML 2016, Nara, Japan, September 22-23, 2016* (Electronic Proceedings in Theoretical Computer Science). Volume 285, 23–58 (cited on page 131).

Nicolas Koh, Yao Li, Yishuai Li, et al. 2019. From C to Interaction Trees: Specifying, Verifying, and Testing a Networked Server. In *Proceedings of the 8th ACM SIGPLAN International Conference on Certified Programs and Proofs - CPP 2019* (CPP 2019). ACM Press, 234–248 (cited on page 132).

John Launchbury. 1993. A natural semantics for lazy evaluation. In *Proceedings of the 20th ACM SIGPLAN-SIGACT Symposium on Principles of Programming Languages* (POPL '93). ACM, 144–154 (cited on page 134).

Daan Leijen. 2016. Algebraic Effects for Functional Programming. Technical Report. Microsoft Research, 15 (cited on page 131).

Thomas Letan, Yann Régis-Gianas, Pierre Chifflier, and Guillaume Hiet. 2018. Modular Verification of Programs with Effects and Effect Handlers in Coq. In *Formal Methods* (Lecture Notes in Computer Science). Volume 10951. Springer, 338–354 (cited on page 131).

Sam Lindley, Conor McBride, and Craig McLaughlin. 2017. Do Be Do Be Do. In *Proceedings of the 44th ACM SIGPLAN Symposium on Principles of Programming Languages* (POPL 2017). ACM, 500–514 (cited on page 131).

Conor McBride. 2015. Turing-Completeness Totally Free. In *Mathematics of Program Construction* (Lecture Notes in Computer Science). Volume 9129. Springer, 257–275 (cited on page 131).

Andrew McCallum, Karl Schultz, and Sameer Singh. 2009. FACTORIE: Probabilistic Programming via Imperatively Defined Factor Graphs. In *Advances in Neural Information Processing Systems 22*. Curran Associates, Inc., 1249–1257 (cited on page 57).

Stefan Mehner, Daniel Seidel, Lutz Straß burger, and Janis Voigtländer. 2014. Parametricity and Proving Free Theorems for Functional-Logic Languages. In *Proceedings of the 16th International Symposium on Principles and Practice of Declarative Programming* (PPDP '14). ACM, 19–30 (cited on pages 73, 136).

Yaron Minsky, Anil Madhavapeddy, and Jason Hickey. 2013. *Real World OCaml: Functional Programming for the Masses*. O'Reilly Media, Inc. (cited on page 1).

Eugenio Moggi. 1989. Computational lambda-calculus and monads. In *Proceedings of the Fourth Annual Symposium on Logic in Computer Science* (LICS '89). IEEE Computer Society, 14–23 (cited on page 7).

Arun Nampally, Timothy Zhang, and C. R. Ramakrishan. 2018. Constraint-Based Inference in Probabilistic Logic Programs. *Theory and Practice of Logic Programming*, 18, 3-4, 638–655 (cited on page 76).

Ulf Norell. 2009. Dependently Typed Programming in Agda. In *Proceedings of the 4th International Workshop on Types in Language Design and Implementation* (TLDI '09). ACM, 1–2 (cited on page 98).

Brooks Paige and Frank Wood. 2014. A Compilation Target for Probabilistic Programming Languages. In *Proceedings of the 31st International Conference on Machine Learning*. PMLR, 1935–1943 (cited on page 57).

Judea Pearl. 1988. *Probabilistic Reasoning in Intelligent Systems: Networks of Plausible Inference*. Morgan Kaufmann Series in Representation and Reasoning. Morgan Kaufmann (cited on page 56).

Tomas Petricek. 2012. Evaluation strategies for monadic computations. In *Proceedings Fourth Workshop on Mathematically Structured Functional Programming* (EPTCS). Volume 76, 68–89 (cited on page 134).

Simon Peyton Jones. 2002. *Haskell 98 Language and Libraries: The Revised Report* (cited on pages 1, 3).

Benjamin C Pierce, Chris Casinghino, Marco Gaboardi, et al. Software Foundations. (2010) (cited on page 18).

Maciej Piróg, Tom Schrijvers, Nicolas Wu, and Mauro Jaskelioff. 2018. Syntax and Semantics for Operations with Scopes. In *Proceedings of the 33rd Annual ACM/IEEE Symposium on Logic in Computer Science* (LICS '18). ACM, 809–818 (cited on page 130).

Gordon Plotkin and Matija Pretnar. 2009. Handlers of Algebraic Effects. In *Programming Languages and Systems* (Lecture Notes in Computer Science). Volume 5502. Springer, 80–94 (cited on page 131).

Matija Pretnar. 2015. An Introduction to Algebraic Effects and Handlers. Invited tutorial paper. *The 31st Conference on the Mathematical Foundations of Programming Semantics (MFPS XXXI)*. Electronic Notes in Theoretical Computer Science 319, 19–35 (cited on page 131).

2019. ReasonML. https://reasonml.github.io/. (2019) (cited on page 1).

Colin Runciman, Matthew Naylor, and Fredrik Lindblad. 2008. SmallCheck and Lazy SmallCheck: Automatic Exhaustive Testing for Small Values. In *Proceedings of the First ACM SIGPLAN Symposium on Haskell* (Haskell '08). ACM, 37–48 (cited on pages 1, 89).

Adam Ścibior, Zoubin Ghahramani, and Andrew D. Gordon. 2015. Practical Probabilistic Programming with Monads. In *Proceedings of the 2015 ACM SIGPLAN Symposium on Haskell* (Haskell '15). ACM, 165–176 (cited on page 89).

Robert Sedgewick. 1977. Permutation Generation Methods. *ACM Computing Surveys*, 9, 2, 137–164 (cited on page 51).

Antal Spector-Zabusky, Joachim Breitner, Christine Rizkallah, and Stephanie Weirich. 2018. Total Haskell is Reasonable Coq. In *Proceedings of the 7th ACM SIGPLAN International Conference on Certified Programs and Proofs* (CPP 2018). ACM, 14–27 (cited on page 133).

Nikhil Swamy, Catalin Hritcu, Chantal Keller, et al. 2016. Dependent Types and Multi-Monadic Effects in F*. In *Proceedings of the 43rd Annual ACM SIGPLAN-SIGACT Symposium on Principles of Programming Languages* (POPL '16). ACM, 256–270 (cited on page 131).

Wouter Swierstra. 2008. Data Types à la Carte. *Journal of Functional Programming*, 18, 4, 423–436 (cited on pages 9, 10, 25, 44, 103, 105, 131, 151).

Hale F. Trotter. 1962. Algorithm 115: Perm. *Communications of the ACM*, 5, 8, 434–435 (cited on page 51).

Tarmo Uustalu and Niccolò Veltri. 2017. Partiality and Container Monads. In *Programming Languages and Systems* (Lecture Notes in Computer Science). Volume 10695. Springer, 406–425 (cited on page 102).

Marko van Eekelen and Maarten de Mol. 2006. Proof Tool Support for Explicit Strictness. In *Implementation and Application of Functional Languages*. Lecture Notes in Computer Science. Volume 4015. Springer, 37–54 (cited on page 133).

Twan van Laarhoven et al. 2007. *Haskell Mailing List* (cited on page 52).

Daniele Varacca and Glynn Winskel. 2006. Distributing Probability over Non-determinism. *Mathematical Structures in Computer Science*, 16, 1, 87–113 (cited on page 56).

Philip Wadler. 1997. How to declare an imperative. *ACM Computing Surveys*, 29, 3, 240–263 (cited on page 5).

Philip Wadler. 1985. How to Replace Failure by a List of Successes: A method for exception handling, backtracking, and pattern matching in lazy functional languages. In *Proceedings of the International Conference on Functional Programming Languages and Computer Architecture* (Lecture Notes in Computer Science). Volume 201. Springer, 113–128 (cited on page 89).

Philip Wadler. 2015. Propositions as Types. *Communications of the ACM*, 58, 12, 75–84 (cited on page 94).

Philip Wadler. 1989. Theorems for free! In *Proceedings of the Fourth International Conference on Functional Programming Languages and Computer Architecture* (FPCA '89). ACM, 347–359 (cited on page 53).

Frank Wood, Jan Willem Meent, and Vikash Mansinghka. 2014. A New Approach to Probabilistic Programming Inference. In *Proceedings of the Seventeenth International Conference on Artificial Intelligence and Statistics*. Volume 33. PMLR, 1024–1032 (cited on page 57).

Nicolas Wu, Tom Schrijvers, and Ralf Hinze. 2014. Effect handlers in scope. In *Proceedings of the 2014 ACM SIGPLAN Symposium on Haskell* (Haskell '14). ACM, 1–12 (cited on pages 130, 135).

Li-yao Xia, Yannick Zakowski, Paul He, et al. 2019. Interaction Trees: Representing Recursive and Impure Programs in Coq. *Proceedings of the ACM Programming Languages*, 4, POPL (cited on page 132).

# Appendix A

The appendix provides an auxiliary proof used in Chapter 3, omitted definitions and type class instances of Section 2.1, implementations we used for benchmarks comparisons in Chapter 4 as well as omitted proof scripts for lemmas discussed in Chapter 5.

## A.1 Auxiliary Proof

The following proof is used for the inequation in Section 3.2.3. We show by induction that for $n \geqslant 7$, we have $n \leqslant 2^{\frac{n-1}{2}}$.

- Case $n = 7$.
$$n = 7 < 8 = 2^3 = 2^{\frac{6}{2}} = 2^{\frac{n-1}{2}}$$

- Case $n+1$ with induction hypothesis $n < 2^{\frac{n-1}{2}}$.
$$n + 1 < n + 2^{\frac{1}{2}} < n \cdot 2^{\frac{1}{2}} < 2^{\frac{n-1}{2}} \cdot 2^{\frac{1}{2}} = 2^{\frac{(n-1)+1}{2}} = 2^{\frac{(n+1)-1}{2}}$$

## A.2 Functor Type Class and Instances

In Section 2.1.3 we used a functor constraint for the type parameter `f` in order to define a monad instance `Free`. The following code completes these examples as it shows the definition of the functor type class and defines instances for the functors we used.

```
class Functor f where
  fmap :: (a -> b) ->  f a -> f b

instance Functor Zero where
  fmap f z = case z of

instance Functor One where
  fmap f One = One

instance Functor Choice where
  fmap f (Choice x y) = Choice (f x) (f y)

instance Functor (Const e) where
  fmap f (Const y) = Const y
```

## A.3 Proof Sketch: List is not a Free Monad

Swierstra (2008) states that there is no functor `f` such that type `Free f a` is isomorphic to `[a]`. Using the representation of free monads that we introduce in Section 5.1.3, we present a proof sketch on why the list monad cannot be represented as free monad.[1] Given this

---
[1] The initial idea of this sketch comes from a StackOverflow post by Reid Barton: https://stackoverflow.com/a/24918234/458384 (last access: April 15th, 2020).

representation we have pure computation constructed using pure and effectful computations
are constructed using impure. We call the former computations *trivial* and the latter *non-
trivial*. The proof sketch is now based on the following observation: when composing effectful
computations, we know that if one component is non-trivial, the resulting computation
is non-trivial as well. This property needs to hold for all monads that are isomorphic to
Free Shape Pos for some Shape and Pos.

In Coq, we define the necessary propostions as follows.

```
Inductive IsNonTrivial (A : Type) : Free Shape Pos A -> Prop :=
| IsNonT : ∀ s pf, IsNonTrivial (impure s pf).

Lemma trivialCompBind : ∀ (A B : Type)
  (fx : Free Shape Pos A) (f : A -> Free Shape Pos B),
  IsNonTrivial fx -> IsNonTrivial (fx >>= f).
Proof.
  intros A B fx f IsNonTriv.
  inversion IsNonTriv; subst; simpl.
  apply IsNonT.
Qed.
```

That is, composing a non-trivial computation that is constructed using impure always
leads in a resulting value constructed with impure as well. For lists, the trivial case is the
singleton list and all other values are non-trivial, that is, the empty list as well as all lists
with at least two elements are non-trivial lists. Again, we can define propositions in Coq to
reflect these properties.

```
Inductive IsTrivialList (A : Type) : list A -> Prop :=
| IsTList : ∀ x, IsTrivialList (cons x nil).

Inductive IsNonTrivialList (A : Type) : list A -> Prop :=
| IsNil  : IsNonTrivialList nil
| IsCons : ∀ x y ys, IsNonTrivialList (cons x (cons y ys)).
```

For lists, we can find an example with non-trivial list and a function f that yields a trivial
list as result. In Haskell, the example computation looks as follows.

```
example = do
  b <- [True,False]
  if b then return [1] else []
```

That is, we can construct a trivial result, namely `[1] = return 1`, although we use an
inital non-trivial computation `[True,False]`. We can construct the counterexample in Coq
as follows. Recall that the bind-operation for the list monad is `concat . map`.

```
Definition list1 : list bool := cons true (cons false nil).

Lemma isNonTrivialList1 : IsNonTrivialList list1.
Proof.
  apply IsCons.
Qed.
```

```
Example IsTrivialCounter :
  IsTrivialList
    (concat (map (fun x : bool => if x then cons 1 nil else nil) list1)).
Proof.
  simpl. apply IsTList.
Qed.

Example IsTrivialCounter2 :
  not (IsNonTrivialList
    (concat (map (fun x : bool => if x then cons 1 nil else nil) list1))).
Proof.
  simpl. intro IsNonTriv. inversion IsNonTriv.
Qed.
```

## A.4 Example Programs in Probabilistic Programming Languages

The following code snippets present the translation of the Curry code discussed in Section 4.6 to WebPPL and ProbLog. The code is used to perform the benchmarks that we discuss in the corresponding sections.

### WebPPL

**Bayesian Network**

```
var raining = function() {
    return flip(0.2);
};

var sprinkler = function(r) {
    if (r) {
        return flip(0.01)
    } else {
        return flip(0.4);
    }
};

var grassWet = function(s,r) {
    if (s && r) {
        return flip(0.99);
    } else {
        if (s && !r) {
            return flip(0.9);
        } else {
            if (!s && r) {
                return flip(0.8);
```

```
            } else {
                return flip(0.0);
            };
        }
    };
};

var grassModel = function () {
    var r = raining();
    var s = sprinkler(r);
    var g = grassWet(s,r);
    return { isRaining : r, isSprinklerOn : s, isGrassWet : g };
}

var grassWetWhenRaining = function() {
    var g = grassModel();
    return (g.isRaining && g.isGrassWet);
};

Infer({ model: grassWetWhenRaining
      , method: 'enumerate'
      , maxRuntimeInMS: Infinity});
```

**Replicate Die**

```
var die = function() {
    return uniformDraw([1,2,3,4,5,6]);
};

var replicateDist = function (n, dist) {
    if (n == 0) {
        return uniformDraw([[]]);
    } else {
        return [dist()].concat(replicateDist(n-1,dist));
    }
};

var allSix = function(n) {
    var list = replicateDist(n,die);
    all(function(x) { isSix(x) }, list);
}

var isSix = function(dist) {
    var x = dist;
    return x == 6 ? true : false;
};

Infer({ model: function() { allSix(5) }
      , method: 'enumerate'
      , maxRuntimeInMS: Infinity});
```

**Strings**

```
var pickChar = function () {
    return uniformDraw(['a','b']);
};

var replicateDist = function (n, dist) {
    if (n == 0) {
        return uniformDraw([[]]);
    } else {
        return [dist()].concat(replicateDist(n-1,dist));
    }
};

var randomString = function(n) {
    replicateDist(n, pickChar);
};

var isPalindrome = function(str) {
    return (JSON.stringify(str) == JSON.stringify(str.reverse()));
};

var hasConsecutiveBs = function(str) {
    if (str.length < 2) {
        return false;
    } else {
        return helper(str,0,str.length-1);
    }
}

var helper = function (str,n,max) {
    if (n == max) {
        return false;
    } else {
        if (str[n] == 'b' && str[n+1]  == 'b') {
            return true;
        } else {
            return helper(str,n+1,max);
        }
    }
};
```

**Strings (efficient)**

```
var pickChar = function () {
    return uniformDraw(['a','b']);
};

var palindrome = function(n) {
```

```
        return helper(1,n);
};

var helper = function (n,m) {
    if (n > m) {
        return true;
    } else if (n == m) {
        let x = pickChar();
        return true;
    } else {
        let x1 = pickChar();
        let x2 = pickChar();
        if (x1 == x2) {
            return helper(n+1,m-1);
        } else
            return false;
    }
};

Infer({ model: function () { isPalindrome(randomString(5)) }
      , method: 'enumerate', maxRuntimeInMS: Infinity});
Infer({ model: function () { hasConsecutiveBs(randomString(5)) }
      , method: 'enumerate', maxRuntimeInMS: Infinity});
Infer({ model: function () { palindrome(5) }
      , method: 'enumerate', maxRuntimeInMS: Infinity});
```

**Secret Santa**

```
  var santaGame = function(n) {
  return fromTo(1,n);
};

var fromTo = function(n,m) {
  if (n > m) {
    return [];
  } else if (n == m) {
    return [n];
  } else {
    return [n].concat(fromTo(n+1,m));
  }
};

var remove = function (x,arr) {
  var remove2 = function (x,arr,i) {
    if (i >= arr.length) {
        return [];
    } else {
      var y = arr[i];
      if (x == y) {
```

```
          return remove2(x,arr,i+1);
        } else {
          return [y].concat(remove2(x,arr,i+1));
        }
      }
    }
    return remove2(x,arr,0);
};

var pPicks = function (p,hat) {
  if (hat.length == 0) {
    return { nothing : true }
  } else {
    var x = uniformDraw(hat);
    var hatNew = remove(x,hat);
    return { nothing : false
           , getter : x
           , hatNew : hatNew
           };
  }
};

var pickRound = function(hat) {
  if (hat.length === 0) {
    return {failed : true};
  } else {
    var pickRound2 = function(ps, hat, arrs, n) {
      if (n >= ps.length) {
        return { failed : false
               , assignments : arrs
               };
      } else {
        var giver = ps[n];
        var assgnmnt = pPicks(giver,hat);
        if (assgnmnt.nothing) {
          return { failed : true };
        } else {
          pickRound2( ps
                    , assgnmnt.hatNew
                    , arrs.concat({ giver : giver
                                  , getter : assgnmnt.getter
                                  })
                    , n+1);
        }
      }
    };
    pickRound2(hat,hat,[],0);
```

```javascript
      }
    };

    var pickRound2 = function(hat) {
      if (hat.length === 0) {
        return {failed : true};
      } else {
        var pickRound3 = function(ps, hat, arrs, n) {
          if (n >= ps.length) {
            return { failed : false
                   , assignments : arrs
                   };
          } else {
            var giver = ps[n];
            var newHat = remove(giver,hat);
            var assgnmnt = pPicks(giver,newHat);
            if (assgnmnt.nothing) {
              return { failed : true
                     , assignments : arrs
                     };
            } else {
              pickRound3( ps
                        , remove(assgnmnt.getter, hat)
                        , arrs.concat({ giver : giver
                                      , getter : assgnmnt.getter
                                      })
                        , n+1);
            }
          }
        };
        pickRound3(hat,hat,[],0);
      }
    };

    var isInvalid = function(assgnmnt) {
      return (assgnmnt.failed ||
              any( function (a) { a.getter === a.giver }
                              , assgnmnt.assignments));
    };

    Infer({ model: function() { isInvalid(pickRound(santaGame(len))) }
          , method: 'enumerate'
          , maxRuntimeInMS: Infinity});
    Infer({ model: function() { isInvalid(pickRound2(santaGame(len))) }
          , method: 'enumerate'
          , maxRuntimeInMS: Infinity});
}
```

# ProbLog

**Bayesian Network**

```
0.2  :: raining.
0.01 :: sprinkler :- raining.
0.04 :: sprinkler :- \+raining.
0.8  :: grassWet :- \+sprinkler, raining.
0.9  :: grassWet :- sprinkler, \+raining.
0.99 :: grassWet :- sprinkler, raining.

grassWetWhenRaining :- raining, grassWet.
query(grassWetWhenRaining).
```

**Replicate Die**

```
% We need to add some kind of ID in order to "disable" memoisation.
(1/6)::die(D,1).
(1/6)::die(D,2).
(1/6)::die(D,3).
(1/6)::die(D,4).
(1/6)::die(D,5).
(1/6)::die(D,6).

isSix(N) :- N == 6.
allSix([]).
allSix([X|XS]) :- isSix(X), allSix(XS).

replicateDie(0,[]).
replicateDie(N,[X|XS]) :- N \== 0, die(N,X), N1 is N-1,
                         replicateDie(N1,XS).

dieSix :- die(42,N), isSix(N).
allRepSix(N) :- replicateDie(N,XS), allSix(XS).

query(allRepSix(N)).
```

**Strings**

```
0.5::pick(N, a) ; 0.5::pick(N,b).

random_string(0,[]).
random_string(N,[X|L]) :-
    N > 0,
    pick(N,X),
    NN is N-1,
    random_string(NN,L).

palindrome(L) :- reverse(L,L).
```

```prolog
reverse(L,R) :-
    reverse(L,[],R).
reverse([],L,L).
reverse([A|B],S,R) :-
    reverse(B,[A|S],R).

twoBs([b,b|_]).
twoBs([_|L]) :-
    twoBs(L).

string_is_palindrome(N) :- string_is_palindrome(N,_).
string_is_palindrome(N,L) :- random_string(N,L),palindrome(L).

string_with_bb(N) :- string_with_bb(N,_).
string_with_bb(N,L) :- random_string(N,L),twoBs(L).

len(5).
query(string_is_palindrome(X)) :- len(X).
```

**Strings (efficient)**

```prolog
0.5::pick(N,a) ; 0.5::pick(N,b).

% a palindrome of length N spans positions 1 to N
palindrome(N) :- palindrome(1,N).

% base case for even length: left and right crossed
palindrome(A,B) :- A > B.
% base case for uneven length: arbitrary middle character
palindrome(N,N) :- pick(N,X).
% recursive case: add same character at both ends and move
%  positions towards the middle
palindrome(A,B) :- A < B, pick(A,X), pick(B,X),
                AA is A+1, BB is B-1, palindrome(AA,BB).

bb(N) :- Max is N-1, between(1,Max,I), pick(I,b),
        II is I+1, pick(II,b).

len(30).
query(palindrome(X)) :- len(X).
```

**Secret Santa**

```prolog
:- use_module(library(lists)).

% reuse select_uniform as it's not trivial to define
% uniform/2 with the expected behaviour
uniform([X|XS],Y) :- select_uniform(42,[X|XS],Y,ZS).
```

```prolog
pPicks(P,Hat,V) :- uniform(Hat,V), V \== P.

pickRound(Hat,Arrs) :- pickRound(Hat,Hat,Arrs).
pickRound([],_,[]).
pickRound([P|Ps], Hat, [(P,V)|Arrs]) :-
  pPicks(P,Hat,V), delete(Hat,V,HatNew), pickRound(Ps,HatNew, Arrs).

ppPicks(P,Hat,just(V)) :- uniform(Hat,V), V \== P.
ppPicks(P,Hat,nothing) :- uniform(Hat,V), V = P.

ppickRound(Hat,Arrs) :- ppickRound(Hat,Hat,Arrs).
ppickRound([],_,[]).
ppickRound([P|Ps], Hat, [failedGame]) :- ppPicks(P,Hat,nothing).
ppickRound([P|Ps], Hat, [(P,V)|Arrs]) :-
  ppPicks(P,Hat,just(V)), delete(Hat,V,HatNew), ppickRound(Ps,HatNew, Arrs).

is_pair((X,Y)).

allValid([]).
allValid([X|Xs]) :- is_pair(X), allValid(Xs).

anyFailed(Xs) :- member(failedGame,Xs).

hat(2,[2,1]).
hat(N,[N|Xs]) :- N > 1, M is N-1, hat(M,Xs).

% clever pick: a person cannot pick herself
% invalid games are just `false`
pickRound2(Hat,Arrs) :- pickRound2(Hat,Hat,Arrs).
pickRound2([],_,[]).
pickRound2([P|Ps], Hat, [(P,V)|Arrs]) :-
  delete(Hat,P,HatTemp), pPicks(P,HatTemp,V), delete(Hat,V,HatNew),
  pickRound2(Ps,HatNew, Arrs).

% clever pick: a person cannot pick herself
% invalid games are tracked as well
ppickRound2(Hat,Arrs) :- ppickRound2(Hat,Hat,Arrs).
ppickRound2([],_,[]).
ppickRound2([P],[P],[failedGame]).
ppickRound2([P|Ps], Hat, [(P,V)|Arrs]) :-
  delete(Hat,P,HatTemp), pPicks(P,HatTemp,V), delete(Hat,V,HatNew),
  ppickRound2(Ps,HatNew, Arrs).

isValid([],true).
isValid([X|Xs],false) :- X == failedGame.
isValid([X|Xs],Bool) :- X \= failedGame, isValid(Xs,Bool).
```

```
santa(N,Bool) :- hat(N,Xs), ppickRound(Xs,Arrs), isValid(Arrs,Bool).
santa2(N,Bool) :- hat(N,Xs), ppickRound2(Xs,Arrs), isValid(Arrs,Bool).

query(santa(3,X)).
query(santa2(3,X)).
```

## A.5 Formal Reasoning

The following code provides proof scripts of lemmas that we omitted in the corresponding sections.

### A Container Extension is a Functor

Given the following definition of `fmap` for a container extension `Ext`

```
Definition fmap (Shape : Type) (Pos : Shape -> Type) (A B : Type)
  (f : A -> B) (e: Ext Shape Pos A) : Ext Shape Pos B :=
  match e with
  | ext s pf => ext s (fun p => f (pf p))
  end.
```

we show that `fmap` fulfils the functor laws.

```
Section FunctorLaws.

  Variable Shape : Type.
  Variable Pos : Shape -> Type.
  Variable A B C : Type.

  Lemma fmap_id : ∀ (e : Ext Shape Pos A),
      fmap (fun x => x) e = e.
  Proof.
    intros [ s pf ]; reflexivity.
  Qed.

  Lemma fmap_compose : ∀ (f : B -> C) (g : A -> B) (e : Ext Shape Pos A),
      fmap f (fmap g e) = fmap (fun x => f (g x)) e.
  Proof.
    intros f g [ s pf ]; reflexivity.
  Qed.

End FunctorLaws.
```

### Non–Determinism as Effect

We discussed the representation of non–determinism as well as the definition of the corresponding container extension in Section 5.3.1. As we have seen isomorphism proofs for `Zero` and `One` in preceding sections, we omitted the proof for the isomorphism between the functor `ND` and the container extension. The following code proves that the functions `from_ND` and `to_ND` indeed form an isomorphism.

```
Lemma from_to_ND : ∀ (A : Type) (e : Ext ND_S ND_P A),
   from_ND (to_ND e) = e.
Proof.
  intros A e. destruct e as [s pf]; simpl.
  dependent destruction s; simpl.
  - f_equal; extensionality p; destruct p;
    reflexivity.
  - f_equal; extensionality p; destruct p.
Qed.

Lemma to_from_ND : ∀ (A : Type) (nd : ND A),
   to_ND (from_ND nd) = nd.
Proof.
  intros A nd. destruct nd; reflexivity.
Qed.
```

## Proving Properties About Non–deterministic Functions

We discussed a variety of exemplary propositions about non–deterministic programs in Section 5.3.1. The following code provides the omitted proofs for these examples if not already discussed right away.

```
Lemma even_oneOrTwo :
  even (liftM2 mult oneOrTwo (pure 2)) = TTrue ? TTrue.
Proof.
  simpl even.
  f_equal; extensionality p.
  destruct p; simpl liftM2, even; reflexivity.
Qed.

Lemma even_oneOrTwo_allND :
  AllND (fun fb => fb = TTrue) (even (liftM2 mult oneOrTwo (pure 2))).
Proof.
  simpl; constructor.
  intros p; destruct p; repeat constructor.
Qed.

Lemma pulltab_inc : ∀ (fx fy : FreeND nat),
   inc (fx ? fy) = inc fx ? inc fy.
Proof.
  intros fx fy; simpl.
  f_equal; extensionality p.
  destruct p; reflexivity.
Qed.

Lemma pulltab_f_strict : ∀ (A B : Type)
  (f : FreeND A -> FreeND B) (fx fy : FreeND A)
  (Hstrict : forall fz, f fz = fz >>= fun z => f (pure z)) ->
```

```
    f (fx ? fy) = f fx ? f fy.
Proof.
  intros A B f fx fy Hstrict.
  rewrite (Hstrict (fx ? fy)); simpl.
  f_equal; extensionality p.
  destruct p; rewrite Hstrict; reflexivity.
Qed.

Lemma ndInsert_inc :
  AllND (fun fxs => length fxs = pure 3)
        (nfList (ndInsert (pure 1) (Cons (pure 1) (Cons (pure 2) Nil)))).
Proof.
  simpl; constructor.
  intros p; destruct p; simpl.
  - constructor; reflexivity.
  - constructor; simpl.
    intros p; destruct p; simpl;
      constructor; reflexivity.
Qed.
```

## Sharing as Effect

We present the missing proof scripts for lemmas regarding to modelling sharing as discussed in Section 5.3.2.

```
Lemma doublePlus_inline :
  doublePlus oneOrTwo = liftM2 plus oneOrTwo oneOrTwo.
Proof.
  reflexivity.
Qed.

Lemma even_doublePlus_Choice :
  even (doublePlus oneOrTwo) = (TTrue ? FFalse) ? (FFalse ? TTrue).
Proof.
  simpl. f_equal; extensionality p.
  destruct p; simpl.
  - f_equal; extensionality p.
    destruct p; reflexivity.
  - f_equal; extensionality p.
    destruct p; reflexivity.
Qed.

Lemma doubleMult_inline :
  doubleMult oneOrTwo = pure 2 ? pure 4.
Proof.
  simpl. f_equal. extensionality p.
  destruct p; reflexivity.
Qed.
```

```
Lemma even_doublePlus_Failed :
  even (doublePlus Failed) = Failed.
Proof.
  simpl. f_equal; extensionality p.
  destruct p.
Qed.

Lemma even_doubleSharePos_Choice :
  even (doubleSharePos oneOrTwo) = TTrue ? TTrue.
Proof.
  simpl. f_equal; extensionality p.
  destruct p; reflexivity.
Qed.

Lemma even_doubleShare_Choice :
  even (doubleShare oneOrTwo) = TTrue ? TTrue.
Proof.
  simpl. f_equal; extensionality p.
  destruct p; reflexivity.
Qed.

Lemma share_with_const : ∀ (A : Type) (fx : FreeND A),
  (shareStrict oneOrTwo >>= fun fy => const (const fx fy) fy)
  = fx ? fx.
Proof.
  intros A fx. simpl.
  f_equal; extensionality p; destruct p; reflexivity.
Qed.
```

# Erklärung

Die vorliegende Dissertation ist nach Inhalt und Form meine eigene Arbeit. In den jeweiligen Kapiteln wurden die Publikationen vermerkt, auf dessen Grundlage die entsprechenden Inhalte entstanden sind. Die vorliegende Dissertation wurde weder ganz noch in Teilen an anderer Stelle im Rahmen eines Prüfungsverfahrens vorgelegt und ist unter Einhaltung der Regeln guter wissenschaftlicher Praxis der Deutschen Forschungsgemeinschaft entstanden.

_______________________________

M.Sc. Sandra Dylus